%% file: copiathesis.tex
\def\re#1{(\ref{#1})}
\def\beq{\begin{equation}}
\def\eeq{\end{equation}}
\def\beeq{\begin{eqnarray}}
\def\beeqn{\begin{eqnarray*}}
\def\eeeq{\end{eqnarray}}
\def\eeeqn{\end{eqnarray*}}

\def\de{\delta}                 \def\D{\Delta}

\def\l{\lambda}                 
\def\m{\mu}
\def\n{\nu}

\newcommand{\OO}{{\cal O}}

\newcommand{\WW}{{\cal W}}

\newcommand{\lp}{\left(}
\newcommand{\rp}{\right)}
\renewcommand{\lq}{\left[}
\renewcommand{\rq}{\right]}

\newcommand{\no}{\nonumber}

\def\tr{\,\mbox{Tr}\,}
\def\frac#1#2{ {{#1} \over {#2} }}

\def\p{\partial}

\newcommand{\unity}{1\kern-.65mm \mbox{\form l}}
\newcommand{\ks}{\mbox{\form l}\kern-.6mm \mbox{\form K}}
\newcommand{\A}{A \raise0.5mm\hbox{\kern-1.8mm /}}
\def\pmb#1{\leavevmode\setbox0=\hbox{$#1$}\kern-.025em\copy0\kern-\wd0
\kern-.05em\copy0\kern-\wd0\kern-.025em\raise.0433em\box0}
\def\D{\hbox{\hbox{${D}$}}\kern-1.9mm{\hbox{${/}$}}}
\def\kbar{\hbox{$k$}\kern-0.2true cm\hbox{$/$}}
\def\nbar{\hbox{$n$}\kern-0.23true cm\hbox{$/$}}
\def\pbar{\hbox{$p$}\kern-0.18true cm\hbox{$/$}}
\def\nhbar{\hbox{$\hat n$}\kern-0.23true cm\hbox{$/$}}

\documentclass [12pt]{book}
\usepackage{amsfonts}
\usepackage[dvips]{graphicx}
\usepackage{latexsym,amsmath,amssymb,graphics,stmaryrd}
\usepackage{graphics}
\usepackage{subfigure}
\usepackage{fancyhdr}
\usepackage[dvips]{color}
\include{colors}
\newcommand{\ce}{\centerline}
\newcommand{\me}{\medskip}
\newcommand{\bi}{\bigskip}
\begin{document}
\thispagestyle{empty}
\centerline{\bf Alessandro Torrielli}

\centerline{\rm Department of Physics ``Galileo Galilei'', via F. Marzolo 8, 35131 Padova}

\centerline{INFN, sezione di Padova. E-mail: torrielli@pd.infn.it}

\bi

\centerline{\it Noncommutative perturbative quantum field theory: Wilson} 

\centerline{\it loop in two-dimensional Yang-Mills, and unitarity from string theory}\rm

\bi

\centerline{\bf PhD thesis} 

\ce{\rm Universit\`a degli Studi di Padova} 

\ce{December 2002} 

\bi

\ce{\bf Advisor: \rm Prof. A. Bassetto \ \ \ \ \ \ \ \ \ \bf Collaborator: \rm Dr. G. Nardelli}

\bigskip

\bi

\ce{\bf Abstract \rm}

The results of our research on noncommutative perturbative quantum field theory and its relation to string
       theory are exposed with details. 1) We give an introduction to noncommutative quantum field theory and its
       derivation from open string theory in an antisymmetric background. 2) We perform a perturbative Wilson loop
       calculation for 2D NCYM. We compare the LCG results for the WML and the PV prescription. With WML
       the loop is well-defined and regular in the commutative limit. With PV the result is singular. This is intriguing:
       in the commutative theory their difference is related to topological excitations, moreover PV provides a
       point-like potential. 3) Commutative 2D YM exhibits an interplay between geometrical and U(N) gauge
       properties: in the exact expression of a Wilson loop with n windings a scaling intertwines n and N. In the NC
       case the interplay becomes tighter due to the merging of space-time and ``internal'' symmetries. Surprisingly,
       in our up to ${\cal{O}}(g^6)$ (and beyond) crossed graphs calculations the scaling we mentioned occurs for large n, N
       and theta. 4) We discuss the breakdown of perturbative unitarity of noncommutative electric-type QFT in the
       light of strings. We consider the analytic structure of string loop two-point functions suitably continuing them
       off-shell, and then study the Seiberg-Witten limit. In this way we pick up how the unphysical
       tachyonic branch cut appears in the NC field theory.

\me
DFPD 03/TH/04

\vfill{\break}

\thispagestyle{empty}

\vfill{\break}
\tableofcontents

\chapter{Introduction and Summary of Results}

\section{Introduction}
Noncommutative geometry\cite{Connes} has received, in the last years, a lot of attention and a renewed interest from the community of theoretical physicists. The idea that, at the scale of unification of gravity with quantum theory, the structure of space-time could be no longer that of a differentiable manifold, but rather a more complicated one, matches with the possibility that coordinates become noncommuting operatorial degrees of freedom, describing a non-local fundamental scenario\cite{dopli}. Such concepts have received a boost, when it was realized that noncommutative field theories can be derived as an effective description of string/M theory in antisymmetric backgrounds\cite{cds,dh,sw}. Since string theory proposes itself as the most powerful candidate for a unified theory of fundamental interactions, its connection with noncommutative QFT enhanced its relevance in describing the crucial aspects of quantum space-time at the Planck scale. 

Since then, a big amount of literature has been devoted to study the properties of noncommutative field theories, both in connection with the low-energy sector of string theory, and in their own, as particular examples of exotic QFT. We refer the reader to the beautiful existing reviews\cite{Harveyrev,Douglasrev,Szaborev}, and to the introductive chapters of this thesis, where some of the relevant aspects of the derivation from open string theory, and some peculiar features of perturbation theory, are summarized. What one immediately learns is the fact that these theories contain a number of properties which, in a sense, reveal a more stringy-like character rather than a QFT-like one. They are non-local and reveal a very rich quantum spectrum, which includes, or, better, is more properly interpreted as made of extended string-like objects. These features, together with the relative ease in building the perturbative series using Feynman diagrams with the standard machinery of QFT, make them a unique opportunity to test ``deformations'' of quantum field theory. These modifications typically lead to so dramatic effects on the internal consistency of QFT, that one usually has to face the loss of most basic requirements.

One realizes, therefore, that, in its own, noncommutative quantum field theory does provide a greatly manageable and extraordinarily powerful setting in which string theory can be studied from a simplified point of view, maintaining many important aspects of the original setup. Still, it may run into inconsistencies. Our work has been inspired by the twofold attempt to find the boundary to which noncommutative field theory can be pushed inside the framework of QFT (and we will provide an example of a consistent calculation which can reveal the great geometrical richness of even the simple two-dimensional noncommutative Yang-Mills theory), and then to definitely turn to string theory, in order to clarify the reasons for such unusual behaviours. The main themes and keys of reading are, on one hand, the perturbative investigation of the structure of the larger gauge group that underlies noncommutative field theories, in which space-time transformations and internal symmetries merge. Among the features inherited from strings, the most striking one is probably that, like in any theory which should unify gravity with Yang-Mills theories, a large symmetry group is at work. It is a very lucky circumstance that this can be analysed in a field theory context. On the other hand, we devote our attention to infer completely from string theory the features of the spectrum of noncommutative quantum field theory. We think we have found some interesting ingredients for the explanation of long-standing problems, looking at some one-loop open string scattering amplitudes, along with their continuation off-shell and to their field theory limit.      

\section{Summary of Results}
We have explored the noncommutative $U(N)$ Yang-Mills theory using the tool provided by the Wilson loop. In our analysis, we chose to work in a two-dimensional noncommutative space-time, because, in this situation, one can perform a comparison with exact nonperturbative commutative results, and also because in two dimensions some symmetries are necessarily preserved even in the presence of noncommutativity, such as Lorentz invariance, and also invariance under area-preserving diffeomorphisms is into play. This allowed us to explore some general geometrical features; in particular, the larger gauge group underlying noncommutative theories should contain in two dimensions the group of area-preserving diffeomorphisms itself. This is therefore a privileged setting where the effects of the merging of space-time and internal symmetries can be studied.

Our first step was to calculate the first non-trivial order of the perturbative expansion of the Wilson loop, concentrating our attention on the contribution from the crossed diagram, the one which is affected by noncommutativity. We chose the light-cone gauge, and compared the results obtained when using the Wu-Mandelstam-Leibbrandt (WML) and the Cauchy principal value (PV) prescriptions for the vector propagator, remembering that, in ordinary theories on compact manifolds, the difference between the two cases can be traced back to the contribution of topological excitations. With the WML prescription, the $\theta$-dependent term is well-defined and regular, and continuous in the limit $\theta \to 0$, where the commutative theory is recovered. This is at odds with the usual behaviour in higher dimensions, where one typically finds singularities when going to the commutative limit. However, this region is approached in a non-analytic way, revealing that $\theta=0$ remains a singular value. Conversely, it was immediately noticed that the natural variable in which to express the dependence on noncommutativity is $1/\theta$. The loop has a finite value at $1/\theta=0$, corresponding to maximal noncommutativity: this was an explicit proof that in $d=2$ the non-planar diagrams are not suppressed at large $\theta$, at odds with what happens in higher dimensions. This feature has also been confirmed in subsequent works appeared in literature. 

Our WML results provided an example of a consistent non-trivial calculation when non-commutativity involves the time variable. But the most striking point was that, unexpectedly, the result obtained with the PV prescription for the gauge propagator differs from the WML one only by the addition of two terms, with a trivial $\theta$-dependence, but singular: 
$$
{\cal{W}}_{PV} \, = \, \infty
$$
Actually, the series expansions in the variable $1/\theta$ coming from the PV and the WML prescriptions coincide, apart from these two trivial (infinite) contributions\footnote{We refer the reader to the remarks we make about this point in Section 4.3.2}, which anyway can be ascribed to the delta-like term which distinguishes the two prescriptions, and which represents a contact interaction. This therefore brought us to conclude that noncommutativity of space-time is incompatible with an instantaneous (\it point\rm-like) potential, like the one of the PV prescription. 

We then extended the previous results to the case of $n$-windings around the loop. In this way, it was possible to introduce a procedure which relates transformations over the noncommutative space-time base manifold, and transformations over the fiber of the internal group. This was the key in order to explore the structure of the merging we mentioned, from the point of view of perturbation theory. An interesting feature of commutative Yang-Mills theories in 1+1 dimensions was already a non-trivial interplay between the geometrical properties and $U(N)$ gauge structures: in the exact expression of a Wilson loop with $n$ windings, a symmetric scaling intertwines $n$ and $N$. The WML expression has instead a trivial abelian-like dependence on $n$, which does not involve the rank of the gauge group $N$. In the non-commutative case the interplay must be present in the very deep structure of the theory. We pushed our calculations to the next order, where technical difficulties grow exponentially, and found that in the contribution from the crossed graphs (the genuine non-commutative terms), an intricate mixing exists between $n$ and $N$ already in the perturbative WML context, which is a clear signal of this merging. We were able to work out the large $\theta$- limit, and found again that it has a finite value. Rather surprisingly, the symmetric scaling we mentioned is recovered for large $n$ and $N$ in the limit of maximal non-commutativity:
$$
{\cal{W}}(n, N, A) \, = \, {\cal{W}}(N, n, A n/N)
$$
Winding $n$ times around an area $A$ for a group $U(N)$ is like to wind $N$ times around a rescaled area for a group $U(n)$. We presented arguments in favour of the persistence of such a scaling in this regime at any perturbative order, and succeeded in summing the related perturbative series. The leading behaviour of the $\theta$-dependent term is provided by the diagrams with a single crossing and then a ladder of non intersecting propagators. 

During the last year we focused on the direct relation between noncommutative field theories and string theory, and analysed the problem of unitarity. This problem was faced since the appearance of these theories as an effective description for open string theory amplitudes in an antisymmetric constant background. The absence of a straightforward Hamiltonian formulation for the case of an electric-type noncommutativity (in which the time variable is involved) and the bizarre dynamical features of its scattering amplitudes have soon casted doubts on the consistency of this kind of theories. One of the first confirmations of the breakdown of perturbative unitarity was given for a noncommutative scalar theory: Cutkoski's rules were found to hold only in the magnetic case; in the electric case an additional tachyonic branch cut is present. An analogous result was found for noncommutative gauge theories. New possible states were considered, under the claim that unitarity could be recovered provided the appropriate intermediate exchanges are added, but these states are necessarily tachyonic. This is related to the fact that in the electric case the Seiberg-Witten limit does not succeed in decoupling all or a part of the intermediate string exchanges. From the point of view of the behaviour of open strings in electric backgrounds one sees that there is a critical value of the field beyond which the string becomes unstable. The limit of Seiberg and Witten precisely forces the electric field to overcome this critical value; therefore the corresponding field theory seems to be related to an unstable string. 

Our aim was to clarify this situation from the point of view of the analytical structure of the full string theory two-point function, precisely looking at what happens to the branch cuts when one performs the Seiberg-Witten limit. We took as an example the scalar amplitude, which can be derived as a limit of the two-point tachyon amplitude in open string theory, suitably extended off-shell. One also needs to go to sufficiently low values of the dimension $d$ of the brane at which the open string is attached, in order to deal with non-negative masses. In particular, we went to $d=2$: the effective brane-worldvolume theory is a two-dimensional NC $\phi^3$ theory, and in two dimensions again we have necessarily electric-type noncommutativity. We are therefore in the right situation to study the above mentioned phenomena. We explained why it is necessarily problematic to go down from the string theory amplitude to the field theoretical one: in the full string theory amplitude, one has two branch cuts in the complex plane, both positive below the critical value of the electric field; however one of them is parameterized by a quantity that changes sign when the electric field overcomes its critical value, as it happens in the effective limit. This practically produces the appearance of an unphysical tachyonic branch cut in the noncommutative field theory limit. 

Once explained this mechanism, one can interpret these results in the light of the nature of the two branch cuts. One of them is driven by the open string metric and connected with the open string exchange, while the other one is reminiscent of the closed string channel. Our example confirms and cleanlily enforces arguments already appeared in the literature about the role of the closed string sector in the interpretation of noncommutative thresholds. We also calculated the discontinuities across these cuts, in a suitable approximation. 

\chapter{String Theory Derivation}

\section{Introduction}
In this section we show one way in which noncommutative field theories can be derived as an effective low-energy description of a string theory in an antisymmetric background. Let us start considering the sigma-model action for an open string in the presence of a constant background:
\begin{eqnarray}
\label{sigma}
S={{1}\over{4\pi {\alpha}'}}\int_{\Sigma}d^2 \sigma \, (\sqrt{\gamma} \, \gamma^{a b} \, g_{\mu\nu} \, {\partial}_a X^{\mu} {\partial}_b X^{\nu} - 2 i \pi {\alpha}' B_{\mu\nu} \, {\epsilon}^{ab}{\partial}_a X^{\mu} {\partial}_b X^{\nu}).   \end{eqnarray}
Here $\Sigma$ represents the two-dimensional string worldsheet, parameterized by coordinates $\sigma^a$. In the following, we will interpret this action as describing a bosonic open string coupled with a constant closed-string background. We report for completeness the sigma-model action for a bosonic open string coupled with the massless closed sector:
\begin{eqnarray}
\label{sigmacompl}
S&=&{{1}\over{4\pi {\alpha}'}}\int_{\Sigma}d^2 \sigma \, (\sqrt{\gamma} \, \gamma^{a b} \, g_{\mu\nu}(X) \, {\partial}_a X^{\mu} {\partial}_b X^{\nu} \nonumber - \, 2 i \pi {\alpha}' B_{\mu\nu}(X) \, {\epsilon}^{ab}{\partial}_a X^{\mu} {\partial}_b X^{\nu} \\
&& \qquad \qquad \qquad \qquad \qquad \qquad \qquad  \qquad \qquad + \, \, \sqrt{\gamma} \, R^{(2)} \, \phi(X)),     
\end{eqnarray}
where $g_{\mu\nu}(X)$ is the graviton field, $B_{\mu\nu}(X)$ the antisymmetric tensor and $\phi(X)$ is the dilaton field. $R^{(2)}$ is the curvature of the two-dimensional worldsheet metric $\gamma$. For constant backgrounds, which means taking the expectation value of the background fields, the first two pieces give the action in (\ref{sigma}), while the third one gives as usual the string coupling constant of the topological expansion of string diagrams. Finally, one should also add the coupling to the background representing the massless open sector: we will consider this term later, when we will show how to derive noncommutative gauge theories. 

There are a series of assumptions we will make now\cite{sw}: first, we will work at tree-level in all this chapter, therefore we are allowed to fix the gauge $\gamma_{a b} = \eta_{a b}$, where $\eta_{a b}$ is the two-dimensional flat Minkowski metric. Then, we will consider the presence of a $Dp$-brane lying in the first $p+1$ dimensions of the target space, which is 26-dimensional for us\footnote{See \cite{tasi} }. We will use the following notation:
\begin{eqnarray}
\label{static}
\mu, \nu =0, \cdots, 25 \qquad i,j = 0,\cdots,p \qquad m,n =p+1, \cdots, 25.
\end{eqnarray}
We will take the closed metric $g_{\mu\nu}$ to be block-diagonal with respect to this splitting\cite{dgo}. Furthermore, we take the antisymmetric field $B_{\mu\nu}$ to be zero outside the brane. This does not lead to a loss of generality, as we will see in a short. On the contrary, one could ask if it is possible to get rid even of the $B$-field on the brane, by making a gauge transformation. We recall that on the brane-worldvolume one has a $U(1)$ gauge field $A^i$\cite{Witten} (we will return on this point in the following), which couples to the string such that the action has a gauge invariance $A \rightarrow A + \Lambda$, $B \rightarrow B - d\Lambda$ for any one-form $\Lambda$. Trying to eliminate $B$ would produce anyway an $A$ on the brane.     
   
We will mention two different approaches one can follow, in order to prove that, in this situation, the worldvolume of the brane is effectively noncommutative. The first one is based on the path-integral approach, and shows that the low-energy effective field theory for the string in this background on the worldvolume of the brane is a noncommutative field theory. The second one is based instead on the canonical quantization of the string worldsheet, and shows that the commutation relations between the coordinate operators of the ending points of the string get modified by the presence of the term containing the $B$-field in the action, and become different from zero. Since these points are attached to the brane, this means that the brane itself becomes a noncommutative manifold.

\section{Path Integral Approach}
In this section we describe how one can derive tree-level amplitudes of noncommutative field theory from our open string theory in the path integral approach. The discovery of this fact is traced back to the paper \cite{sw}, which, along with other previous papers\cite{cds,dh}, renewed the interest in noncommutative geometry applied at quantum theory, opening to it new powerful perspectives.

\subsection{Tree-Level Amplitudes}
Let us begin considering a tree-level open string amplitude described by the action (\ref{sigma})\cite{sw}. We will conformally map the disc worldsheet to the complex upper plane by using the following relations:
\begin{eqnarray}
\label{mappa}
&&z = e^{\tau + i \sigma} \qquad  \bar{z} = e^{\tau - i \sigma} \qquad \Im{z}\geq 0, \nonumber\\
&&\partial_{\tau} = {{1}\over {2}} (z\partial_z + \bar{z}\partial_{\bar{z}}) \qquad \partial_{\sigma} = -{{1}\over {2i}} (z\partial_z - \bar{z}\partial_{\bar{z}}), 
\end{eqnarray}
where we have called as usual $\sigma^0 = \tau$, $\sigma^1 = \sigma$.

We will now restrict ourselves to the indices on the brane, disregarding what happens outside. One immediately realizes that the term containing the $B$-field can be written as a boundary term, namely, in the old $\sigma^i$ coordinates, as $- {{i}\over {2}} \int_{\partial'\Sigma} B_{i j} \, X^{i}\, \partial_{\tau} X^{j}$, where we have kept only the contribution along the boundary $\partial'\Sigma :\{\sigma =0,\pi\}$. This would not be the case for a non-constant $B$, and this also shows that a $B$ outside the brane would be irrelevant, since we choose $\partial_{\tau} X^{m} = 0$ in $\partial'\Sigma$. A term like $- {{i}\over {2}} \int_{\partial'\Sigma} B_{i j} \, X^{i}\, \partial_{\tau} X^{j}$ does not modify the equations of motions, which remain
\begin{eqnarray}
\label{eom}
g_{ij}\, \partial_z \, \partial_{\bar{z}}  X^{j} \, = \, 0,
\end{eqnarray}  
but it modifies the boundary conditions in the following way:
\begin{eqnarray}
\label{warning}
g_{ij} \, (\partial_z \, - \, \partial_{\bar{z}})  X^{j} \, + \, 2 \pi \alpha' \, B_{ij} \, (\partial_z \, + \, \partial_{\bar{z}})  X^{j} \, \, = \, \, 0 
\qquad \, z \, = \, \bar{z}.
\end{eqnarray}
We have in fact that the boundary of the string worldsheet $\partial'\Sigma :\{\sigma =0,\pi\}$ is mapped through the (\ref{mappa}) on the real axis of the complex $z$-plane. The boundary conditions (\ref{warning}) are a deformation of the Neumann's ones, and reduce to Neumann's when the $B$-field is absent. Outside the brane one has of course Dirichlet boundary conditions.

The propagator obtained from these equations of motions, and satisfying these new boundary conditions, is the following: \footnote{The reader is referred for this to \cite{ft,acny,clny}} 
\begin{eqnarray}
\label{prop}
<X^i (z) \, X^j (z')> \, &=& \, - \alpha' \Big[ D^{ij} \, + \, {[g^{-1}]}^{ij} \log |z-z'| \, - \, {[g^{-1}]}^{ij} \log |z-\bar{z}'| \, + \nonumber \\
&&{[G^{-1}]}^{ij} \, \log {(z- \bar{z}')}^2 + \, {{{{\theta}^{ij}}}\over {2\pi\alpha'}} \log {{z- \bar{z}'}\over {\bar{z} - z'}}\Big],
\end{eqnarray} 
where $G$ is the ``open string metric''
\begin{eqnarray}
\label{openmetric}
G = (g-2\pi {\alpha}' B)g^{-1} (g+2\pi {\alpha}' B)
\end{eqnarray}
and $\theta$ is the ``noncommutativity parameter'' 
\begin{eqnarray}
\label{noncomm}
\theta = -{(2\pi{\alpha}' )}^2 {(g+2\pi {\alpha}' B)}^{-1}B {(g-2\pi {\alpha}' B)}^{-1}.
\end{eqnarray}
We will assume that the branch cut of the logarithm is drawn along the positive real axis, so that the string worldsheet is half of the Riemann sheet. 

In order to compute scattering amplitude, it is necessary to insert suitable vertex operators, representing states of incoming and outgoing particles belonging to the string spectrum, at the boundary of the string worldsheet, which, in this case, as we have already mentioned, is the real axis. What will be relevant for string amplitudes will be therefore the propagator (\ref{prop}) evaluated at the boundary. We set $z=s$ and $z'=s'$ with $s$ and $s'$ real, and get
\begin{eqnarray}
\label{propbound}
<X^i (s) \, X^j (s')> \, = - \alpha'  {[G^{-1}]}^{ij} \, \log {(s-s')}^2 + {{i}\over {2}} \, {\theta}^{ij} \Theta (s-s'),
\end{eqnarray} 
where the constant $D^{ij}$ has been set for convenience equal to ${{- i}\over {2 \alpha'}} \, {\theta}^{ij}$, and $\Theta$ is the sign-function. 

One remark one can do about this result is the following: imagine you try to evaluate the quantity $<[X^i (s) , X^j (s')]>$ when $s'$ is very near to $s$, let us say $s = s'+\delta$ with a small positive $\delta$. This has to do with the OPE between the coordinates on the brane\cite{Schom}. Then we have from (\ref{propbound})
\begin{eqnarray}
\label{OPE}
<[X^i , X^j ]> \, = \, i \, \theta^{\, ij}. 
\end{eqnarray}
This is the first evidence that the boundary of the string worldsheet. i.e. the brane worldvolume, is described by a noncommutative manifold.

Another remark is that, in the propagator (\ref{propbound}), the open string metric (\ref{openmetric}) has replaced the metric $g$ in front of the coefficient of the logarithm. This means that the free mass-shell condition is now fixed by the open string metric $G$, through the usual considerations about the anomalous dimension of the vertex operators\footnote{For precisations we refer the reader to Chapter 6}.

Once we know the propagator, we can build the generic tree-level string amplitude in the usual way: we consider a vertex operator of the form 
\begin{eqnarray}
\label{vertex}
V(k) = \int ds \, \, P(\partial X )\, e^{\, i k X},
\end{eqnarray}
where $P$ is a suitable polynomial in the derivatives of the $X$'s with suitable polarization tensors. The coordinates and the momentum will be taken along the brane, and normal ordering will be understood. The scalar product in the exponent and saturation inside $P$ is always made using the open string metric, and integration is along the boundary. One has to take the vacuum expectation value of the correct number of such operators, weighted with the string action, namely
\begin{eqnarray}
\label{ampl}  
< \int ds_1 \, \cdots \, ds_n \, \prod_{m=1}^n \, P_m (\partial X) e^{\, i k_m X}{>}_{(G,\theta)}.
\end{eqnarray}
When weighted with the action (\ref{sigma}), and after standard path integral manipulations, the correlator (\ref{ampl}) becomes exponentials of the propagator (\ref{propbound}) multiplied by some sources. The subscript $(G,\theta)$ signals the two relevant parameters of the propagator (\ref{propbound}). From this consideration, and from the fact that the propagator (\ref{propbound}) is a sum of two pieces, we see immediately that the amplitude factorizes as follows:
\begin{eqnarray}
\label{amplfact}
&&\int ds_1 \, \cdots \, ds_n \, e^{\, - {{i}\over {2}} \sum_{m_1 <m_2 } \, k_{m_1} \theta k_{m_2} \, sign(s_{m_1} - s_{m_2})} \nonumber \\
&& \, \times \, <\prod_{m=1}^n \, P_m (\partial X) e^{\, i k_m X}{>}_{(G,\, \theta=0)}. 
\end{eqnarray}
The amplitude in the presence of the antisymmetric background is therefore the same as an amplitude without antisymmetric background, but evaluated with the open string metric (\ref{openmetric}), plus a phase factor which depends on the momenta and on the noncommutativity parameter. This is a key observation and will turn out to be very important in what follows.
\subsection{The Seiberg-Witten Limit}
We will try now to perform a zero-slope limit of the amplitude (\ref{amplfact}), looking for an effective description of string processes at low-energy. When one performs the limit in which the string constant $\alpha'$ goes to zero, or equivalently one restricts himself to a particular sector of the domain of integration of the string amplitudes, one expects to decouple from the string spectrum all nonzero masses, which depends on $1/\alpha'$, and the amplitude to reduce to a field theory amplitude for the particles of the lowest-energy sector. We will study the limit proposed by Seiberg and Witten\cite{sw}. In this limit, one has
\begin{eqnarray}
\label{SWlimit} 
\alpha' \, \sim \, \epsilon^{{{1}\over {2}}} \qquad \qquad g \, \sim \, \epsilon  \qquad \qquad \epsilon \, \to 0.
\end{eqnarray}
The peculiarity of the scaling (\ref{SWlimit}) is that, when the string constant goes to zero, then the closed string metric approaches zero as well, but in such a way that the open string parameters tend to constant finite values. In fact, from (\ref{openmetric}), one has 
\begin{eqnarray}
\label{metrlim}
G^{-1} \, \to \, - {{1}\over {4 \pi^2}} \, {{1}\over {B}} \, {{g}\over {{\alpha'}^2}} \,  {{1}\over {B}}\, \to \, const,
\end{eqnarray}
while, from (\ref{noncomm}), one has
\begin{eqnarray}
\label{nonclim}
\theta \, \to \, {{1}\over {B}}.
\end{eqnarray}
One can immediately see from comparison with the action (\ref{sigma}) that the Seiberg-Witten limit can be viewed as a limit of strong B-field. This provides a connection with all the literature which studies noncommutativity in phenomena involving for examples electrons in strong magnetic fields, and the Landau problem\footnote{One notices that the action basically tends to $- {{i}\over {2}} \int_{\partial'\Sigma} B_{i j} \, X^{i}\, \partial_{\tau} X^{j}$}. We refer the reader to \cite{Magro} for an explanation of how it works in that case; one should also notice stimulating analogies with the tractation we present in Section 2.3.

Some remarks are in order about this limit. First, one immediately sees that it does not commute with $B \to 0$; in other words, it is different to perform the low-energy Seiberg-Witten limit and then to send $B$ to $0$, or to get first rid of $B$ and then to take the zero slope limit afterwards. This is a first signal that in the noncommutative effective field theory description, $\theta=0$ could be \it a priori \rm a singular point. A great amount of papers has shown that this is indeed the case: when one compute quantum corrections to the noncommutative tree-level amplitudes, one finds typically non-analytic behaviours at small $\theta$. We will discuss this feature later, in the chapter of this thesis devoted to the noncommutative field theories, where also the relevant literature will be mentioned. We will then exhibit a particular and subtle example of non-analyticity in the part devoted to our calculations, for dimension of the brane-worldvolume equal to two. 

The second remark is that the Seiberg-Witten limit (\ref{SWlimit}) is a good candidate for a decoupling limit at least from the point of view of the string mass-shell condition: the masses of the string spectrum are determined by the piece proportional to the open string metric in the propagator. Therefore, to decouple all the massive string states, one has at least to send to zero this first piece, which is proportional to $\alpha' G^{-1}$. The limit (\ref{SWlimit}) succeeds in doing this, the propagator tending to ${{i}\over {2}} {{1}\over {B}} \, \Theta (s-s')$. However, this is not the end of the story when considering also string loops, since one should actually see if the massive string states really disappear from the actual string amplitudes. This point will be crucial in our later developments. For the moment, let us rely on the previous naive consideration, and suppose that all massive states indeed decouple. 

If we perform the Seiberg-Witten limit of the amplitude (\ref{amplfact}), we see that the phase factor remains essentially the same with the constant noncommutativity parameter (\ref{nonclim}), while the other piece tends to the string amplitude as computed without the $B$-field boundary term in (\ref{sigma}), but evaluated with the metric (\ref{metrlim}), in the low-energy limit $\alpha' \to 0$. This last term is nothing but the low-energy limit of the usual string amplitude, with metric $G_{lim}$ obtained by Eq.(\ref{metrlim}); one knows that it is described by a field theory action 
\begin{eqnarray}
\label{field}
S_{eff} \, = \, \int dx \, \sqrt{G_{lim}} \, \, \, L(\Phi, \partial \Phi),
\end{eqnarray}
where we have indicated briefly with $\Phi$ the set of fields which describe the various particles of the massless sector, with $\partial \Phi$ their derivatives, and the indices are contracted using the metric $G_{lim}$. The integration is understood over the worldvolume of the brane we are considering. Therefore, the low-energy limit of (\ref{amplfact}) is an amplitude which is derived from (\ref{field}) and multiplied in momentum space by the noncommutative phase. This is exactly what we mean by a noncommutative amplitude, in the way it is defined in the standard approach to noncommutative quantum field theory. The effective action for this situation is just the same action (\ref{field}) but with noncommutative phases, i.e. precisely
\begin{eqnarray}
\label{NCfield}
S_{eff} \, = \, \int dx \, \sqrt{G_{lim}} \, \, \, L_{\star} (\Phi, \partial \Phi),
\end{eqnarray}
where the star means that all the products of the fields and their derivatives in $L$ must be replaced by the noncommutative (Groenewold-Moyal) star-product\cite{G,M} 
\begin{eqnarray}
\label{starproduct}
\phi_1(x)\star\phi_2(x)=\exp \lq \frac{i}2 \, \theta^{\mu\nu}\frac{\partial}
{\partial x_1^\mu}
\frac{\partial}{\partial x_2^\nu}\rq \phi_1(x_1)\phi_2(x_2)|_{x_1=x_2=x}. 
\end{eqnarray}
This shows that, in order to obtain the effective description for bosonic open string theory in a $B$-field background, it is sufficient to modify the standard effective field theory, which should describe the process in the absence of $B$, in order to transform it into its noncommutative analogue, and this amounts to a replacement of all the products of the fields in the standard action with ``star-products''. Analogously, it is enough to transform the brane worldvolume over which the standard effective field theory is defined, in the analogous noncommutative manifold, obtained modifying the algebra of the functions over it into the noncommutative algebra defined by (\ref{starproduct}).

\section{Canonical Quantization}
Now we want to describe the other approach to string theory in antisymmetric backgrounds that we mentioned, which points towards a noncommutative description of the brane worldvolume, and relies on canonical quantization\cite{ch,chka,aas,ch2,c}. Let us consider again the sigma model action for the bosonic string in a closed constant background. After suitably rescaling the $B$-field to get it dimensionless, one can write 
\begin{eqnarray}
\label{sigmaChu}
S={{1}\over{4\pi {\alpha}'}}\int_{\Sigma}d^2 \sigma \, \sqrt{\gamma} \, \gamma^{a b} \, g_{\mu\nu} \, {\partial}_a X^{\mu} {\partial}_b X^{\nu} \, + \, {{1}\over{4\pi {\alpha}'}}\int_{\partial'\Sigma} B_{\mu\nu} \, X^{\mu}\, \partial_{\tau} X^{\nu}.
\end{eqnarray}
We have already derived the equation of motions and boundary conditions from this action, we only have to update our rescaled factors. When expressed in the worldsheet variables $\sigma$ and $\tau$, they become on the brane:
\begin{eqnarray}
\label{eomChu}
e.o.m. \qquad \qquad \qquad \qquad g_{ij} \, \, \square  \, X^{j} \, = \, 0,\qquad \qquad \qquad \qquad 
\end{eqnarray}  
\begin{eqnarray}
\label{warningChu}
b.c \qquad \qquad \qquad \qquad \partial_{\sigma}  X^{i} \, + \, {F_j}^i \, \partial_{\tau} \, X^{j} \, \, = \, \, 0 
\qquad \sigma = 0,\pi.
\end{eqnarray}
We have defined ${F_j}^i = B_{j\, l}\, g^{li}$, and the symbol $\square$ represents the worldsheet Dalembertian. One thing we notice is the curious fact that when the $B$-field is sent to infinity, one ends with Dirichlet-like boundary conditions also on the brane. This looks like the endpoints of the string would become fixed to something similar to a zero-brane. 

One can solve this set of equations as an infinite series of harmonic oscillators, and obtain  
\begin{eqnarray}
\label{solut}
X^{i} = {x_0}^i + {a_0}^i \tau - {a_0}^j {F_j}^i \sigma + \sum_{n\neq0} {{e^{- i n \tau}}\over {n}} \Big( i {a_n}^i \cos n\sigma - {a_n}^i {F_j}^i \sin n\sigma \Big).
\end{eqnarray}
The canonical momentum on the brane is given by 
\begin{eqnarray}
\label{momentum}
P^i &=& {{1}\over {2 \pi \alpha'}} \Big( \partial_{\tau} X^i + {F_j}^i \partial_{\sigma} X^j \Big) \nonumber \\
&=& {{1}\over {2 \pi \alpha'}} \Big({a_0}^j + \sum_{n\neq0} {a_n}^j e^{- i n \tau} \cos n\sigma \Big) {M_j}^i, 
\end{eqnarray}
where ${M_j}^i = {\delta_j}^i - {F_j}^l {F_l}^i$. The total momentum $\Pi^i$ is a conserved quantity:
\begin{eqnarray}
\label{momentumtot}  
\Pi^i \, = \, \int_0^{\pi} \, d\sigma \, P^i \, =\, {{1}\over {2 \alpha'}} \, {a_0}^j \, {M_j}^i,
\end{eqnarray}    
while of course the total momentum outside the brane is not conserved, due to the breaking of translational symmetry by the choice (\ref{static})\cite{dgo}.
To proceed to the quantization program one should promote $X$ and $P$ to quantum operators, and impose the canonical commutation relation between them. But, in so doing, namely if one tries to impose simultaneously the standard canonical commutation relation, one runs immediately into an \it absurdum \rm. In fact, from (\ref{warningChu}) and (\ref{momentum}), one has   
\begin{eqnarray}
\label{incons}
2 \pi \alpha' \, P^j (\tau , 0/\pi) \, {F_j}^i \, = \, - \, \partial_{\sigma} \, X^j (\tau , 0/\pi) \, {M_j}^i,
\end{eqnarray}
from which one derives 
\begin{eqnarray}
\label{incons2}
2 \pi \alpha' \, \Big[ P^j (\tau , 0/\pi) , P^l (\tau , \sigma') \Big] \, {F_j}^i \, = \, - \, \Big[ \partial_{\sigma} \, X^j (\tau , 0/\pi) , P^l (\tau , \sigma') \Big]\, {M_j}^i.
\end{eqnarray}
From formula (\ref{incons2}) it is clear that to impose simultaneously 
\begin{eqnarray}
\label{standard}
\Big[ X^j (\tau , \sigma) , P^l (\tau , \sigma') \Big] \, = \, i g^{jl} \delta (\sigma - \sigma') \, \, , \, \, \Big[ P^j (\tau , \sigma) , P^l (\tau , \sigma') \Big] \, = \, 0 
\end{eqnarray}
is not consistent.

One should use here Dirac's quantization method for constrained systems\cite{dir, nakoji}. One takes (\ref{standard}), together with 
\begin{eqnarray}
\label{stand2} 
\Big[ X^j (\tau , \sigma) , X^l (\tau , \sigma') \Big] \, = \, 0, 
\end{eqnarray}
as equal time classical Poisson bracket among the canonical variables, and considers the constraints: from the boundary conditions (\ref{warningChu}) one reads the primary constraints (\ref{incons})
\begin{eqnarray}
\label{primary}
2 \pi \alpha' \, P^j \, {F_j}^i \, + \, \partial_{\sigma} X^j \, {M_j}^i \, = \, 0  \qquad \qquad on \, the \, boundary.
\end{eqnarray}
By imposing the Poisson bracket of the primary constraints (\ref{primary}) with the Hamiltonian
\begin{eqnarray}
\label{hamilt}
H \, = \, {{1}\over {4 \pi \alpha'}} \, \int_0^{\pi} \, d\sigma \, \Big[ {(\partial_{\tau} \, X)}^2 \, + \, {(\partial_{\sigma} \, X)}^2 \Big] 
\end{eqnarray}
to vanish, which means that the constraints must be conserved in time, one obtains secondary constraints. One should again impose vanishing Poisson bracket between this new constraints and the Hamiltonian, and continue in this way until one finds the set of all possible constraints, such that no further conditions need to be added in order to reach consistency\footnote{Many subtleties underlie the correct application of this procedure, for which we refer the reader to the already cited papers and to more complete treatments}. Let us consider now the matrix obtained by taking the Poisson bracket among all the constraints with themselves. This matrix is invertible if the constraints are all \it second class \rm, as we will assume. Then, the Dirac bracket between two dynamical variables is defined as the old Poisson bracket, plus a term which involves this matrix of constraints, as follows:
\begin{eqnarray}
\label{diracb}     
&&{\Big[ A (\sigma) , B (\sigma') \Big]}_{D} \, = \, {\Big[ A (\sigma) , B (\sigma') \Big]}_{P} -  \sum_{1,2,n,m} {\Big[ A (\sigma) ,  \varphi^n (\sigma_1) \Big]}_{P} \nonumber \\
&&\qquad \qquad \qquad \times \, {(C^{-1})}_{nm} (\sigma_1, \sigma_2)  {\Big[ A (\sigma_2) ,  \varphi^n (\sigma') \Big]}_{P},
\end{eqnarray}
where $\varphi^n = 0$ is the $n$-th constraint, and $C$ is the matrix of constraints. Quantum commutator is obtained as $- i$ times the Dirac bracket.

After performing the calculation in the present case, it was found that two of the three quantum commutators are unchanged, namely
\begin{eqnarray}
\label{diracunch}
{\Big[ X^j (\tau , \sigma) , P^l (\tau , \sigma') \Big]}_D \, = \, i g^{jl} \delta (\sigma - \sigma') \, \, , \, \, {\Big[ P^j (\tau , \sigma) , P^l (\tau , \sigma') \Big]}_D \, = \, 0, 
\end{eqnarray}
while the third one undergoes a crucial modification:
\begin{eqnarray}
\label{stand2dirac} 
\Big[ X^j (\tau , 0) , X^l (\tau , 0) \Big] \, = \, - \Big[ X^j (\tau , \pi) , X^l (\tau , \pi) \Big] \, = \, 2 \pi i \alpha' \, {\Big( M^{-1} \, F \Big)}^{jl},
\end{eqnarray}
all other combinations still vanishing\footnote{We can notice that the commutator (\ref{stand2dirac}) is proportional to $\alpha'$}. This means that, on the boundary, the coordinate quantum operators do no longer commute, revealing that the brane worldvolume has assumed the features of a noncommutative manifold. 

One can express the new commutation relations derived from (\ref{diracunch}) and (\ref{stand2dirac}) in terms of creation and annihilation operators:
\begin{eqnarray}
\label{zeromodes}
[{x_0}^i ,{x_0}^j ] \, = \, 2 \pi i \alpha' \, {\Big( M^{-1} \, F \Big)}^{ij},
\end{eqnarray}
\begin{eqnarray}
\label{a0}
[{a_0}^i ,{a_0}^j ] \, = \, 0, 
\end{eqnarray}
\begin{eqnarray}
\label{x0a0}
[{x_0}^i ,{a_0}^j ] \, = \, 2 i \alpha' \, {M^{-1}}^{ij}, 
\end{eqnarray}
\begin{eqnarray}
\label{anx0}
[{a_n}^i ,{x_0}^j ] \, = \, [{a_n}^i ,{a_0}^j ] \, = \, 0, 
\end{eqnarray}
\begin{eqnarray}
\label{anam}
[{a_n}^i ,{a_m}^j ] \, = \, 2 \alpha' \, n \, {M^{-1}}^{ij} \, \delta_{n+m}.   
\end{eqnarray}
The crucial difference with respect to the case without the $B$-field is that the zero modes of the string do not commute anymore: this is the mechanism through which the static brane inherits noncommutativity (see \cite{kar}, and \cite{witt} for ideas on how to go beyond the zero-mode approximation). What obtained here in the quantization of the free spectrum shows that the effective theory for the worldvolume of the brane is a field theory defined on a set of noncommuting coordinates, with noncommutativity parameter $\theta^{ij}$ given by Eq. (\ref{zeromodes}).  

As a final remark, one notices that the mass-shell condition can be derived from the Virasoro generators, which in turn can be extracted from our previous formulas. In this condition, the matrix $M$ plays a role which is the same as the open string metric of the previous section (see the remark  under Eq. (\ref{OPE}))\cite{ssto}. The central charge is instead unchanged; one repeats from here on all the considerations about a consistent quantization of the free spectrum. The critical dimension is unchanged, and the physical Fock space of states has to be built up using the suitable Virasoro generators. 

\section{Gauge Fields}
As already announced, we end this chapter by making some remarks on gauge fields. The complete sigma model for the bosonic open string coupled with the massless sector should contain also an interaction term 
\begin{eqnarray}
\label{gaugebrana}
- i \int_{\partial' \Sigma} \, A_i (X) \, \partial_{\tau} X^i.
\end{eqnarray}

At the classical level, the term in Eq.(\ref{gaugebrana}) is invariant under the gauge transformation
\begin{eqnarray}
\label{gaugtr}
\delta A_i \, = \, \partial_i \Lambda,
\end{eqnarray}
since the variation is a total derivative, and we assume suitable conditions at $\tau = \pm \infty$. However\cite{sw}, at the quantum level one finds for the path integral
\begin{eqnarray}
\label{pathint}
\delta \exp \Big[ i \int_{\partial' \Sigma} \, A_i \, \partial_{\tau} X^i \Big] \, \sim \, i \int d\tau \partial_{\tau} \Lambda \, - \, \int d\tau \, A_i \, \partial_{\tau} X^i \int d\tau' \partial_{\tau'} \Lambda  
\end{eqnarray}
plus higher orders in the gauge field. Then, the authors in \cite{sw} raised the problem that, if one considers the second term in (\ref{pathint}) operatorially, one has to regularize it when $\tau$ get $\tau'$ are close to each other. If one makes use of a point-splitting method
\begin{eqnarray}
\label{pointspl}
\int_{- \infty}^{+\infty} d\tau' \, = \, \lim_{\epsilon \to 0} \, \, \int_{- \infty}^{\tau - \epsilon} \, + \, \int_{\tau + \epsilon}^{+\infty} \, d\tau' ,  
\end{eqnarray}
and one performs the OPE using the antisymmetric part of the propagator (\ref{propbound}) (the one that survives in the field theory limit), the result is that this term can be written as
\begin{eqnarray}
\label{starspl}
\int d\tau \, : \Bigg[ A_i \Big(x(\tau)\Big) \star \Lambda \Big(x(\tau)\Big) \, - \, \Lambda \Big(x(\tau)\Big) \star A_i \Big(x(\tau)\Big) \Bigg] \partial_{\tau} X^i :\, ,
\end{eqnarray}
therefore one has invariance if one asks 
\begin{eqnarray}
\label{stargaugeinvariance} 
\delta A_i\, =\, \partial_i \Lambda \, - \, i (A_i \star \Lambda \, - \, \Lambda \star A_i ). 
\end{eqnarray}
This form of the gauge transformation is consistent also at higher orders in the gauge field in the expansion (\ref{pathint}). This is precisely the form of noncommutative \it star-\rm gauge invariance, as it is well known in noncommutative gauge theories, in its infinitesimal version. Actually, the same result is obtained when computing string amplitudes for external open string massless vectors, in the way shown in section (1.2). The effective field theory action is the noncommutative version of the Yang-Mills theory
\begin{eqnarray}
\label{NCYM}
\int d^{p+1}x \, \, \sqrt{- \det G} \, G^{i j} \, G^{r s} \, F_{i r} \star F_{j s}, 
\end{eqnarray}
where
\begin{eqnarray}
\label{fieldstr}
F_{i j}&=&\p_i A_j -\p_j A_i -i (A_i \star A_j - A_j \star A_i)\nonumber \\
&\equiv&\p_i A_j -\p_j A_i -i [A_i \, \, ,\star \, \, A_J]
\end{eqnarray}
is the ``noncommutative field strength''. 

\chapter{Noncommutative Quantum Field Theory} 
\section{Introduction}
Once we have understood a possible way to derive noncommutative field theories from string theory as an effective low-energy description, we turn our attention to their particular properties. They show themselves to be very exotic quantum field theories, and the aim of this chapter is mainly to explain their features in the light of their stringy origin.  

We have seen that, in order to turn a standard action into its noncommutative counterpart, it is sufficient to modify all the products among fields in the classical action such that they become star-product (\ref{starproduct}). Equivalently, we consider as a basis of the geometrical setting of our theory the commutation relation
\begin{eqnarray}
\label{commrel}
[x^{\mu} , x^{\nu}] \, = \, i \theta^{\mu \nu}, 
\end{eqnarray}
which we saw emerge from the canonical quantization of the string coordinate operators, or from Eq.(\ref{OPE}). Let us see now what happens strictly in the effective theory, starting from the perturbative expansion\cite{mvs}. 

\section{Noncommutative Perturbation Theory}
As a first property, one realizes that the free propagators of the new theory are the same as the old ones, if one has trivial boundary conditions. We will consider in our treatment a perturbative noncommutative theory as defined in terms of ordinary fields, which live on an ordinary manifold, but multiplied by the star-product. We mention here the existence of an operatorial description of noncommutative theories, which is very useful for example for nonperturbative studies, like the analysis of soliton solutions or of particular geometrical properties\cite{alvawadia,gronekra}. We will not make use of this approach throughout this thesis. The quadratic part of the noncommutative action, the one which drives the free perturbative propagator, feels the presence of the star-product only through a total derivative, which we set to zero if the boundary conditions allow it. We will restrict our discussion to the cases in which this is possible. 

The part of the action which is heavily modified by the replacement with the star-product is the interactive part. Here, the vertices of the theory are modified by the presence of the noncommutative phases in momentum space, which are the ones we have encountered in Eq.(\ref{amplfact}). This changes a lot the nature of the interaction. First, one generally loses invariance under arbitrary permutation of the momenta. The best thing to do when one deals with a noncommutative Feynman diagram, is to draw it in 't Hooft double-line notation, very much as one does in gauge theories in order to keep track of the nonabelian character of the theory. Here, we can say that the relation (\ref{commrel}) introduces as well a general \it nonabelianity \rm in the theory, even if one starts from an action which does not manifestly contain gauge fields. We will see that one of the main point will be actually the existence of a large generalized gauge group, which will contain internal as well as space-time transformation, and will include  and unify both internal and space-time nonabelianity.  

\medskip
\centerline{\includegraphics[width=8.5cm]{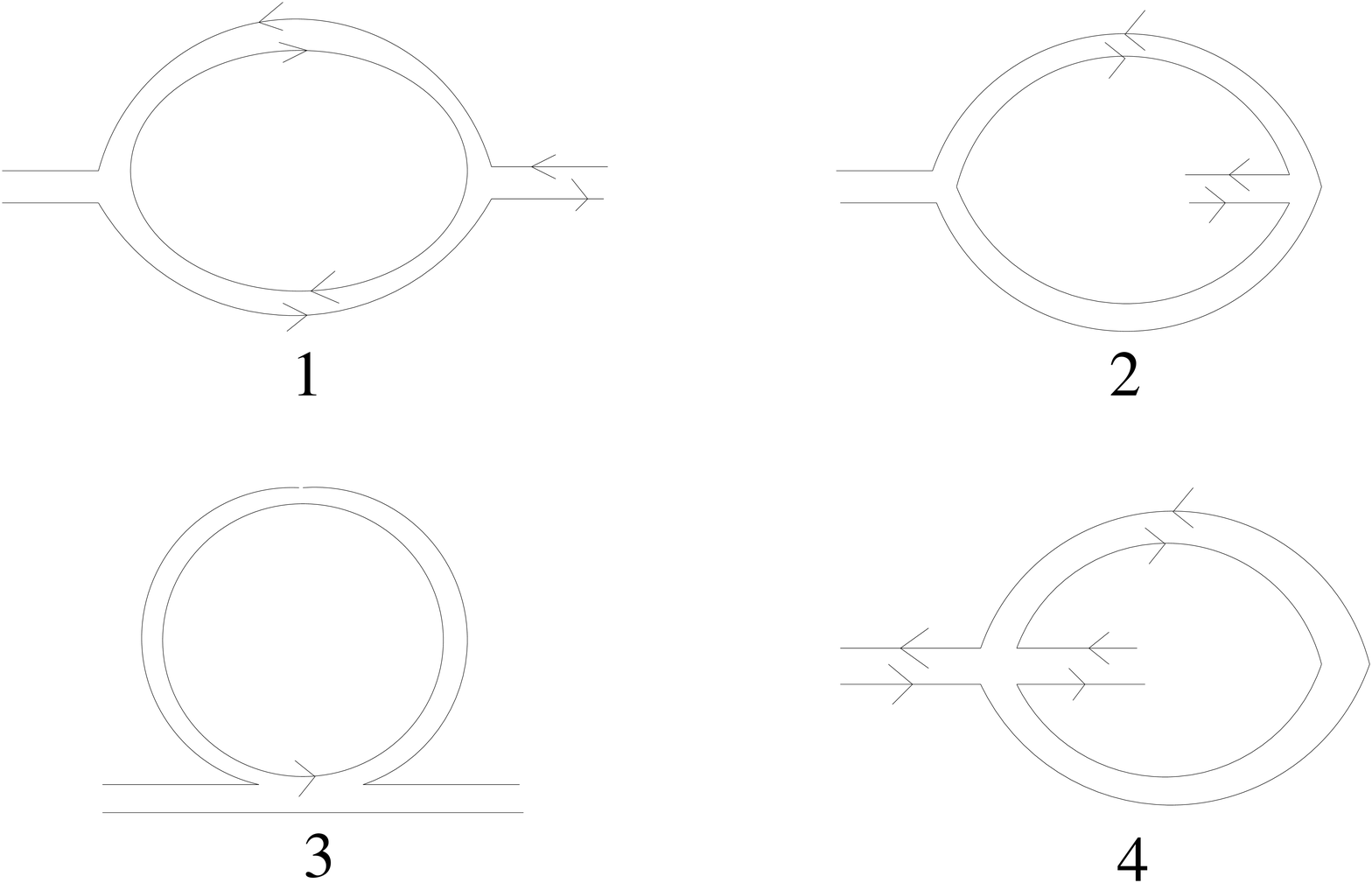}}

\bigskip

Here above, we have drawn four particular diagrams in the double-line notation: 1 and 2 belong to the expansions of the $NC{\Phi}^3$, 3 and 4 to the one of $NC{\Phi}^4$, respectively. One is naturally brought to distinguish between planar and non-planar diagrams: 1 and 3 are planar, while 2 and 4 are non-planar. In planar diagrams, conservation of momenta at each vertex produces strong cancellation among the noncommutative phases, and only an overall phase survives, which depends on the external momenta as $\exp[- {{i}\over {2}} \sum_{n<m} p_n \theta p_m]$, and therefore factorizes out of the loop integrals. Planar graphs are essentially the same as the commutative ones, in particular they have the same divergences and pattern of singularities. In non-planar diagrams, this cancellation is no longer at work, and some of the noncommutative phases survive, which depend on the momenta running in the loops, and therefore change substantially the behaviour of the amplitude. Precisely, one has inside the integral a supplementary $\exp[- {{i}\over {2}} \sum_{r<s} c_{rs} k_r \theta k_s]$, where $c_{rs}$ is the ``intersection matrix'', namely a matrix which has $0$ as an entry when two propagators do not cross, and $1$ if they do cross. We show the fourth diagram, in single line notation: 

\bigskip

\bigskip

\centerline{\includegraphics[width=3cm]{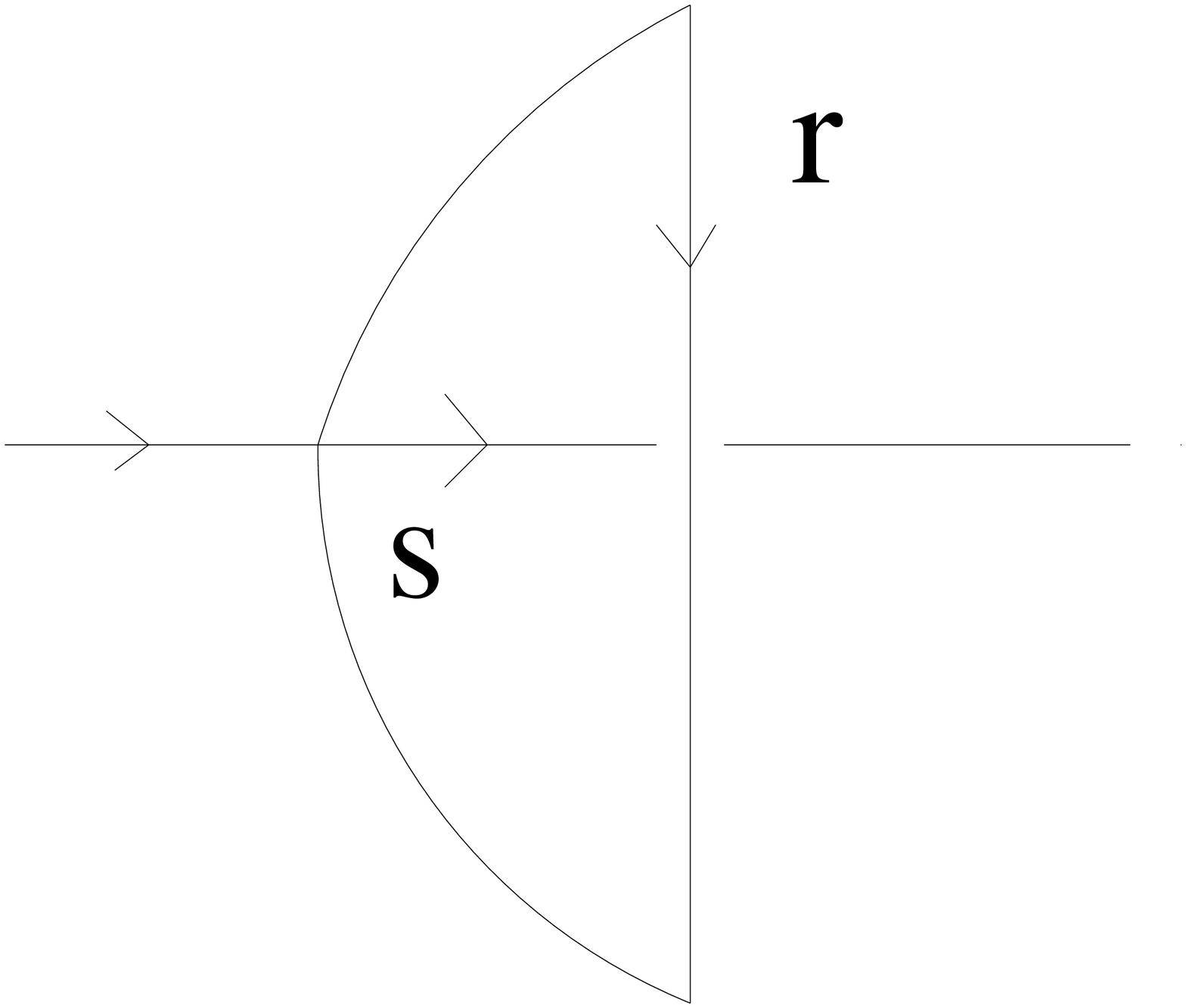}}

\medskip

\bigskip

Nonplanar graphs are normally cutoff in the ultraviolet by the presence of this phase, and in most cases they turn out to be ultraviolet convergent and finite. This observation was at the root of the earliest attempts to introduce possible non-local deformation in the space time over which a quantum field theory is defined, in order to make it finite: in fact, the ultraviolet divergences of quantum field theory are connected with local interactions, tipically resulting in multiplication of distributions in the same point, which is ill-defined from a mathematical point of view. Physically, the absence of a short distance cutoff implies an infinite momentum singularity. Introducing a sort of minimal distance, or, otherwise stated, a sort of coordinate uncertainty principle, can sometimes solve this problem. Now we see that here this is not sufficient: noncommutativity is able to cutoff only nonplanar diagrams, but planar loops remain infinite\cite{Filk}. Even if it inherits nonlocality from string theory, noncommutative quantum field theory suffers from ultraviolet divergences on its own. But this is not the end of the story: in fact, also nonplanar diagrams are sources of divergences in noncommutative theories. They produce infinities of another kind, precisely infrared divergences, that show up when nonplanar loop diagrams are inserted as subgraphs in more complicated graphs in the perturbative expansion. This phenomenon is called UV/IR mixing\cite{mvs,mst}, because it consists of infrared singularities that are produced when one integrates very high momenta in the loops. The above diagram 4 of the NC$\Phi^4$ theory is a classical example of how this mechanism works at one loop. In the following we draw it again and calculate its value, along with its planar partner 3:
\bigskip

\bigskip

\bigskip

\centerline{\includegraphics[width=3cm]{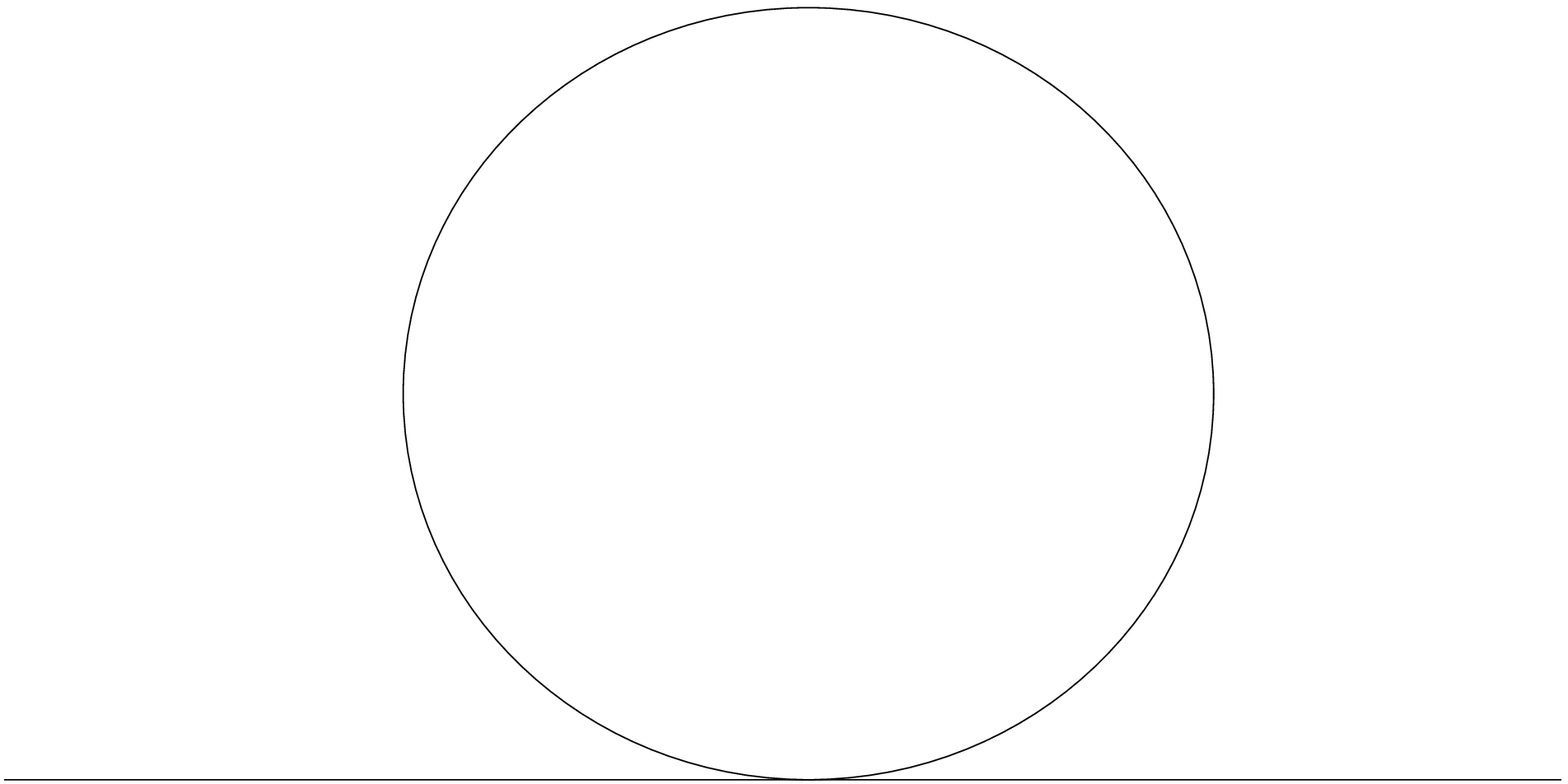} $ \qquad \, \, \,  \propto \, \, \, \, \, {{1}\over {2}} \, \int {{d^4k }\over {(k^2 + m^2)}} \, \, \propto \, \, \Lambda^2,$} 
  
\bigskip

\bigskip

\bigskip

\bigskip

\centerline{\includegraphics[width=3cm]{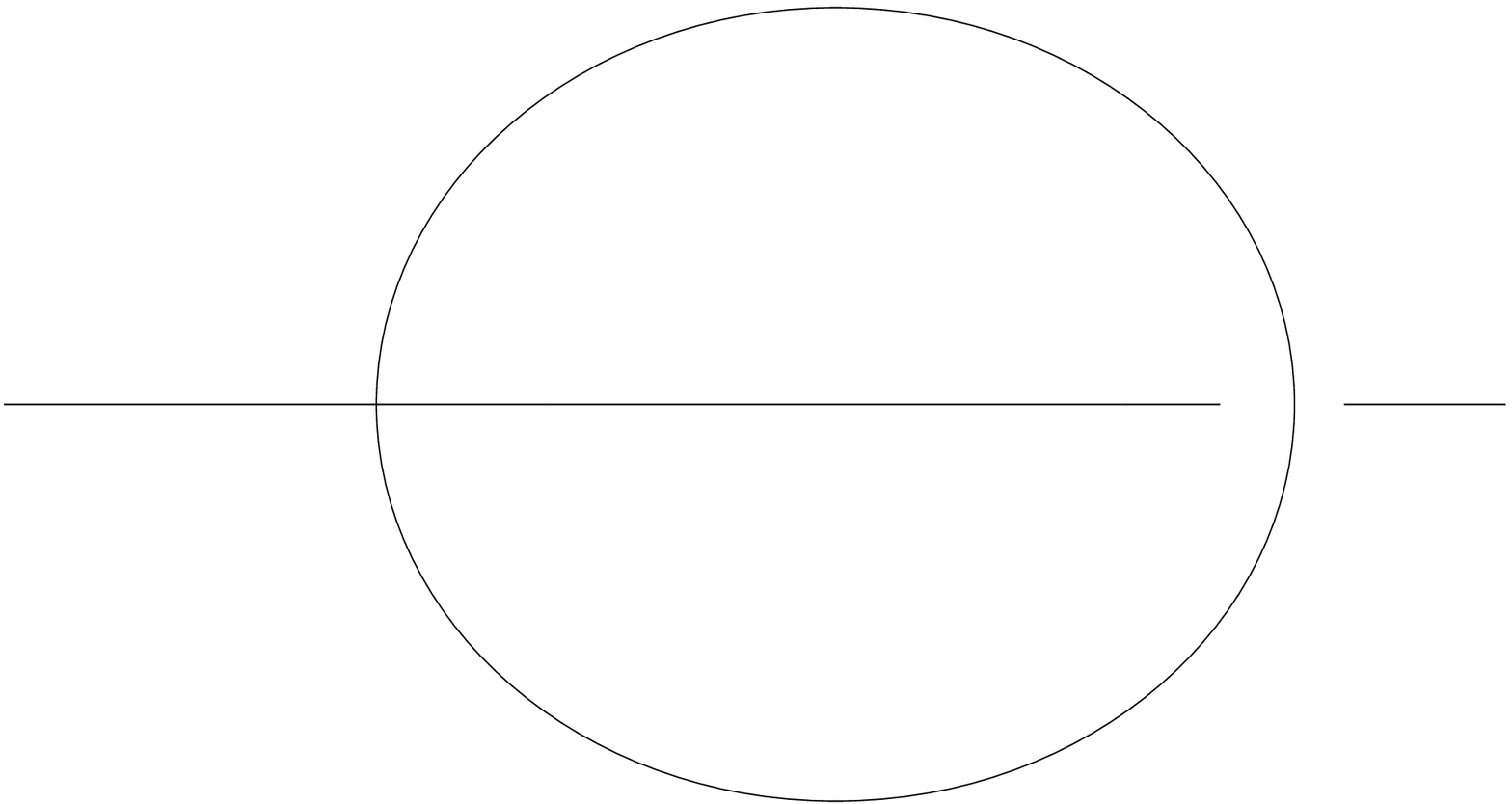} $ \qquad \, \, \, \propto \, \, \, \, \, {{1}\over {2}} \, \int {{d^4k}\over {k^2 + m^2}}\, e^{i k \theta p}\, \, \propto \, \, {{\Lambda^2}\over {1 + \Lambda^2 p\,\theta^2 p}} \, \to {(p\, \theta^2 p)}^{- 1},$} 
  
\bigskip

\bigskip

Here $\Lambda$  is some ultraviolet cutoff. The two diagrams give an equal amplitude in the commutative case, that means that their value is the same if one performs the limit $\theta \to 0$ \it inside \rm the integral. Together, they contribute to the value of the graph which, in the commutative case, is undistinguishably drawn as \includegraphics[width=0.8cm]{bubblepl.eps} . Commutative scalar particles do not distinguish between planar and nonplanar diagrams. Noncommutativity makes the two amplitudes different, and, in particular, the second diagram is finite, due to the noncommutative phase which damps the large values of $k$, as we have already mentioned. But this finite value is proportional to ${{1}/ {p\, \theta^2 p}}$, i.e, it is infrared divergent, and, when inserted in higher order loops, it can give rise to new unexpected infinities. As a side remark, we notice that, after performing the integral, it is no longer possible to send $\theta$ to zero, which means that in this case the limit to the commutative theory is discontinuous\cite{armoni}. 

The phenomenon of UV/IR mixing is a very interesting effect which noncommutative quantum field theories inherits from strings, and shows up precisely when quantum loop corrections enter the game. We will soon see how string loops can explain this phenomenon, which is pretty nonlocal and different from what one finds in usual QFT. In particular, this stringy residue ruins the standard perturbative order-by-order renormalizability program, which is based on the decoupling between infrared and ultraviolet regimes, running the analysis of divergences of the perturbative series out of control. All the attempts made in the literature to say something about the perturbative renormalizability of these theories have encountered great difficulties in going beyond the first few loop orders\cite{Sheikh,Chepelev,shmic,Chepelev2,bonorasali,martin}. For approaches relying on the renormalization group see \cite{lucamass,lucamass2,Becchi}. 

\section{Nonlocality and Star Product}The mixing of the scales can be simply justified when looking at the commutation relation (\ref{commrel}): roughly speaking, small distances along $x^{\mu}$ force long distances along $x^{\nu}$. Another way to understand it from the basic definitions is to analyse the features of the star-product (\ref{starproduct})\cite{mvs}. Let us consider the star-product between two functions $f$ and $g$, which are well localized in space-time around two certain points, as depicted in the following figure: 

\bigskip

\medskip

\centerline{\includegraphics[width=6.5cm]{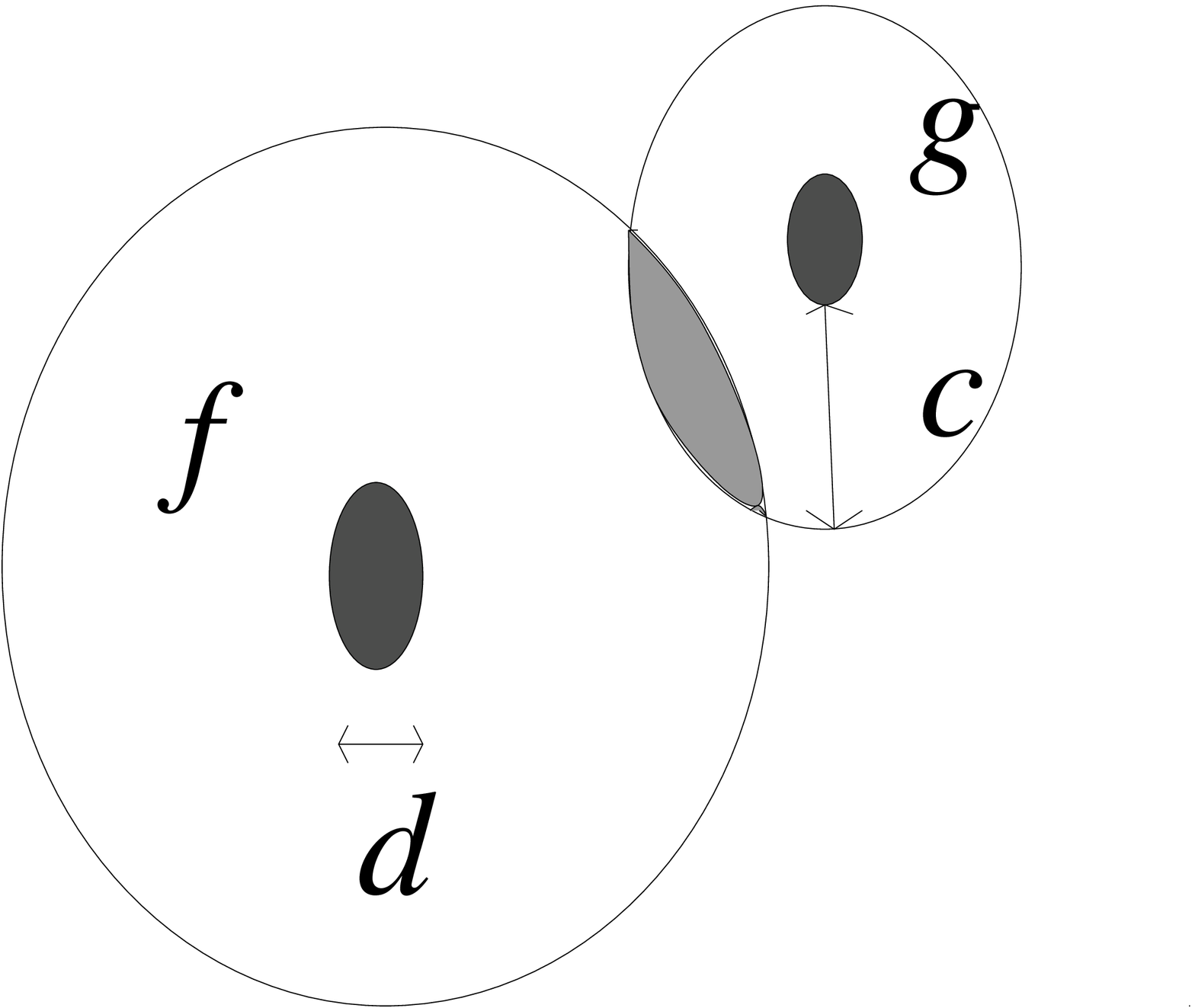}}

\bigskip

\medskip

In this figure, the dark regions are those in which $f$ and $g$ respectively are different from zero. One can rewrite the expression (\ref{starproduct}) in the following manner, provided that the matrix $\theta$ is non-singular:
\begin{eqnarray}
\label{alternstar}
f \star g \, (z) \, = \, \int d^d x \, d^d y \, \, f(x) \, g(y) \, {{1}\over {\det \theta}} \, \exp [2 i (z - x) \, {{\theta}^{-1}} (z - y)]
\end{eqnarray}
If $d$ is the size of $f$ in the variable $x_1$, then, since the integral contains a factor 
\begin{eqnarray}    
\label{suppressed}
\int dx_1 \, f(x_1 , \cdots) \, \exp [- 2 i x_1 {{(z_2 - y_2 )}\over {\theta}}],
\end{eqnarray}
then $f \star g$ is suppressed by oscillations if $ |d (z_2 - y_2 )| >> \theta$. This means that the star-product is non-zero in a width $c \propto \theta / d$ in the variable $z_2$. The same applies to $g$, therefore, the non-local star-product between $f$ and $g$ is different from zero only in the intersection of this two region, which is indicated in grey in the figure. One can immediately imagine, in a very schematic way, the situation in which these functions are two wave packets, interacting in a noncommutative theory through the star-product multiplication in space-time. To the extent in which this description is a good effective one for the interaction process, we expect that incoming distribution localized in small regions will spread out very far, which mixes small distances with long distances. Equivalently, since the main effects of the star-product are felt in (nonplanar) loops, we can also imagine this producing a mixing of momentum scales such that particles circulating into quantum loops with high energy $E$ do have effects at very low energies too, of order $1 / \theta E$. 

These considerations are mainly based on intuition. They succeed in catching the fundamental reason for the UV/IR phenomenon in noncommutative field theories. What emerges is that it is ultimately linked with the nonlocality one introduces in the theory, by means of the commutation relation (\ref{commrel}), or of the star-product (\ref{starproduct}). Since these theories derive from strings, one expects that those features should already be present in some way in the original string theory. In fact, this is the case\cite{gsw}: the phenomenon of UV/IR mixing has its stringy counterpart in the duality one has in viewing diagrams, such as the double twist diagram which originates the one-loop non-planar contribution in the field theory limit, both as one-loop diagrams for the open string, and as tree-level diagrams for the closed string. The former has an UV singularity when the proper time of propagation of the open string is small, the latter turns it into an IR singularity at large proper time of propagation of the closed one. This fact is at the basis of the unitarity analysis of open string theory, and we will see an important example of this phenomenon in our subsequent calculations. The fact that an UV divergence, which, if not renormalizable, leads to inconsistency of the theory, could be interpreted as IR, which has a physical meaning, teaches a lot about the possibility of absence of ultraviolet difficulties in string theory. Furthermore, if one add supersymmetry to the string, one can get rid of infrared divergences from physical principles, therefore getting rid of divergences at all! 

This beautiful landscape becomes horrible if truncated to the noncommutative perturbative effective theory. Despite this stringy interpretation, the problems produced by the UV/IR mixing phenomenon are quite big, if one looks at the field theory in itself. We have already mentioned the troubles it generates in the renormalizability program. Now we look at some other problems, which appear already from the quite rough scheme of the example above. The fact that one loses locality has strong consequences\cite{sst}. If the noncommutative parameter is of ``magnetic'' type, i.e. one can choose a frame in which time is not involved in the noncommutativity, then the theory is non-local in space. Such a scattering reminds the interaction among extended objects, of length proportional to $l = \theta p$, where $p$ is their momentum:  

\bigskip

\centerline{\includegraphics[width=9cm]{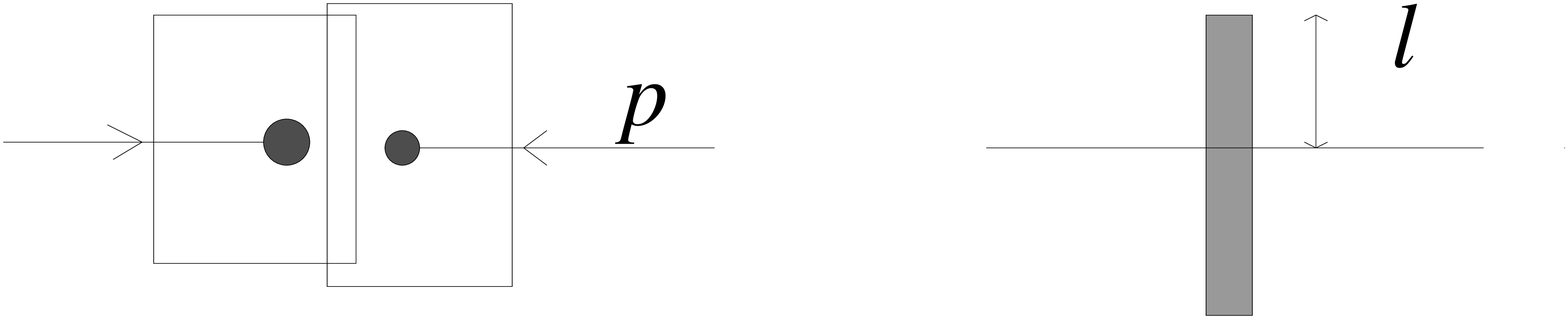}}

\bigskip
One could say that this is another hint at the fact that the real objects in the spectrum of these theories are extended string-like or dipole-like objects. Noticeably, the relation $l = \theta p$ is the same one finds for the gauge-invariant open Wilson lines we will shortly define as the true observables of any noncommutative theory. They can be seen to interact joining and splitting, very much like open strings (see \cite{rey}, along with \cite{rey1, rey2, rey3}).

When $\theta^{\mu \nu}$ is of ``electric'' type, instead, it means that time is necessarily involved in the noncommutativity: there, the extended objects are seen as extended \it in time! \rm From this consideration one can argue that causality is lost, since one expect effects to precede their causes. This is a pathology of the field theory that is difficult to master. 

There are a lot of attempts in the literature which face this kind of problems directly in the field theory, trying to reconcile these effects in a consistent field theoretical \it paradigma\rm, without relying on strings, but implementing noncommutativity as a property of space-time, in regimes where general relativity effects are as important as quantum ones. The field theory that emerges from these approaches was described in \cite{dopli}, and some recent achievements are, for example, in \cite{sibold}. Our approach will be instead, as we have already mentioned, to rely on the string mother theory and its low-energy limit, in order to understand why the effective action has such pathologies. We will check in various examples what happens when performing the limit down to the field theory in the magnetic and electric case, and will explain what is responsible for all these behaviours. We follow the idea that noncommutative theories are a wonderful tool to test our understanding of high-energy physics in regimes where space-time looses its conventional features, and they are able to catch, in a simplified version, many crucial points of what should be the space-time in the string regime, keeping, at odds with any other standard field theory, a rather stringy character. But, just for this reason, they should be seen as an effective description for strings, from which to learn what string theory foresees for space-time at the energies of gravity unification. We will dedicate therefore a great part of our work in understanding what is the reason of the difference between magnetic and electric case in the light of string theory.

\section{Unitarity}
Apart from the intuitive considerations above, the first computation which showed concretely this difference and gave a precise proof of these expectations in the field theory can be considered the analysis of perturbative unitarity\cite{gm,bgnv}. 

Reference \cite{gm} examined a noncommutative scalar theory, and studied the optical theorem for this kind of amplitudes. They started analysing the noncommutative $\lambda \Phi^3$ two point function at one loop, corresponding to the following diagram

\bigskip

\centerline{\includegraphics[width=7cm]{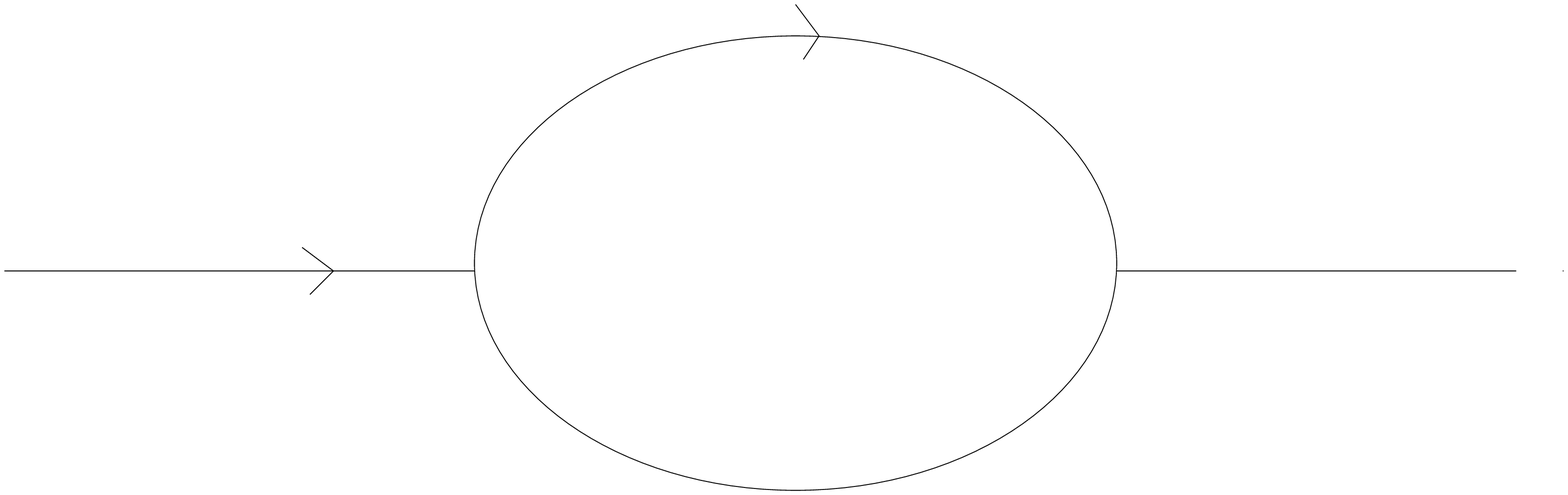}}

\bigskip

This diagram is the same we exhibited in double line notation in our first example on how noncommutative Feynman graphs look like (see Diagram 1, Section 3.2): there, it was drawn to emphasize its planar character. We have learned how to draw its non-planar partner (Diagram 2, Section 3.2): for conciseness, we will indicate with the above figure the sum of both contributions.

The Feynman rule for the vertex in this theory is the following:

\bigskip

\centerline{\includegraphics[width=2cm]{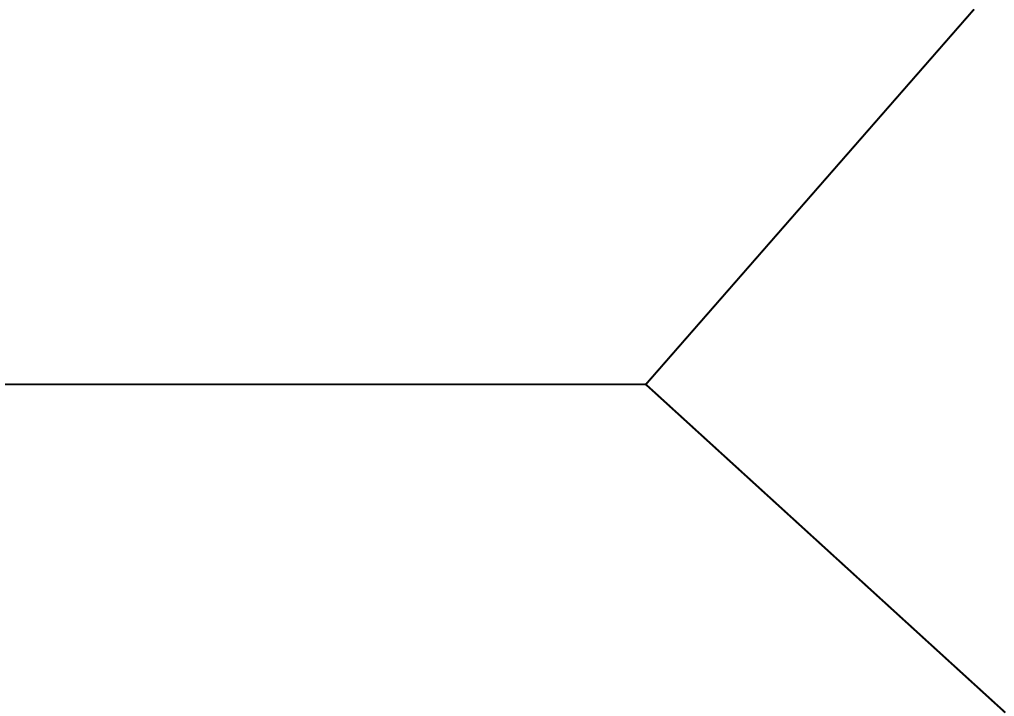} $ \qquad \qquad = \, - \, i\,  \lambda \, \cos [ {{1}\over {2}} \,  k \, \theta \, q ].$}

\bigskip

The amplitude relative to the diagram is thereby
\begin{eqnarray}
\label{Gomis}
i M \, = \, {{\lambda^2}\over {2}} \, \int {{d^d k}\over {{(2 \pi)}^d}} \, {{(1 + \cos [p\, \theta\, k])}\over {2}} \, {{1}\over {k^2 - m^2 + i \epsilon}} \, {{1}\over {{(k + p)}^2 - m^2 + i \epsilon}},  
\end{eqnarray}
where $p$ is the external momentum, and $m^2$ is the positive squared mass of the scalar particle. In writing Eq.(\ref{Gomis}), one has used that ${\cos}^2 p\, \theta\, k$, coming from the two vertices, is equal to $(1 + \cos 2 p\, \theta\, k)/ 2$. This is done in order to separate the planar and the nonplanar part: the $\cos 2 p \theta k$-part is the nonplanar part, which becomes equal to the planar one if we send $\theta \to 0$ before integrating. Since the planar part presents no surprises with respect to the commutative case, we will restrict ourselves to the nonplanar one in the following. 

One can use standard tricks to compute noncommutative Feynman diagrams, for example Feynman and Schwinger parameters. When evaluated, for instance, in $d=4$ dimensions, the diagram in scrutiny turns out to give
\begin{eqnarray}
\label{Gomis4d}
M_{d=4} \, =  \, {{\lambda^2}\over {2}} \, \int_0^1 dx \, K_0 (\sqrt{p \circ p (m^2 - p^2 x (1 - x) - i \epsilon)}),
\end{eqnarray}
where $K_0$ is a modified Bessel function, and one defines $p \circ p = p \theta G \theta p$, $G$ being the field theory background metric. Here we choose this metric to be the usual Minkowski one $diag(1, -1, -1, -1)$.

If one looks at the analytical structure of such an amplitude, one immediately realizes that it has a branch cut for $p^2 > 4 m^2$, and another one for $p \circ p <0$. The key observation is that one can write $p \circ p$ in the following way:
\begin{eqnarray}
\label{pallino}
p \circ p \, = \, \theta_E^2 (p_0^2 - p_1^2) \, + \theta_M^2 (p_2^2 + p_3^2),  
\end{eqnarray}
where we have chosen a frame in which the only nonzero entries of $\theta$ are $\theta^{01} = - \theta^{10} = \theta_E$ and $\theta^{23} = - \theta^{32} = \theta_M$. Therefore, we can also write 
\begin{eqnarray}
\label{pallinonew}
p \circ p \, = \, \theta_E^2 p^2 \, + (\theta_M^2 + \theta_E^2) (p_2^2 + p_3^2),  
\end{eqnarray}
and keep fixed the variable $p_{\perp}^2 = (p_2^2 + p_3^2)$. If $\theta_E$ is zero, we are in the magnetic case, and no further branch cut is present with respect to $p^2 > 4 m^2$. Otherwise, if $\theta_E$ is different from zero, we are in the electric case, and there is a new branch cut. To make things simpler, let us take $p_{\perp}$ equal to zero for a moment. In this case, the first branch cut is for $p^2 < 0$, the second one for $p^2 > 4 m^2$, both in the physical region. One of these two branch cuts is necessarily tachyonic, and perturbative unitarity is necessarily lost in this case. The authors analysed also the noncommutative $\Phi^4$ four point scattering amplitude, finding an analogous result.   

Another computation was then performed by the authors in \cite{bgnv} for noncommutative Yang-Mills theory, where facts related to the gauge symmetry mix together with the previously discussed phenomena. The theory under analysis is a $U(N)$ gauge theory: one of the main features of the noncommutative star-gauge invariance (\ref{stargaugeinvariance}) is that it imposes strong restriction on the allowed gauge groups. For example, groups like $SU(N)$ are not permitted, since their algebras do not close under star-commutation\cite{Matsubara, BonoraSchnabl, Jurco,Bars,Chaichian}. This fact produces a lot of problems if one desires to build up the noncommutative analogue of phenomenological commutative gauge theories, such as the Standard Model or QCD\cite{Schucker,StMod}. $U(N)$ groups are instead allowed. We notice that another striking feature of noncommutative pure gauge theory is that, being, as we mentioned, space-time itself ``nonabelian'' in a sense, even $U(1)$ is an interacting theory, with three and four point vertices determined by the non-vanishing of the star-commutator $[A_i \, \, ,\star \, \, A_j]$ in Eq.(\ref{fieldstr}). Perturbatively, its structure constants can be considered to depend on the momenta flowing into the vertices, a sign of the already mentioned merging of space-time and internal symmetries\cite{Sheikh}. The $U(1)$ part mixes actively with the $SU(N)$ one, and consistency with the Ward identities requires at least $U(N)$\cite{bonorasali}. 

Let us consider the analytical structure of the vacuum polarization tensor $\Pi_{i j}(p)$\footnote{We continue using $i,j=0,\dots,p$ for the space-time indices of the worldvolume of the brane, on which the effective noncommutative field theory is defined (see Chapter 2). Sometimes however, in the following chapters, we will use as well greek indices, where no confusion is allowed by the context}. Noncommutative gauge symmetry entails the Ward identity $p^{i}\Pi_{i j}(p) = 0$, therefore one can decompose the two point function as follows:

\bigskip

\medskip

\centerline{\includegraphics[width=6cm]{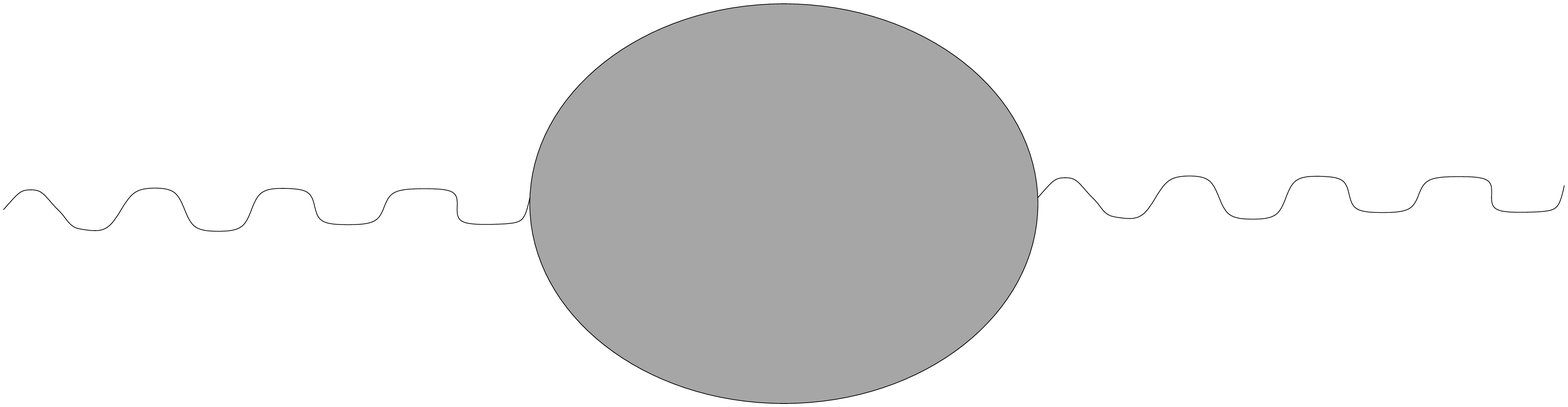}}

\bigskip

\begin{eqnarray}
\label{Ward}
\Pi_{i j}(p) \, \, \, = \,  \, (G_{i j} p^2 \, - \, p_{i} p_{j}) \, \Pi_1 \, + \, {\tilde{p}}_{i} {\tilde{p}}_{j} \, \Pi_2,     
\end{eqnarray}
where we have defined ${\tilde{p}}_{i} = \theta_{i j} p^{j}$. Of course, $\Pi_2$ will come exclusively from the nonplanar part of the diagram, while $\Pi_1$ will receive contributions from both the planar and the nonplanar part. One can write it as
\begin{eqnarray}
\label{pi1}
\Pi_1 \, = \, \Pi_1^{pl} \, + \, \Pi_1^{nonpl}. 
\end{eqnarray}
Here we report the values for all these quantities, as was calculated at one loop by the authors in \cite{bgnv} in Feynman gauge\footnote{We refer the reader to the paper mentioned for a detailed description of the Feynman rules in noncommutative Yang-Mills, as well as for an accounting of the various diagrams, including ghosts' contributions, that come into play in the one-loop calculation}:
\begin{eqnarray}
\label{pi1pl}
\Pi_1^{pl} \, = \, {{i g}\over {16 {\pi^2}}} \, {\Big[ {{- p^2}\over {4 \pi {\mu^2}}} \Big]}^{\omega - 2} \, {{\Gamma (2 - \omega ) {{\Gamma}^2} (\omega - 1)}\over {\Gamma (2 \omega - 2)}},
\end{eqnarray} 
\begin{eqnarray}
\label{pi1nonpl}
\Pi_1^{nonpl} \, &=& \, {{- i g}\over {4 {\pi^2}}} \, {\Big[ {{p^2}\over {16 {\pi^2} {\mu^4} {{\tilde{p}}^2}}} \Big]}^{{{\omega}\over {2}} - 1} \, \int_0^1 dx {[x (1 - x)]}^{{{\omega}\over {2}} - 1} \nonumber \\
&& \, \times \, [2 - (\omega - 1){(1 - 2x)}^2 ] \, K_{2 - \omega} (\sqrt{- p^2 x (1 - x) p \circ p}),
\end{eqnarray} 
\begin{eqnarray}
\label{pi2}
\Pi_2 \, &=& \, {{i g}\over {4 {\pi^2}}} \,  {[4 \pi \mu^2 ]}^{2 - \omega} \, {\Big[ {{4 p^2}\over {{{\tilde{p}}^2}}} \Big]}^{{{\omega}\over {2}}} \, \int_0^1 dx {[x (1 - x)]}^{{{\omega}\over {2}}} \, \nonumber \\
&&\, \qquad \qquad \times \, K_{\omega} (\sqrt{- p^2 x (1 - x) p \circ p}).
\end{eqnarray} 

The quantity $\omega$ is one half of the space-time dimensions.  A first comment one can make about this result is that, once again, UV/IR mixing is manifest. While the planar part has a pole in $\omega = 2$, namely when approaching four space-time dimensions, and is of course as commutative, the nonplanar part is finite in this limit, since the Moyal phase has regularized the UV. Anyway, the reult is IR singular when $\tilde{p}$ goes to zero: $\Pi_1$ has a logarithmic singularity $\log p \circ p$, while $\Pi_2$ has a pole ${(p \circ p)}^{- 2}$. 

From the point of view of perturbative unitarity, once again we notice the appearance of a branch cut driven by the noncommutativity parameter, in the same combination $p \circ p$ as before. Here, the first cut starts necessarily from zero, since the basic quanta of the noncommutative Yang-Mills field are massless. The same argument as before reveals that a tachyonic branch cut appears in the electric case. 
\section{Observables}
In this section, we will describe one of the features that mostly joins noncommutative quantum field theory and string theory. When looking for observables in noncommutative gauge theories, one has to face the problem that no gauge invariant local operators exist. This is easily understood, if we recall that the actual gauge group contains also space-time transformations; therefore, if an operator has to be gauge-invariant, he must also be invariant under space-time symmetries, and cannot be local. We can be more specific, in particular the following relation holds for the star-product:
\begin{eqnarray}
\label{trasl}
\exp [i \, k \, x] \, \star \, f(x) \, \star \, \exp [- i \, k \, x] \, = \, f(x \, + \, \theta k). 
\end{eqnarray}
Thus, $\exp [i \, k \, x]$ act as a generator of global translations. But, for a field transforming formally in the adjoint representation, Eq.(\ref{trasl}) is a star-gauge transformation, of the form
\begin{eqnarray}
\label{gaugeadj}
U(x) \, \star \, f(x) \, \star U^{\dagger} (x) \, = \, f'(x).
\end{eqnarray}
Therefore translations are a subset of the gauge transformations\footnote{Similar formulae hold for global rotations and dilatations for example}. A gauge-invariant operator must be at least translation-invariant, hence it must be global. 

The absence of local probes can be quite puzzling. One can still define objects which are gauge-invariant, and indeed carry a momentum different from zero, but they are to be constrained in a specific way. These are the Open Wilson Lines (OWL)\cite{iikk,amns,amns2,amns3,Gross,bv}. They were originally found following the duality with supergravity, since, from the SUGRA side, one has observables carrying non-zero momentum. The definition of this objects is as follows:
\begin{eqnarray}
\label{OWLines}
W (k, C) \, = \, \int d^d x \, \tr W (x, C) \, \star \, \exp [i \, k \, x],
\end{eqnarray} 
\begin{eqnarray}
\label{OWcore}
W (x, C) \, = \, P_{\star} \, \exp [i \, g \, \int_C A_i \Big( x \, + \, \zeta \Big) \, d\zeta^i].  
\end{eqnarray}
In these formulae, $C$ is a line in space-time, parameterized by $\zeta$, such that $\zeta^i (0) - \zeta^i (1) = l^i$, and $P_\star$ denotes the noncommutative path-ordering along $\zeta$ from left to right with respect to increasing $s$ of $\star$-products of functions. To prescribe the ordering is essential, since the star-product is not commutative. Finally, the symbol $\tr$ indicates the trace over the color indices of the internal $U(N)$ gauge group. One can give the explicit expression of (\ref{OWcore}) in terms of a power series, which provides also the natural link to perturbation theory calculations of correlators of these quantities:
\begin{eqnarray}
\label{pertexp}
W (x, C)&=&\sum_{n=0}^\infty (i g)^n \int_{0}^1 ds_1 \ldots \int_{s_{n-1}}^{1} ds_{n} \, \dot{\zeta^{i_1}} (s_1)\ldots\dot{\zeta^{i_n}}(s_{n}) \no\\
&&A_{i_1} \Big( x \, + \, \zeta (s_{i_1})\Big) \, \star \, \cdots \, \star \, A_{i_n} \Big( x \, + \, \zeta (s_{i_n})\Big). 
\end{eqnarray}
We now impose the gauge invariance of $W (k, C)$, observing that, for the local operator $W (x, C)$, one has the following rule:
\begin{eqnarray}
\label{OWcorerule} 
W' (x, C) \, = \, U (x) \, \star \, W (x, C) \, \star \, U^{\dagger} (x + l). 
\end{eqnarray}
Inserting Eq.(\ref{OWcorerule}) into Eq.(\ref{OWLines}), and using Eq.(\ref{trasl}), one immediately finds that, in order to recover gauge invariance, one has to impose
\begin{eqnarray}
\label{lenght}
l \, = \, \theta k.
\end{eqnarray}
The length of the OWL is proportional to the momentum carried by the line itself, which is another striking example of UV/IR mixing. As a remark, one notices that a quick way to turn an otherwise gauge-dependent local operator, such as for example $\tr F^2$, into its noncommutative gauge invariant version, is to attach to it an OWL $\int dx \tr F^2 \star W(x,C) \star \exp[ikx]$, provided again (\ref{lenght}) is satisfied. 

Aside from this highly constrained configuration, which is quite different from having a true \it local \rm observable\footnote{The operators $W (k, C)$ should be actually seen as a set of genuinely different operators at different momenta, rather then the same operator at different $k$'s}, one has a great novelty with respect to the commutative case, where only closed loops were gauge invariant\cite{test}. Here one recovers a closed loop for the particular value $k = 0$. The crucial point is that now the lines are a complete set of gauge invariant operators, like the loops were in the commutative case\cite{azam1,azam2,azam3}. They are the real observables of the theory. The fact that they are string-like has much to say about the stringy nature of noncommutative theories. This argument receives a further boost from the merging of space-time and internal symmetries in a larger group, in the following way: one can define a kind of OWL also for noncommutative scalar theories, or, \it vice versa \rm, one cannot avoid considering also connection-like quantities even starting with pure scalars. This consideration brought some authors to formulate the noncommutative scalar effective action entirely in terms of these lines\cite{rey,rey1,rey2,rey3}. These lines are shown to interact, joining and splitting, a fact which could have far-reaching consequences for the open string field theory\cite{rey}. 

\chapter{Wilson Loop in 2D NCYM: First Order}
\section{Introduction}
From what we said in the previous chapter, in order to understand completely the real nature of noncommutative field theories, it appears to be crucial to explore, first, the features of noncommutative space-time through the field-theory probes and computable quantities, trying to unravel the effects of the unknown unifying gauge structure. Then, in order to bring to a fully clarifying point the comparison with string theory, it is necessary to determine precisely the link between the building blocks of these theories, namely the scattering amplitudes. This last study will be done in the next chapter. Now, we concentrate on calculations which could reveal the response of noncommutative space-time to gauge probes. These probes have been defined in Chapter 3. In this chapter, we start presenting our results obtained in the analysis of the Wilson loop in a Yang-Mills theory defined on a two-dimensional noncommutative manifold. Some remarks are in order about this choice: first, in two dimensions the Lorentz symmetry is necessarily preserved. Of course, $\theta$ is necessarily of electric type: after a few steps, however, we will devote most of our treatment to an Euclidean formulation. In any case, it could be a right place where to study the consequences of a noncommuting time. Second, in the commutative case both the partition function and the Wilson loop on a two-dimensional manifold are exactly computable, thanks to an additional symmetry one has in this case, namely the invariance under area-preserving diffeomorphisms\cite{Witten2d}. Therefore, the commutative \it paradigma \rm is perfectly under control, included instantons' contribution, and one can rely on exact expressions for comparison. This will turn out to be extremely precious in trying to extrapolate a resummed expression from our perturbative computations, and to identify the correct symmetries to look for. Of course, no exact forms are achieved in the noncommutative case, even if we have persistence of the area-preserving invariance, because of the complicacy brought by the Moyal phase, which in turn reveals extremely interesting new effects when coupled to the loop. One warning is that finally, as it always happens, the two-dimensional theory could represent not only a too simplified version of the higher dimensional cases, but, as we will notice, it could develop autonomous features, that one does not find anymore if $d \neq 2$. Our idea is that these calculations can shed some light on the crucial points which one has to explore, and show some of the mechanisms that could occur, \it mutatis mutandis \rm, also in higher dimensions, and definitely serve as a paradigmatic exploration, in order to point out the directions in which to look. 

This matter is reported in \cite{nostro1} and \cite{nostro2}. For a comparison with previous Wilson loop calculations in four dimensional noncommutative gauge theory the reader can see \cite{bgnv}.
   
\section{Instruments}
We start recalling the expression of the Yang-Mills theory on a two-dimensio\-nal (brane worldvolume of dimension two)
non-commutative space for gauge group $U(N)$. We remind that, in two dimensions, the constant antisymmetric noncommutativity parameter is necessarily proportional to the antisymmetric tensor $\epsilon$.
According to what was showed in Chapter 2, our classical action can be written as 
\begin{eqnarray}
\label{action}
S=-\ \frac1{4} \int d^2x\, \tr \Big( F_{\m\n} \star  F^{\m\n} \Big),
\end{eqnarray}
where the noncommutative field strength $F_{\m\n}$ is given by Eq.(\ref{fieldstr}),
\begin{eqnarray}
F_{\m\n}=\p_\m A_\n -\p_\n A_\m -ig (A_\m\star A_\n - A_\n\star A_\m),
\end{eqnarray}
and $A_\m$ is a $N\times N$ matrix. We have rescaled here and in the following the gauge field in favour of perturbation theory. Our convention for the Casimir operator is $\tr T_N^a T_N^b = {\delta}^{a b}$, $T_N^a$ being the generators of the fundamental.

The $\star$-product was defined in Eq.(\ref{starproduct}).
We know that the action in Eq.~\re{action} is invariant under $U(N)$
non-commutative gauge transformations 
\begin{eqnarray}
\label{gauge}
\de_\l A_\m= \p_\m \l -ig (A_\m\star\l -\l\star A_\m) \,.
\end{eqnarray}
We choose the light-cone gauge $n^{\mu}A_{\mu}\equiv A_{-}=0$,
the vector $n_{\mu}$ being light-like, $n^{\mu}\equiv\frac{1}{\sqrt 2}(1,-1).$
The gauge condition can be imposed either adding to the action the gauge-fixing term
\begin{eqnarray}
S_{g.f.}=\frac1{\alpha} \int d^2x\,Tr\Big( (A_{-} \star  A_{-})(x) \Big)=
\frac1{\alpha} \int d^2x\,Tr \Big( (A_{-}(x))^2 \Big),
\end{eqnarray}
or by means of a Lagrange multiplier.

It is known that
Faddeev-Popov ghosts decouple even in non-commutative theories \cite{sheikh},
whereas the field tensor has only one component $F_{-+}=\p_{-} A_{+}.$

As we know, the free propagator coincides with the one of the ordinary theory. In two dimensions, two different prescriptions can be chosen for the vector propagator in momentum space, namely
\begin{eqnarray}
\label{hooft}
D_{++}=i\  PV\ [k_{-}^{-2}]
\end{eqnarray}
and
\begin{eqnarray}
\label{mand}
D_{++}=i\ [k_{-}+i\epsilon k_{+}]^{-2},
\end{eqnarray}
$PV$ standing for the Cauchy principal value.
The two expressions above are usually referred in the literature as 't Hooft \cite{hoo}
and Wu-Mandelstam-Leibbrandt ($WML$) \cite{w,m,le} propagators. They correspond to 
two different ways of quantizing the theory, namely by means of a light-front
or of an equal-time algebra \cite{bbg}, respectively. Their Fourier transforms to coordinate space are the following exchange potentials, respectively:
\begin{eqnarray}
\label{hoonellex}
D_{++}^{'t Hooft} \, = \, - \, {{i}\over {2}} \, |x_+ | \, \delta ({x_{-}} ),
\end{eqnarray}
\begin{eqnarray}
\label{mandnellex}
D_{++}^{WML} \, = \, - \, {{1}\over {2 \pi}} \, {{x_+ }\over {(x_- \, - i \epsilon \, x_+ )}}.
\end{eqnarray}
Here, $x_+ = x^- = {{1}\over {\sqrt{2}}} (x^0 - x^1 )$ and $x_- = x^+ = {{1}\over {\sqrt{2}}} (x^0 + x^1 )$. 

The loop for the commutative $U(N)$ theory, computed according to 't Hooft, coincides 
with the exact one, which can be obtained 
on the basis of
geometrical considerations \cite{boul,daul},
\begin{equation}
\label{looft}
{\cal W}= \exp (\frac{i}{2}g^2 N {\cal A}),
\end{equation}
and exhibits the Abelian-like exponentiation one expects on the basis of unitarity
arguments\cite{test},
whereas the $WML$ propagator leads to a different, genuinely perturbative
result\cite{stau}, in which topological effects are disregarded \cite{anlu},
\begin{equation}
\label{wml}
{\cal W}_{WML}= {{1}\over {N}} \, \exp (\frac{i}{2}g^2 {\cal A}) L^{(1)}_{N-1}(-i g^2 {\cal A}),  
\end{equation}
$L^{(1)}_{N-1}$ being a Laguerre polynomial\footnote{In two dimensions, the combination $g^2 A$ is dimensionless}.
The $WML$ propagator can be Wick-rotated, thereby
allowing for an Euclidean treatment. This is indeed the normal procedure followed in
order to obtain the solution above. The continuation of the propagator
is instead impossible when using the $PV$ prescription.
One should however bear in mind that, in the Euclidean formulation, the original
Minkowski contour of integration is changed to a (complex) contour, according to the rule $(x_0,x_1) \to (ix_2,x_1)$. The area ${\cal A}$
is thereby converted into $i {\cal A}$. Furthermore, one has to consistently rotate the noncommutativity parameter $\theta \rightarrow i \theta$. From what we have said, the Euclidean formulation 
cannot be defined if we choose
the 't Hooft propagator\footnote{As a side remark, we notice that the expression (\ref{looft}) leads trivially to confinement, since the area law for the Wilson loop reveals a linearly rising potential\cite{test}. Expression (\ref{wml}) has a confining character as well, but not in the 't Hooft limit of infinite number of colours $N$ with 't Hooft coupling $g^2 N$ fixed\cite{stau,QCD} (see the comments at formula (\ref{wuplanar}) in Section 5.3)}.

Even in two noncommutative dimensions, when quantized in axial gauge, the theory looks indeed free,
in particular the action
remains trivially invariant under area preserving diffeomorphisms. However, we must remember that this apparent triviality is due to the particular gauge choice, such that, in much the same way it happened in the commutative case, the nontrivial aspects of the theory are encoded in the singular nature of the light-cone propagator. Moreover, the coupling with the contour inside the correlation functions of the Wilson loop (and Wilson lines) provides a highly non-trivial behaviour due to the presence of the star-product, which we employ as a test for noncommutative space-time.

We recall the expression of the non-commutative Wilson loop (see Section 3.5): 
\begin{eqnarray}
\label{wloop}
\WW[C]=\frac{1}{N}<\int d^2x\,\tr {\cal T} P_{\star} \exp \lp
ig \int_C A_{+} 
(x+\xi(s))\, d\xi^{+}(s)\rp >\, ,
\end{eqnarray}
where $\tr$ is the $U(N)$ color trace, $C$ is a closed contour in non-commutative space-time
parameterized by $\xi(s)$, with $0 \leq s \leq 1$, and $P_\star$
denotes non-commutative path ordering along $x(s)$ from left to right
with respect to increasing $s$ of $\star$-products of functions. 

The perturbative expansion of $\WW [C]$,
expressed by Eq.~\re{wloop}, reads 
\begin{eqnarray}
\label{loopert}
&&\WW   [C]=\frac{1}{N}
\sum_{n=0}^\infty (-g^2)^n \int_{0}^1 ds_1 \ldots
\int_{s_{2n-1}}^{1} ds_{2n} 
\, \dot{\xi}_{-}(s_1)\ldots\dot{\xi}_{-}(s_{2n})\qquad \qquad \no\\
&&\qquad \int d^2x \langle 0\left|\tr{\cal T}\lq  A_{+}(x + \xi(s_1))
\star\ldots\star    A_{+}(x + \xi(s_{2n}))\rq \right|0\rangle, 
\end{eqnarray}
so that we can write
\begin{eqnarray}
\label{wpert}
\WW   [C]= 1 + g^2 \WW_2+g^4\WW_4+\OO(g^6)\, .
\end{eqnarray}
The way in which the expression in Eq.~\re{loopert} has to to be understood is the following:
\begin{eqnarray}
\label{precise}
\WW   [C]&=&\frac{1}{N}
\sum_{n=0}^\infty (-g^2)^n \int_{0}^1 ds_1 \ldots
\int_{s_{2n-1}}^{1} ds_{2n} 
\, \dot{\xi}_{-}(s_1)\ldots\dot{\xi}_{-}(s_{2n}) \no\\
&&\int d^2{p_1} \ldots \int d^2{p_n} \, \exp[ip_1 \xi (s_1 )] \ldots \exp[ip_n \xi (s_n )] \no \\  
&&{\cal {D}}_{+ \ldots +}^{a_1 , \ldots ,a_n} (p_1 , \ldots p_n ) \, \int d^2x \, \exp[ip_1 x] \star \ldots \star \exp[ip_n x]\no \\
&&\tr (T^{a_1} \ldots T^{a_n}), 
\end{eqnarray}
where ${\cal {D}}_{+ \ldots +}^{a_1 , \ldots ,a_n} (p_1 , \ldots p_n )$ is the correlation function among $n$ fields $A_{+}^a$ in momentum space, as it was evaluated in the commutative case. The following formula is then used:
\begin{eqnarray}
\label{Sheikh}
\int {{d^2x}\over {4\pi^2}} \, \exp[ip_1 x] \star \ldots \star \exp[ip_n x] \, = \delta (\sum_{i=1}^n p_i ) \, \exp[- {{i}\over {2}} \, \sum_{i<j} {\tilde{p}}_j p_i ]  ,
\end{eqnarray} 
where ${\tilde{p}} q = p_{\mu} {\theta}^{\mu \nu } q_{\nu}$.

 Through an explicit evaluation one is convinced that the function
$\WW_2$ in Eq.~\re{wpert} is reproduced by the single-exchange diagram, 
which is exactly as in the ordinary $U(N)$ theory. In fact, ${\cal {D}}_{++} (p_1 , p_2 )$ contains a factor $\delta (p_1 + p_2 )$, and ${\tilde{p}}_1 p_1$ is zero\footnote{Note that the delta functions present in the Green function $D$ always define a domain which is included in the support of the overall delta in Eq.(\ref{Sheikh}), so a global volume of space-time emerges from this conflict. We will ignore it in the sequel, understanding that the loop is defined up to a total volume.}. We can draw a picture of this situation, setting the graphical landscape in which to represent all the perturbative series. One has propagators which carry momentum, and connect different points of the boundary of a closed surface in the two dimensional plane, as in the following picture:

\bigskip

\bigskip

\medskip

\centerline{\includegraphics[width=5cm]{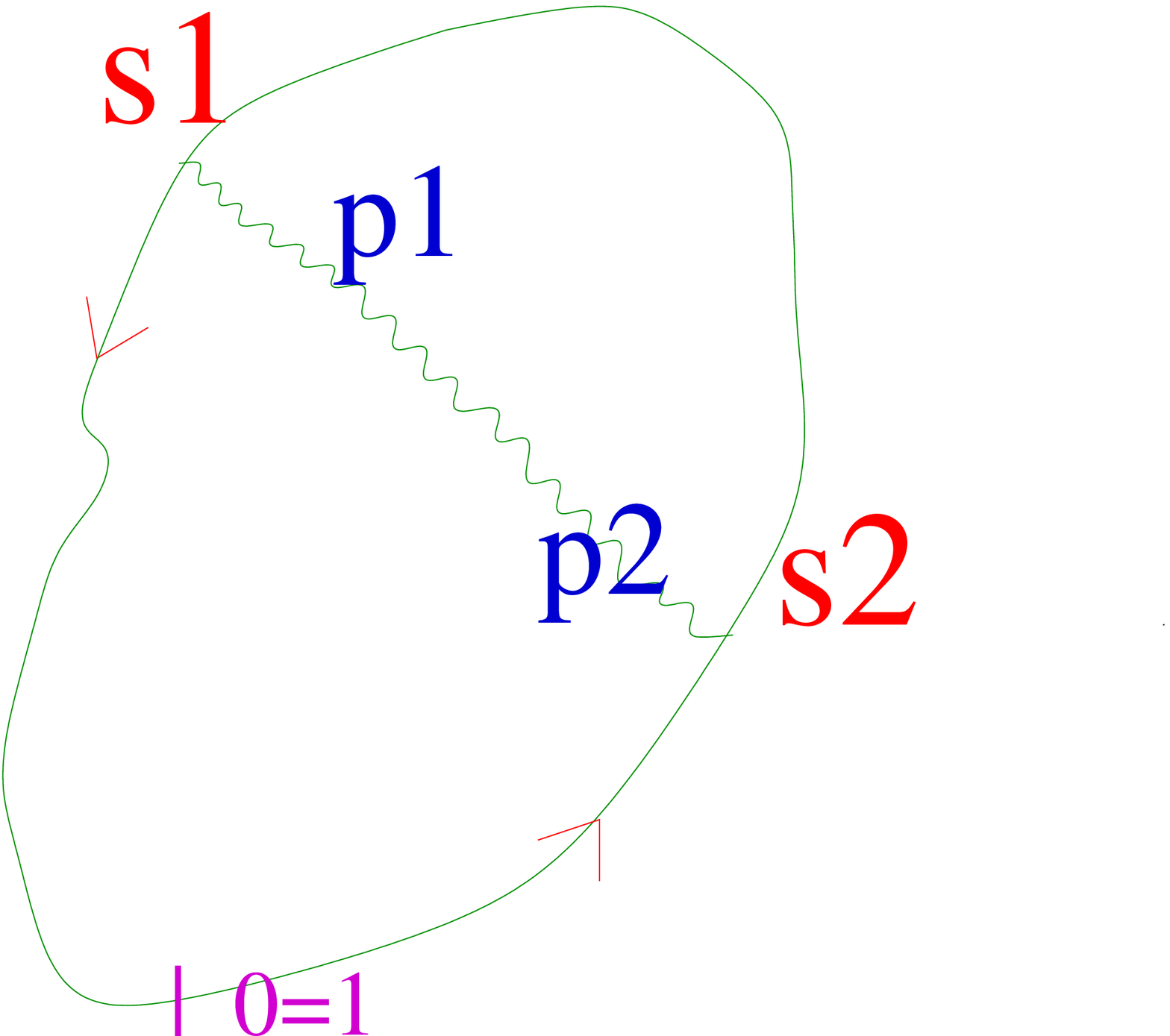}}

\bigskip

\bigskip

The first interesting quantity is $\WW_4$. It is obtained by
the exchange of two propagators, which can either cross or uncross. In
the pairings $(12)(34)$ and $(23)(41)$ they do not cross, whereas
in the pairing $(13)(24)$ they do. Let us denote by $\WW_4^{(1)}$,
$\WW_4^{(2)}$ and $\WW_4^{(cr)}$ the contributions in the three cases:

\bigskip

\bigskip

\centerline{\includegraphics[width=8cm]{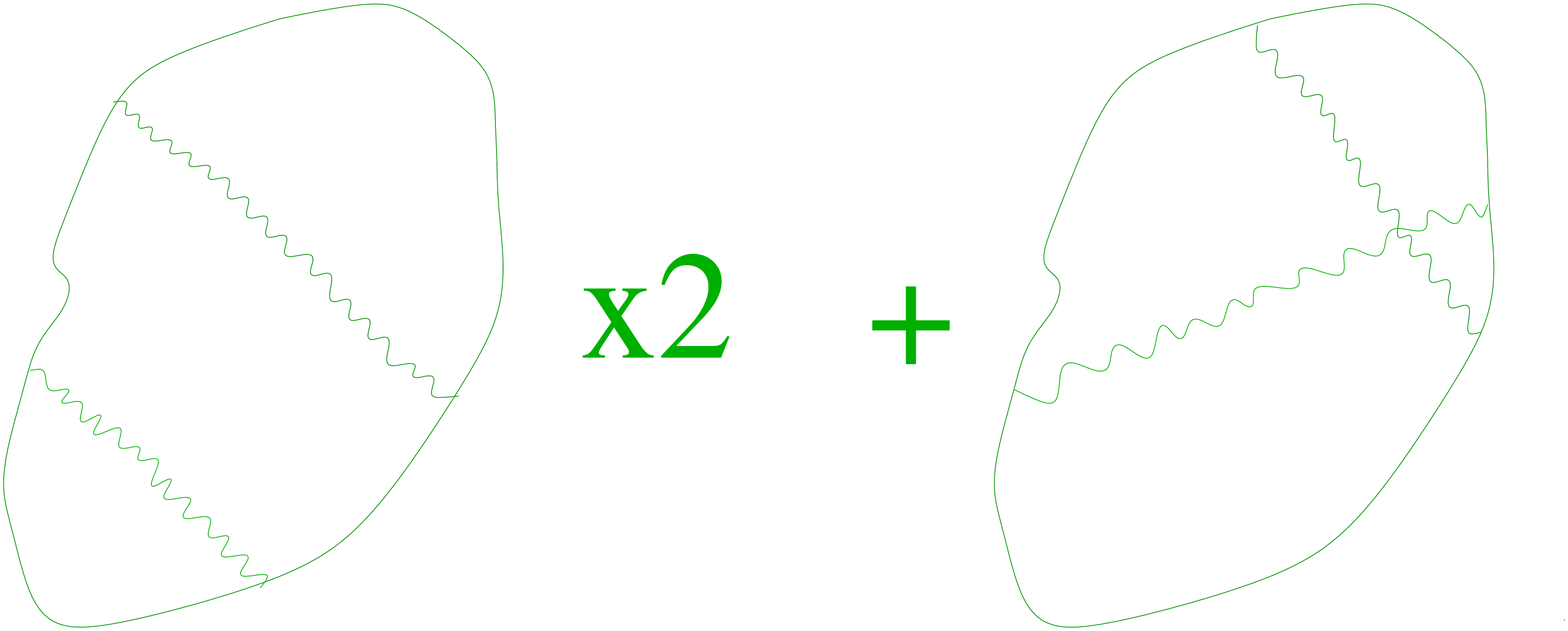}}

\bigskip

\bigskip

It is not difficult to realize that $\WW_4^{(1)}$ and
$\WW_4^{(2)}$ are the same as the corresponding ones in the ordinary 
theory, being unaffected by the Moyal phase. This reveals again that planar diagrams do not feel the noncommutativity. We therefore concentrate 
our attention on $\WW_4^{(cr)}$, which is instead affected by a noncommutative phase factor, as one can see from Eq.(\ref{Sheikh}).   

\section{The Minkowski formulation}
Our first exploration is in the Minkowski space, for the first non-trivial order, namely the first which feels the effects of noncommutativity. We will realize how strong they are, especially in the case of the 't Hooft prescription for the potential.

\subsection{The ${\cal O}(g^4)$ Calculation}
The ${\cal O}(g^4)$ contribution from the crossed diagram reads
\begin{eqnarray}
\label{mink}
\WW_{4}^{(cr)}\ &=& \int [ds] \, \dot{x}_{-}(s_1)\dots\dot{x}_{-}(s_{4})
\int \frac{dp_{+} dp_{-}}{-4\pi^2 [p_{-}^2]} \frac{dq_{+} dq_{-}}{4\pi^2 
[q_{-}^2]}
\, e^{i\theta (p_{-}q_{+}-q_{-}p_{+})} \nonumber \\
&\times& \exp(i[p_{+}(x_{-}(s_1)-x_{-}(s_3))+p_{-}(x_{+}(s_1)-x_{+}(s_3))])\nonumber \\
&\times& \exp(i[q_{+}(x_{-}(s_2)-x_{-}(s_4))+q_{-}(x_{+}(s_2)-x_{+}(s_4))]).
\end{eqnarray}
We see in fact that, with our conventions, the trace factor in formula (\ref{precise}) is $\tr (T_N^a T_N^a T_N^b T_N^b )= N^3$ (for planar diagrams), $\tr (T_N^a T_N^b T_N^a T_N^b )= N$ (for non-planar, remember the factor ${{1}\over {N}}$ in front of the loop). Throughout this and the following chapter we set $\theta^{0 1} = \theta$ for the noncommutativity parameter\footnote{With this convention, $[x_+ , x_- ] = i \theta$}. We notice that the quantity in Eq.(\ref{mink}) is in principle a complex quantity, since the complex conjugate of $\WW_{4}^{(cr)}(\theta)$ is $\WW_{4}^{(cr)}(-\theta)$. For both the prescriptions Eqs.(\ref{hooft}) or (\ref{mand}) we adopt a distributional formula \cite{test}, namely 
\begin{eqnarray}
\label{deriv}
[p_{-}^{- 2}] =  - \, {{\partial}\over {\partial p_{-}}} [p_{-}^{- 1}],
\end{eqnarray}
by which the equation above can be rewritten as
\begin{eqnarray}
\label{mink1}
\WW_{4}^{(cr)}\ &=& \  \int [ds] \, \dot{x}_{-}(s_1)\dots\dot{x}_{-}(s_{4})
\int \frac{dp_{+} dp_{-}}{4\pi^2 [p_{-}]} \frac{dq_{+} dq_{-}}{4\pi^2 
[q_{-}]}
\, e^{i\theta (p_{-}q_{+}-q_{-}p_{+})} \nonumber \\
&\times& \exp(i[p_{+}(x_{-}(s_1)-x_{-}(s_3))+p_{-}(x_{+}(s_1)-x_{+}(s_3))])\nonumber \\
&\times& \exp(i[q_{+}(x_{-}(s_2)-x_{-}(s_4))+q_{-}(x_{+}(s_2)-x_{+}(s_4))])\no\\
&\times&[ (x_{+}(s_1)-x_{+}(s_3)+\theta \ q_{+}][(x_{+}(s_2)-x_{+}(s_4)
-\theta \ p_{+}].
\end{eqnarray}
We now recall the well known relation between the WML distribution and 
the Cauchy principal value 
\begin{eqnarray}
\label{split}
[k_{-}+i\epsilon k_{+}]^{-1}= 
[k_{-}^{-1}]_{PV}-i\pi\ \delta (k_{-})\ sign(k_{+}).
\end{eqnarray}
By plugging this relation into Eq.(\ref{mink1}), we find that it splits naturally into four contributions, the first one being 
nothing but $\WW_4^{(cr)}$ evaluated with 't Hooft's prescription
\begin{eqnarray}
\label{pv1}
\WW_{4}^{(cr,\, PV)}\ &=& \  \int [ds] \, \dot{x}_{-}(s_1)\dots\dot{x}_{-}(s_{4})
\int \frac{dp_{+} dp_{-} dq_{+} dq_{-}}{16\pi^4 [p_{-}]_{PV} [q_{-}]_{PV}} 
\, e^{i\theta (p_{-}q_{+}-q_{-}p_{+})} \nonumber \\
&\times& \exp i[p_{+}(x_{-}(s_1)-x_{-}(s_3))+p_{-}(x_{+}(s_1)-x_{+}(s_3))]\nonumber \\
&\times& \exp i[q_{+}(x_{-}(s_2)-x_{-}(s_4))+q_{-}(x_{+}(s_2)-x_{+}(s_4))]\nonumber \\
&\times&[(x_{+}(s_1)-x_{+}(s_3)+\theta \ q_{+}][(x_{+}(s_2)-x_{+}(s_4)
-\theta \ p_{+}].
\end{eqnarray}
It is easy to realize that the expression above vanishes if we set $\theta = 0$ inside the integral, owing to the measure $[ds]$. In fact, in this case it reduces to
\begin{eqnarray}
\label{pvcomm}
\WW_{4}^{(cr,\, PV)}\ &=& \  \int [ds] \, \dot{x}_{-}(s_1)\dots\dot{x}_{-}(s_{4})
\int \frac{dp_{+} dp_{-}}{4\pi^2 [p_{-}]_{PV}} \frac{dq_{+} dq_{-}}{4\pi^2 
[q_{-}]_{PV}} \nonumber \\
&\times& \exp i[p_{+}(x_{-}(s_1)-x_{-}(s_3))+p_{-}(x_{+}(s_1)-x_{+}(s_3))]\nonumber \\
&\times& \exp i[q_{+}(x_{-}(s_2)-x_{-}(s_4))+q_{-}(x_{+}(s_2)-x_{+}(s_4))]\nonumber \\
&\times&[x_{+}(s_1)-x_{+}(s_3)][x_{+}(s_2)-x_{+}(s_4)].
\end{eqnarray}
The Fourier transform of the distribution $[k_{-}^{-1}]_{PV}$ is easily evaluated:
\begin{eqnarray}
\label{anti}
\int dk_{-} \, e^{i \, a \, k_{-}} \, [k_{-}^{-1}]_{PV} = \pi i \, sign(a).  \end{eqnarray}
What remains of the momentum integration gives just two delta functions, providing the result 
\begin{eqnarray}
\label{commzer}
&&\WW_{4}^{(cr,\, PV)}\ = \, {{(- 1)} \over {4}} \, \int [ds] \, \dot{x}_{-}(s_1)\dots\dot{x}_{-}(s_{4}) \, |x_{+}(s_1)-x_{+}(s_3)|  \qquad \qquad \qquad \nonumber \\
&&\qquad \times |x_{+}(s_2)-x_{+}(s_4)| \delta\Big( x_{-}(s_1)-x_{-}(s_3)\Big) \, \delta\Big(x_{-}(s_2)-x_{-}(s_4)\Big).
\end{eqnarray}
We see that the effect of the delta functions is to equate the $x_{-}$ coordinates of the ending points of the two propagators, thus preventing them from crossing. But the measure $[ds]$ has instead support where propagators do cross, so the result vanishes. We have of course recovered exactly the commutative result that crossed diagrams do not contribute to the loop when propagators are prescribed \it \`a la \rm 't Hooft, as one can easily realize looking at Eq.(\ref{hoonellex}): with 't Hooft's prescription, only parallel (in the $x_-$ variable) propagators are allowed. But this happens just because, by setting $\theta = 0$ \it inside \rm the integral, we are precisely repeating the commutative calculation. On the contrary, by performing the integral and \it then \rm sending $\theta \to 0$, we do not recover the commutative result, as we will see below.

When $\theta \ne 0$ the integrations over the momenta in Eq.(\ref{pv1}) can again be performed starting from Eq.(\ref{anti}), which gives
\begin{eqnarray}
\label{pvfinth}
\WW_{4}^{(cr,\, PV)}\ &=& \,  {{(- 1)}\over {16 {{\pi}^2}}}\, \int [ds] \, \dot{x}_{-}(s_1)\dots\dot{x}_{-}(s_{4}) \int dp_{+} dq_{+} \nonumber \\ 
&\times& \exp i\Big[ p_{+}(x_{-}(s_1)-x_{-}(s_3))\Big] \exp i\Big[ q_{+}(x_{-}(s_2)-x_{-}(s_4))\Big] \nonumber \\
&\times&  \Big[ \vert x_{+}(s_1)-x_{+}(s_3)+ \theta q_{+}\vert \, \vert x_{+}(s_2)-x_{+}(s_4)-\theta p_{+}\vert \Big].
\end{eqnarray}
We now collect a ${\theta}^2$ outside the integral and shift the momenta $p_{+}$ and $q_{+}$ to clean the arguments of the moduli. We are thus left with an expression that is the Fourier transform of the modulus distribution, and can be easily calculated:
\begin{eqnarray}
\label{antianti}
\int dk_{+} \, e^{i \, b \, k_{+}} \, |k_{+}| = (- 2) \, {{[b^{- 2}]}_{PV}},\end{eqnarray}
where the r.h.s. has to be interpreted as derivative of the Cauchy principal value distribution. The result is therefore the following
\begin{eqnarray}
\label{formal}
\WW_{4}^{(cr,\, PV)}\ &=& - \frac{\theta^2}{4\pi^2} \ \int [ds] 
\frac{\dot{x}_{-}(s_1)\dots\dot{x}_{-}(s_{4})}{[\Big(x_{-}(s_1)-x_{-}(s_3)\Big)^2]_{PV}
[\Big(x_{-}(s_2)-x_{-}(s_4)\Big)^2]_{PV}} \nonumber \\
&\times& \exp \frac{i}{\theta}\Big[\Big(x_{-}(s_1)-x_{-}(s_3)\Big)\Big(x_{+}(s_2)-x_{+}(s_4)\Big)\nonumber \, \, \, \, \, \, \, \, \, \, \, \,  \\
&&\, \, \, \, \, \, \, \, \, \, \, \, \, \, \, \, \, \, \, \, \, \, \, \, \, \, - \, \Big(x_{+}(s_1)-x_{+}(s_3)\Big)\Big(x_{-}(s_2)-x_{-}(s_4)\Big)\Big].
\end{eqnarray}

After having computed the first (PV) contribution coming from the splitting (\ref{split}), we turn to the second contribution, which comes from the product of the two $\delta$-distributions, namely
\begin{eqnarray}
\label{delta2}
\WW_{4}^{(cr,\, \delta)}&& =- \frac{1}{16\pi^2}\  \int [ds]\,  
\dot{x}_{-}(s_1)\dots\dot{x}_{-}(s_{4})
\int dp_{+} dq_{+}\ sign(p_{+})\ sign(q_{+})
\nonumber \\
&&\times \exp(i[p_{+}(x_{-}(s_1)-x_{-}(s_3))])\, \exp(i[q_{+}(x_{-}(s_2)-x_{-}(s_4))])
\nonumber \\
&&\times\Big( (x_{+}(s_1)-x_{+}(s_3))+\theta \ q_{+}\Big)\Big( (x_{+}(s_2)-x_{+}(s_4))
-\theta \ p_{+}\Big).
\end{eqnarray}
To solve the integrals over the momenta, we need again Eq.(\ref{antianti}), and its analogous for the $sign(p_{+})$ distribution, which can be easily obtained inverting Eq.(\ref{anti}). The result is 
\begin{eqnarray}
\label{delta22}
\WW_{4}^{(cr,\, \delta )}&=&\frac{1}{4\pi^2}\int [ds]\, \, \, \dot{x}_{-}(s_1)\dots\dot{x}_{-}(s_{4}) \\  
&&\Bigg[\frac{\Big(x_{+}(s_1)-x_{+}(s_3)\Big)
\Big(x_{+}(s_2)-x_{+}(s_4)\Big)}{\Big(x_{-}(s_1)-x_{-}(s_3)\Big)_{PV}
\Big(x_{-}(s_2)-x_{-}(s_4)\Big)_{PV}} \nonumber \\
&-&i\theta \frac{x_{+}(s_1)-x_{+}(s_3)}{\Big[ \Big(x_{-}(s_1)-x_{-}(s_3)\Big)^2
\Big]_{PV} \Big[ \Big(x_{-}(s_2)-x_{-}(s_4)\Big)\Big]_{PV}}\nonumber \\
&+&i\theta \frac{x_{+}(s_2)-x_{+}(s_4)}{\Big[ \Big(x_{-}(s_2)-x_{-}(s_4)\Big)^2
\Big]_{PV} \Big[ \Big(x_{-}(s_1)-x_{-}(s_3)\Big)\Big]_{PV}} \nonumber \\ 
&+&\theta^2 \frac{1}{\Big[ \Big(x_{-}(s_1)-x_{-}(s_3)\Big)^2
\Big]_{PV} \Big[ \Big(x_{-}(s_2)-x_{-}(s_4)\Big)^2 \Big]_{PV}}\nonumber \ \Bigg].
\end{eqnarray}
Finally, there is the term due to mixed products (see again Eq.(\ref{split})),
\begin{eqnarray}
\label{mixed}
\WW_{4}^{(cr,\, mx)}\ &=&- i\pi \int [ds] \dot{x}_{-}(s_1)\dots\dot{x}_{-}(s_{4})
\int \frac{dp_{+} dp_{-}}{4\pi^2} \frac{dq_{+} dq_{-}}{4\pi^2}
e^{i[\theta (p_{-}q_{+}-q_{-}p_{+})]} \nonumber \\
&\times& \Big[ \frac{sign(p_{+})\ \delta(p_{-})}{[q_{-}]_{PV}}+
\frac{sign(q_{+})\ \delta(q_{-})}{[p_{-}]_{PV}}\Big] \nonumber \\
&\times& \exp(i[p_{+}(x_{-}(s_1)-x_{-}(s_3))+p_{-}(x_{+}(s_1)-x_{+}(s_3))])\nonumber \\
&\times& \exp(i[q_{+}(x_{-}(s_2)-x_{-}(s_4))+q_{-}(x_{+}(s_2)-x_{+}(s_4))])\nonumber \\
&\times&\Big( x_{+}(s_1)-x_{+}(s_3)+\theta \ q_{+}\Big) \Big( x_{+}(s_2)-x_{+}(s_4)-\theta \ p_{+}\Big).
\end{eqnarray}
This term vanishes for symmetry reasons. The proof goes as follows: after integrating over $p_{-}$ and $q_{-}$, using again formula (\ref{anti}) where needed , and after some simple manipulations on the remaining momentum variables $p_{+}$ and $q_{+}$, we can rewrite the expression (\ref{mixed}) as the difference of two pieces, namely (omitting some common factors which are irrelevant to the present purpose)
\begin{eqnarray}
\label{dmixed}
\WW_{4}^{(cr,\, mx)}\ &\propto&  \int [ds] \dot{x}_{-}(s_1)\dots\dot{x}_{-}(s_{4})
\int dp_{+} dq_{+} \exp[-ip_{+}(x_{-}(s_1)-x_{-}(s_3))]\nonumber \\
&&\times \exp [i q_{+}(x_{-}(s_2)-x_{-}(s_4))] sign(q_{+} ) \nonumber \\
&&\vert x_{+}(s_1)-x_{+}(s_3)+\theta \ q_{+}\vert \, (x_{+}(s_2)-x_{+}(s_4) + \theta \ p_{+} ) \, - \, \no \\
&&\int [ds] \dot{x}_{-}(s_1)\dots\dot{x}_{-}(s_{4})
\int dp_{+} dq_{+} \exp[-iq_{+}(x_{-}(s_1)-x_{-}(s_3))] \nonumber \\
&&\times \exp [i p_{+}(x_{-}(s_2)-x_{-}(s_4))] sign(q_{+} ) \nonumber \\
&& ( x_{+}(s_1)-x_{+}(s_3)+\theta \ q_{+}) \, \vert x_{+}(s_2)-x_{+}(s_4) + \theta \ p_{+} \vert. 
\end{eqnarray}
What we want to prove now is that a suitable change of variables in the second term of the difference can make it exactly cancel the first one. To reach this goal we perform this sequence of transformations on the integration variables: 1) $s_3 \leftrightarrow s_2$ 2) $s_1 \leftrightarrow s_2$ 3) $s_1 \leftrightarrow s_4$ 4) $s_2 \leftrightarrow s_4$ 5) $s_i \leftrightarrow (1-s_i ) \, \forall i$ 6) $p_{+} \leftrightarrow - p_{+}$ 7) $q_{+} \leftrightarrow - q_{+}$. This should be understood as follows: one has to exchange $s_3 $ with $s_2 $, then in the resulting new integral one exchanges $s_1 $ with the actual $s_2 $, and so on. After doing this sequence of change of variables, one realizes that, to complete the work, it is necessary to specify the behaviour of the quantities $x_{-}(s_a)-x_{-}(s_b)$ and $x_{+}(s_a)-x_{+}(s_b)$ under the fifth transformation $s_i \leftrightarrow (1-s_i ) \, \forall i$: relying on invariance under area preserving diffeomorphism, it is sufficient to find the result for a specific contour, and we choose a simple circle, parameterized by $x_{+} = R \cos 2 \pi s , x_{-} = R \sin 2 \pi s$. With this choice, one immediately proves that the sequence of transformations reduces the second piece to the first one with opposite sign, which completes the proof. In order to have a check we tried with other entire classes of contours, finding the same result.             

To summarize, the results for the WML and the 't Hooft prescriptions differ in the presence of the contribution written in Eq.(\ref{delta22}).

\subsection{'t Hooft against WML}
We look at 't Hooft's expression Eq.(\ref{formal}), and expand the exponential inside the integral in powers of $1 \over {\theta}$. The global effect is to obtain a series starting from $\theta^2$, $\theta$, $\theta^0$ and going on with negative powers of $\theta$. We realize that the $\theta^2$ and the $\theta$ term of the series obtained in this way are divergent. The reason is that, for these two terms, the numerator does not succeed in sterilizing the singularity of the denominator. An intuitive argument for this fact is based on power counting: to reach the number of powers present in the denominator, and therefore to be able to compete with it and to sterilize it, one has to go to the second order in the expansion of the exponential. The orders zero and one cannot compete with the four powers of the denominator, and cannot clean it. This argument has to be confirmed by an explicit proof, because there could be cancellations or hidden symmetries that succeed in making these first two pieces finite. To assure that this is not the case, and that these terms really diverge, was a dreadful effort, and we will not report here the details. We have proved it in two different ways. The first one is a very practical way: we start again from Eq.(\ref{mink}), defined with 't Hooft's prescription for the two propagators, and then we integrate first all the loop variables, choosing, for simplicity, a rectangular contour \footnote{As we will do in another occasion in the sequel, we introduce a rectangular contour, which is not a smooth one, hence it could be unreachable with smooth diffeomorphisms, or it could introduce additional singularities at its vertices. We have copious evidence this is not the case: first of all, we will prove that the Minkowski result for WML prescription on a rectangle is the same (suitably rotated) as the Euclidean value on a circle. Furthermore, we can always consider the rectangle as a limit of smooth curves, and we do not expect any singularity in this limit. Eventually, the rectangle is even more significant than other contours, if we interpret it as in the commutative case, namely as the closure of the wordlines of two infinitely heavy quarks sitting at a certain distance from each other}. We end up with an integral over the four momentum variables of a quite complicated function of the momenta. We integrate over $q_{-}$ and $p_{-}$ using complex analysis to take into account the principal value prescription, ending up with a two dimensional integral of a quite wild function of $q_{+}$ and $p_{+}$. Now, we cutoff the integral everywhere is needed in order to make it explicitly finite, in particular both at high momenta and at small momenta, with two independent regulators (we put the integral in a finite square and moreover extract a hole around the origin). In this way, we have a perfectly finite result which we calculate introducing a suitable parametrization. One realizes at this point that this result does depend on the IR regulator as $(log\epsilon)^2$ with a non zero coefficient, therefore the integral has a divergence which cannot be cured. We will mention later the second proof.

For all the remaining terms of the series, starting from the constant and going to all the negative powers of $\theta$, the numerator is able to cancel the divergence of the denominator. The power counting provides a first evidence of this fact, as we have already seen, but this should be too naive without the further consideration that the numerator goes quickly to zero whenever also the denominator does, just because of the nested measure of integration. In order to corroborate this preliminary argument, we will finally rely on the calculation (see below) of the contribution of the order $\theta^0$ coming from the exponential (which is also the only one surviving in the large-$\theta$ limit) to confirm it is finite. This will give strong support to the argument that at every order in $\theta$ the numerator and the nested measure will regulate the denominator in the WML prescription. 

What we learned is therefore that the Wilson loop at this perturbative order, when calculated using the 't Hooft (PV) prescription, is singular, namely it is infinite for any value of $\theta \neq 0$. This is somehow surprising, especially when compared with the commutative case. In particular, we see that there is of course no continuity of this result in the ``commutative'' limit $\theta \to 0$.

We have developed an argument which may explain the reason of such a singular behaviour. 't Hooft's prescription is a rather particular one, and it is certainly not smooth. We remember that in the commutative case the \it perturbative \rm loop evaluated with 't Hooft's propagator is equal to the exact nonperturbative one, which is quite a strange fact. If we consider 't Hooft's propagator in the coordinate space, we see that it has the form of Eq.(\ref{hoonellex}), i.e. it represents a contact interaction (in the coordinate $x_{-}$). In the case in which the coordinates of the space-time do commute, this potential is certainly peculiar and probably responsible of the strange behaviour we have mentioned, but not so bad, since it produces finite and sensible results. In the situation in which the coordinates no longer commute and in which an uncertainty principle is automatically implemented among the coordinates, such a pointwise potential could be naturally ill-defined. Since noncommutativity does not allow to single out ``points'' in the space-time manifold, such a pointwise force may lose its meaning, and, as a consequence, our divergence might ensue. 

As we have seen, in order to obtain the WML result we have to add Eq.(\ref{delta22}) to Eq.(\ref{formal}). If we do this, we see that all the singular terms we have just described are exactly cancelled by analogous contributions with the opposite sign coming from $\WW_{4}^{(cr, \, \delta)}$, such that we are left with a regular WML loop. This is quite surprising, but could be interpreted along the same line we have sketched above: the WML potential has a very smooth form (\ref{mand}), which has no apparent reasons to conflict with space-time noncommutativity. Such a smooth interaction gives therefore origin to a regular Wilson loop, and the way it works signals that, actually, the entire divergence with the PV propagator comes from the unbalanced delta-like (pointwise) term hidden in the PV propagator. In turn, this provides a singularity which has a very trivial $\theta$-dependence, just a quadratic and a linear term\footnote{This is also easily understood: the delta-like terms appear under derivatives, which in turn have to be brought by parts on the exponentials, which contain $\theta$ in their exponents. In so doing, one recovers trivially $\theta^2$ when both delta's are present, and $\theta$ in the mixed terms. Another thing we notice is that the quadratic term does not depend on the area of the loop} with two infinite coefficients in front of them. This fact is another source of surprise: while, in the commutative case, the two prescriptions were essentially different, since in particular 't Hooft's one contains all instanton contributions, now their difference is striking, because one of them makes the loop divergent, but only through a couple of simple (infinite) terms. If they are disregarded, the two expansions are identical.                    

The complete expression obtained for WML is
\begin{eqnarray}
\label{WML}
&&\WW_{4}^{(cr,\, WML)}=\frac{- \theta^2}{4\pi^2}\int [ds]\frac{\dot{x}_{-}(s_1)\dots\dot{x}_{-}(s_{4})}{\Big[x_{-}(s_1)-x_{-}(s_3)\Big]^2_{PV} \Big[x_{-}(s_2)-x_{-}(s_4)\Big]^2_{PV}} \nonumber \\
&\times& \sum_{n=2}^{\infty} {{{(\frac{i}{\theta})}^n}\over {n!}} \Bigg[\Big(x_{-}(s_1)-x_{-}(s_3)\Big)\Big(x_{+}(s_2)-x_{+}(s_4)\Big)\nonumber \, \, \, \, \, \, \, \, \, \, \, \, \nonumber \\
&& \, \, \, \, \, \, \, \, \, \, \, \, \, \, \, \, \, \, \, \, \, \, \, \, \,- \, {\Big(x_{+}(s_1)-x_{+}(s_3)\Big)\Big(x_{-}(s_2)-x_{-}(s_4)\Big)\Bigg]}^n \, + \\ 
&&\int {{[ds]}\over{4\pi^2}}\dot{x}_{-}(s_1)\dots\dot{x}_{-}(s_{4})\frac{\Big(x_{+}(s_1)-x_{+}(s_3)\Big)\Big(x_{+}(s_2)-x_{+}(s_4)\Big)}{\Big[x_{-}(s_1)-x_{-}(s_3)\Big]_{PV}\Big[x_{-}(s_2)-x_{-}(s_4)\Big]_{PV}}\nonumber.
\end{eqnarray}
The limit at $\theta=0$ of Eq.(\ref{WML}) is best found by considering the sum of Eq.(\ref{formal}), that we have proved to vanish for this value of the noncommutativity parameter, and of Eq.(\ref{delta22}), in which only the first term survives. Therefore we find  
\begin{eqnarray}
\label{deltar}
\WW_{4}^{(cr, \, WML)}(\theta=0)\ &=& 
\  \frac{1}{4\pi^2}\  \int [ds]\ 
\dot{x}_{-}(s_1)\dots\dot{x}_{-}(s_{4})\nonumber \\
&&\frac{\Big(x_{+}(s_1)-x_{+}(s_3)\Big)
\Big(x_{+}(s_2)-x_{+}(s_4)\Big)}{\Big(x_{-}(s_1)-x_{-}(s_3)\Big)
\Big(x_{-}(s_2)-x_{-}(s_4)\Big)}, 
\end{eqnarray}
that is nothing but the usual commutative WML result, as one can see from Eq.(\ref{mandnellex}). We have continuity in the commutative case for WML, and, according to what we said in Chapter 3, this can be considered one of the few examples where it happens. We further refer to the remark we will do on this point in the next, after Eq.(\ref{reintegrato}).

We notice that it is redundant to prescribe the denominators in Eq.(\ref{deltar}) since the measure provides the necessary regularization. We will find that the same is true for the large-$\theta$ limit below. The value of Eq.(\ref{deltar}) is
\begin{equation}
\label{tetazero}
\WW_{4}^{(cr, \, WML)}(\theta = 0) =\ - {{{\cal A}^2 }\over {24}},
\end{equation}
where $\cal A$ is the area of the loop. The large-$\theta$ limit of Eq.(\ref{WML}) is easily found to be
\begin{eqnarray}
\label{formalgg}
\WW_{4}^{(cr, \, WML)}|_{\theta \to \infty}\ &=&\ \  \frac{1}{8\pi^2} \ \int [ds]\ 
\frac{\dot{x}_{-}(s_1)\dots\dot{x}_{-}(s_{4})}{[\Big(x_{-}(s_1)-x_{-}(s_3)\Big)^2]
[\Big(x_{-}(s_2)-x_{-}(s_4)\Big)^2]} \nonumber \\
&\times& \Big(\Big[\Big(x_{-}(s_1)-x_{-}(s_3)\Big)\Big(x_{+}(s_2)-x_{+}
(s_4)\Big)\Big]^2\nonumber \\
&& \, \, \, \, \, \, \, + \, \, \Big[\Big(x_{+}(s_1)-x_{+}(s_3)\Big)\Big(x_{-}(s_2)-x_{-}(s_4)\Big)\Big]^2\Big).
\end{eqnarray}
The integrals in Eq.(\ref{formalgg}) can be fairly easily computed, 
using for instance a rectangular contour, where all integrands become polynomials. In this way one can see it is finite and independent of the prescription used for the denominators; the final outcome is
\begin{equation}
\label{finres}
\WW_{4}^{(cr, \, WML)}|_{\theta \to \infty} = {{{\cal A}^2 }\over {8 {\pi }^2 }}.
\end{equation}
One remark is in order: at odds with what happens in dimension larger than two, where a decrease at large $\theta$ is expected in crossed diagrams due to the exponential oscillations (see for example \cite{mvs}), in two dimensions the contribution of the crossed diagram is not suppressed at large noncommutativity. Indeed it is instead of the same order of magnitude as the planar one. This will be a persistent feature of all our calculations, even at higher orders, and has been confirmed also in other two-dimensional explorations performed in the literature\cite{bv}. 

\section{The Euclidean Formulation}
Since we have realized that 't Hooft's prescription is incompatible with the noncommutative Wilson loop, we concentrate our attention purely on WML and on the Euclidean framework, and choose to generalize our treatment from the beginning to a number $n$ of windings around the loop, for a reason that will become clear later, and that is crucial for our scopes. 

\subsection{The ${\cal{O}}(g^4 )$ Calculation}
We consider the simple choice of a circular contour 
\begin{eqnarray}
\label{cerchio}
x(s)\equiv(x_1(s),x_2(s))=r(\cos(2\pi s),\sin(2\pi s)).
\end{eqnarray}
Then we are led to the expression
\begin{eqnarray}  
\label{crociato}
\WW_4^{(cr)} &=& \, r^4 \, \int_0^{n} d{s_1} \int_{s_1}^{n} d{s_2} \int_{s_2}^{n} 
d{s_3 }  \int_{s_3}^{n} d{s_4} \times \nonumber \\
&&\int_0^{\infty } \, {{dp}\over {p}} {{dq}\over {q}} 
\, \int_0^{2 \pi } d\psi \, d\chi \, \exp({- 2 i (\psi + \chi )})\nonumber \\
&&\exp({2 i p \sin \psi \sin \pi (s_1 - s_3 )}) \, \exp({2 i q \sin \chi \sin \pi (s_2 - s_4 )}) \nonumber \\
&&\exp({i {{\theta }\over {r^2}} 
p \ q\ \sin [\psi - \chi + \pi (s_2 + s_4 -  s_1 - s_3)] })\nonumber \\
&=& {\cal A}^2 \ F(\frac{\theta}{{\cal A}},n),
\end{eqnarray}
where ${\cal A}$ is the Euclidean area enclosed by the loop, and we recall that we have turned $\theta$ into $i \theta$ (see section 4.2).

This formula has been obtained by using the Wick-rotated expression of the propagator (\ref{mand}), namely
\begin{eqnarray}
\label{mandrot}
D_{++} \, = \, - \, {{2}\over {{(k_1 \, - \, i k_2 )}^2 }}
\end{eqnarray}
with the prescription of Wu\cite{w} of ``symmetric integration''. Here, it means that we perform the integrals over the angles \it before \rm the ones over the corresponding radial variables \footnote{We have expressed the two-dimensional momenta in polar coordinates}.

Integrating over the phase $\psi$ and its associate radial variable $p$ as indicated in formula (\ref{prima}) of the Appendix A, we get, after trivially rescaling $n$ in all the $s_i$ and performing the change of variable $z = e^{i \chi}$,
\begin{eqnarray}  
\label{integrato}
\WW_4^{(cr)} \, &=& \, \pi r^4 n^4 \int [ds]_4 \, 
\int_0^{\infty } {{dq} \over {q}} \, 
\ointop_{|z|=1}\, {{dz} \over {i {{z}^3}}} \, \, 
e^{-\ q  \sin [n \pi (s_4 - s_2 )] (z - {{1} \over {z}})} \nonumber \\
&&\, \, \, \, \, \, \, \, \, \, \, \, \, \, \, \, \, \, \, \, \, \, \, \, \, \, \, \, \, \, \, \, \, \, \, \, \, \, \, \, \, \, \, \, \, \, \, \, \, \, \, \, \, \, \, \, \, \, \, \, \, \, \, \, \, \, \, \, \, \, \, \, \, \, \, \, \, \, \, \, \times \, {{1 - {{\gamma } \over {z}} e^{- i n \pi \sigma }} 
\over {1 - \gamma z e^{i n \pi \sigma }}},
\end{eqnarray}
where  $\sigma = s_1 + s_3 - s_2 - s_4 $, and
$$ \gamma = \frac {\theta q }{2 r^2 \sin [n \pi (s_3 - s_1 )]},\qquad\qquad
\int [ds]_4 = \int_0^1 ds_1 \int_{s_1 }^1 ds_2 \int_{s_2 }^1 ds_3 \int_{s_3 }^1 ds_4 .$$ 
One can notice that the fraction appearing in Eq.(\ref{integrato}) is actually a pure phase in the domain of integration of $z$. We can further integrate over $z$ and over $q$: this is quite a complicated procedure, but it can be completely carried out, as described in the formula (\ref{seconda}) of the Appendix A. 
The outcome of this calculation is
\begin{eqnarray}  
\label{reintegrato}
\WW_4^{(cr)} \,
&=&2n^4{\cal A}^2 \int [ds]_4\left[\frac{1}{2}+\frac{2}{\beta^2}
\left(\exp[i\beta \sin\alpha]-1-i\beta \sin \alpha\right)\right] \nonumber \\
  &=&\, 2n^4{\cal A}^2 \int [ds]_4
\left[\frac{1}{2}+\frac{2}{\beta^2}\sum_{m=2}^{\infty}\frac{(i\beta \sin \alpha)^m}
{m!}\right],
\end{eqnarray}
where
\begin{eqnarray}
&&\alpha=n\pi (s_1+s_3-s_2-s_4),\no \\  
&&\beta=\, {{4{\cal A}}\over {\pi \theta}}\, \sin[n\pi (s_4-s_2)] \, \sin[n\pi(s_3-s_1)].
\end{eqnarray}
Let us keep fixed the value of the area $\cal{A}$. It is easy to check from (\ref{reintegrato}) that the function $F$ in Eq.(\ref{crociato}) is continuous at 
$\theta=0$ (which corresponds to large $\beta$), with $F(0)=\frac{n^4}{24}$, exactly corresponding to the value
of the commutative case obtained with the WML propagator \cite{stau}. 

Even if we find continuity, we see that $\theta=0$ ($\beta=\infty$) is a point of non-analyticity of the integrand of (\ref{reintegrato}), in particular the exponential has an essential singularity: continuity is obtained thanks to the denominator. On the contrary, it appears that the natural point of analyticity around which to expand the integrand is $\theta=\infty$ ($\beta=0$), as one can see writing $\WW_4^{(cr)}$ as a series like in Eq.(\ref{reintegrato}). This is a feature we have already encountered in the Minkowski formulation (see Eq.(\ref{formal}) and Eq.(\ref{WML})), and that will appear systematically even in the higher orders calculations. The presence of continuity is connected to the fact that in two dimensions one does not have UV/IR mixing, since the integrals tipically converge in the UV (see for example formula (\ref{13})). But, as we noticed in Chapter 2, $\theta =0$ is still quite a singular point. 

One could wonder whether this behaviour is in some way modified by the integration over the loop variables, which is still to be performed. We are not able to do it analytically for arbitrary values of $\theta$, but we can evaluate the first correction at small $\theta$, and, more easily, find the first two or three orders in the large-$\theta$ expansion. The first order correction in $\theta$ can be singled out by considering that the exponential and the ``$-1$'' in Eq.(\ref{reintegrato}) are more depressed than the term linear in $\beta$:
\begin{equation}  
\label{piccolo}
\WW_4^{(cr)} \, \simeq 2n^4{\cal A}^2 \int [ds]_4
\left[\frac{1}{2}-\frac{2i}{\beta}\sin\alpha\right].
\end{equation}
The  ${{1}\over {2}}$ -part is easy to integrate, while the second piece is equal to
\begin{eqnarray}  
\label{st1}
\WW_\theta&=&-in^4{\cal A}\pi\theta \int_0^1[ds]
{\sin[n \pi (s_1+s_3-s_2-s_4)]\over\sin[n \pi (s_4-s_2)]
\sin[n\pi (s_3-s_1)]}\nonumber\\
&&\equiv -in^4{\cal A}\pi\theta I,
\end{eqnarray}
with measure $[ds]=ds_1 ds_2 ds_3 ds_4 \theta(s_4-s_3)\theta(s_3-s_2
\theta(s_2-s_1)$. Integrating in $ds_1$ and $ds_4$ leads to
\begin{eqnarray}  
\label{st2}
I&=& {1\over n^2 \pi^2}\int_0^1ds_2 ds_3\theta(s_3-s_2)\Bigg( -n \pi 
\cos[2 n \pi (s_3-s_2)]\times \nonumber\\
&&\Big[ s_2 \log |{\sin n\pi s_2\over \sin n \pi(s_3-s_2)}
|+(1-s_3)\log |{\sin n\pi s_3\over \sin n \pi(s_3-s_2)}
|\Big] \nonumber\\
&&+\sin[2 n \pi (s_3-s_2)]\Big[ -n^2 \pi^2 s_2 (1-s_3)+
\log |\sin n \pi (s_3-s_2)| \no \\
&&\times \log |{\sin n \pi (s_3-s_2)
\over\sin n\pi s_2 \sin n\pi s_3}| \nonumber\\
&&+\log |\sin n \pi s_2 |\log |\sin n \pi s_3 |\Big]\Bigg)
= I_1+I_2 ,
\end{eqnarray}
where $I_1$ and $I_2$ refer to the first and second square brackets 
in (\ref{st2}), respectively.
The two integrals in $I_1$ coincide. They can be easily performed, leading to
\begin{equation}
\label{st3}
I_1=-\left( {1\over 6 n \pi}+{1\over 4 n^3 \pi^3}\right)\ .
\end{equation}
Concerning $I_2$, the first term is trivial, and provides us with a factor
equal to $1/(8 n^3 \pi^3) - 1/(12 n\pi)$, whereas in the remaining integrals it is more convenient to integrate first on one variable, and then to add
the integrands together before performing the final integration, i.e.
\begin{eqnarray}
\label{st4}
I_2 &=&{1\over 8 n^3 \pi^3}-{1\over 12 n\pi}-{1\over 2 n^3 \pi^3}
\int_0^1 ds\sin n\pi s\times\nonumber\\
&&\Bigl[2 n \pi s\cos n\pi s\log|\sin n \pi s |+\sin n \pi s 
(\log |\sin n \pi s |-1)\Bigr]\nonumber\\
&=&{1\over 4 n^3 \pi^3}-{1\over 12 n\pi}\ .
\end{eqnarray}
Adding (\ref{st3}) and (\ref{st4}), taking Eq.(\ref{st1}) into account, and finally adding the result coming from the ${{1}\over {2}}$ -part, we get
\begin{equation}
\label{respic}
\WW_4^{(cr)}\simeq \frac{n^4{\cal A}^2}{24}+
i \theta \frac{n^3{\cal A}}{4}.
\end{equation}
The Taylor series in the natural variable $1\over \theta$ starts instead as\footnote{This is tedious but straightforward, since here one has to integrate sinusoidal functions present at the numerator only, at odds with Eq.(\ref{piccolo}) where one has sine functions at the denominator. This is another example of the naturalness of the expansion in $1\over \theta$}  
\begin{equation}
\label{granteta}
\WW_4^{(cr)} \, =-\frac{n^2{\cal A}^2}{8\pi^2}+i\frac{n^3{\cal A}^3}{8\pi^2\theta}+
\frac{8n^4{\cal A}^4}{3\pi^2\theta^2}\left(\frac{1}{256}+\frac{175}{3072}\frac{1}
{n^2\pi^2}\right)+{\cal O}(\theta^{-3}).
\end {equation}

We have integrated numerically the expression (\ref{reintegrato}) for a quite large number of values of $\theta$, intermediate between the two asymptotic regimes: this has fully confirmed that in these regions the behaviours are Eq.(\ref{piccolo}) and Eq.(\ref{granteta}), respectively. In the following we report the plots.    

\centerline{\includegraphics[width=9.2cm,angle=270]{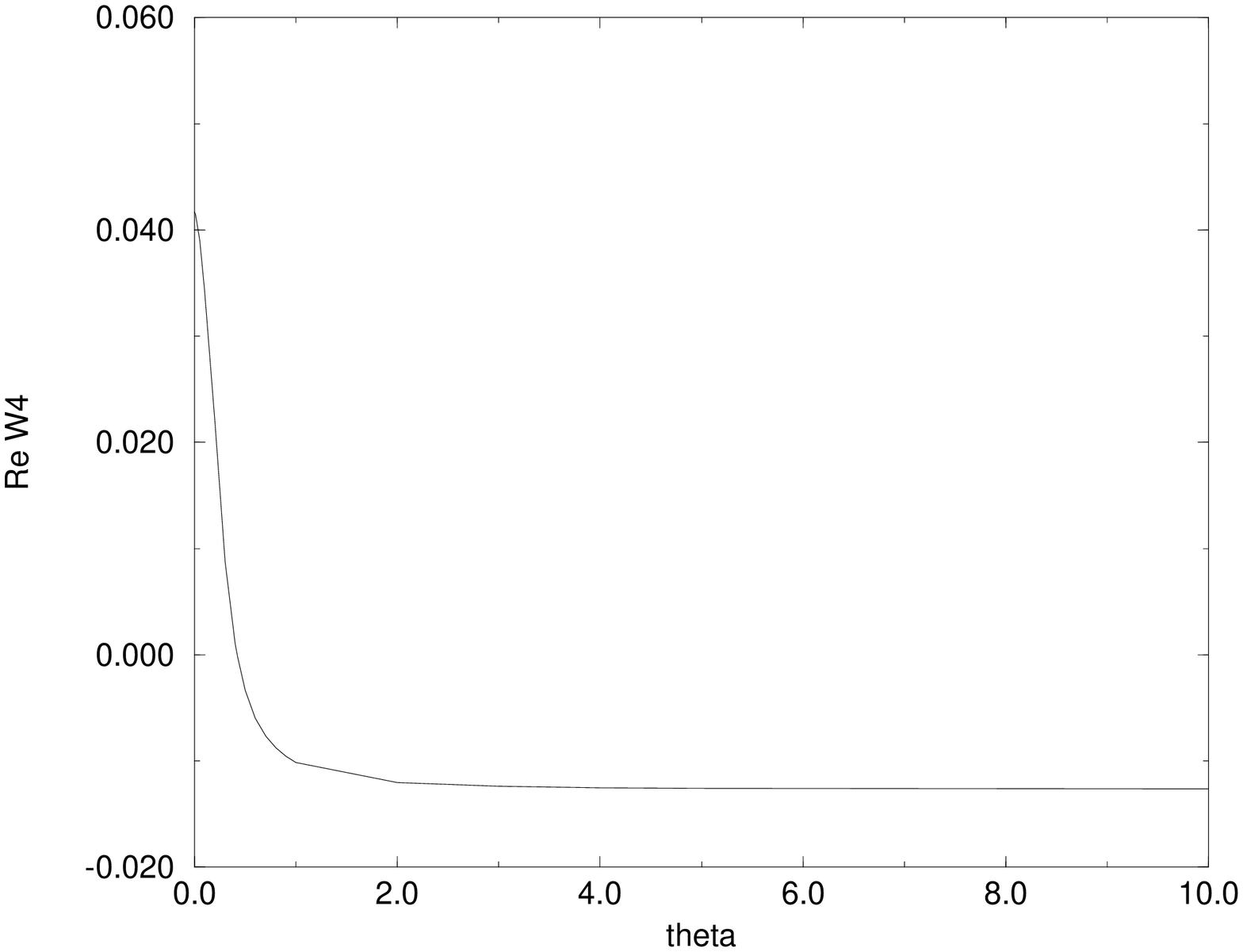}}

\centerline{\includegraphics[width=9.2cm,angle=270]{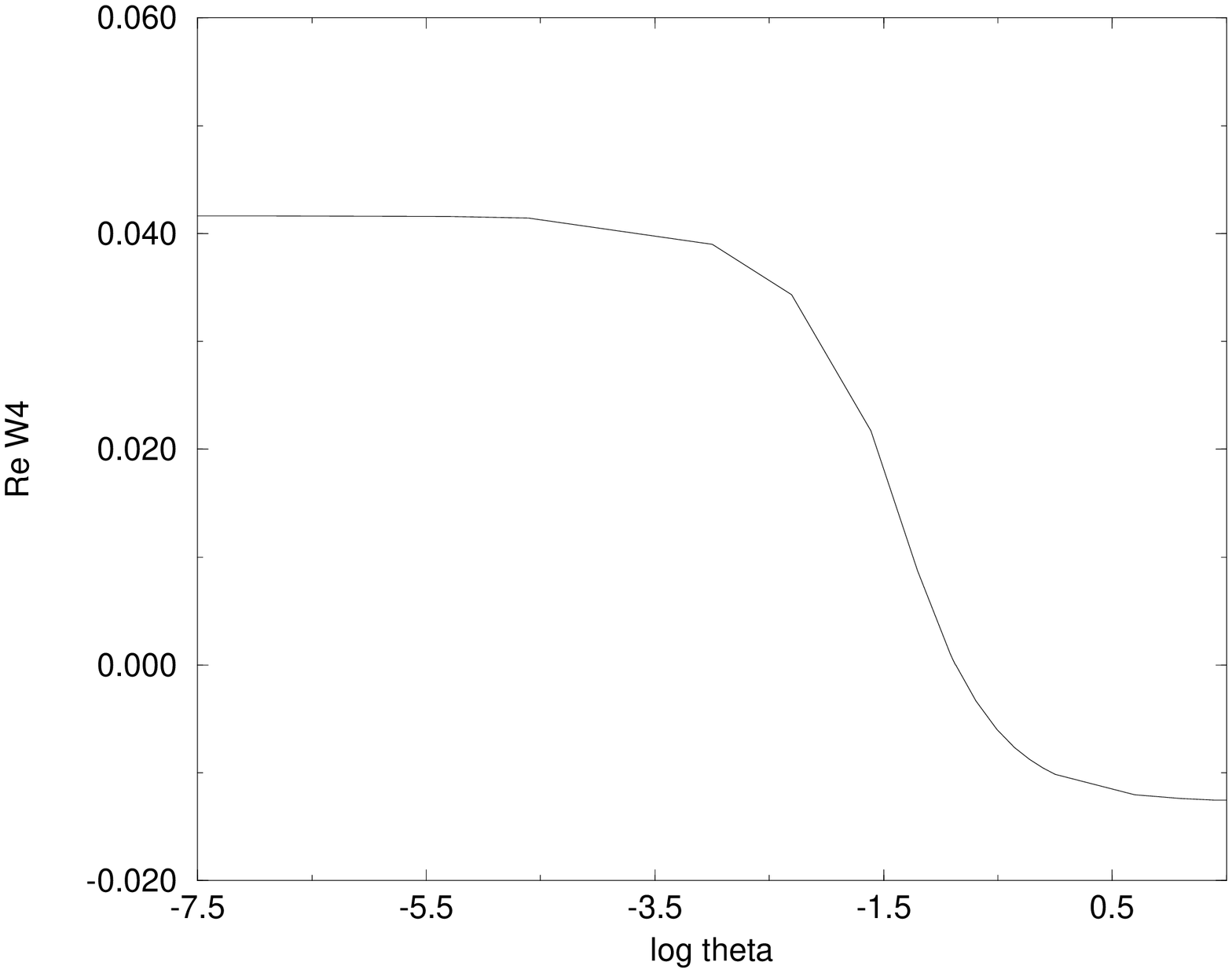}}

\centerline{\includegraphics[width=9.2cm,angle=270]{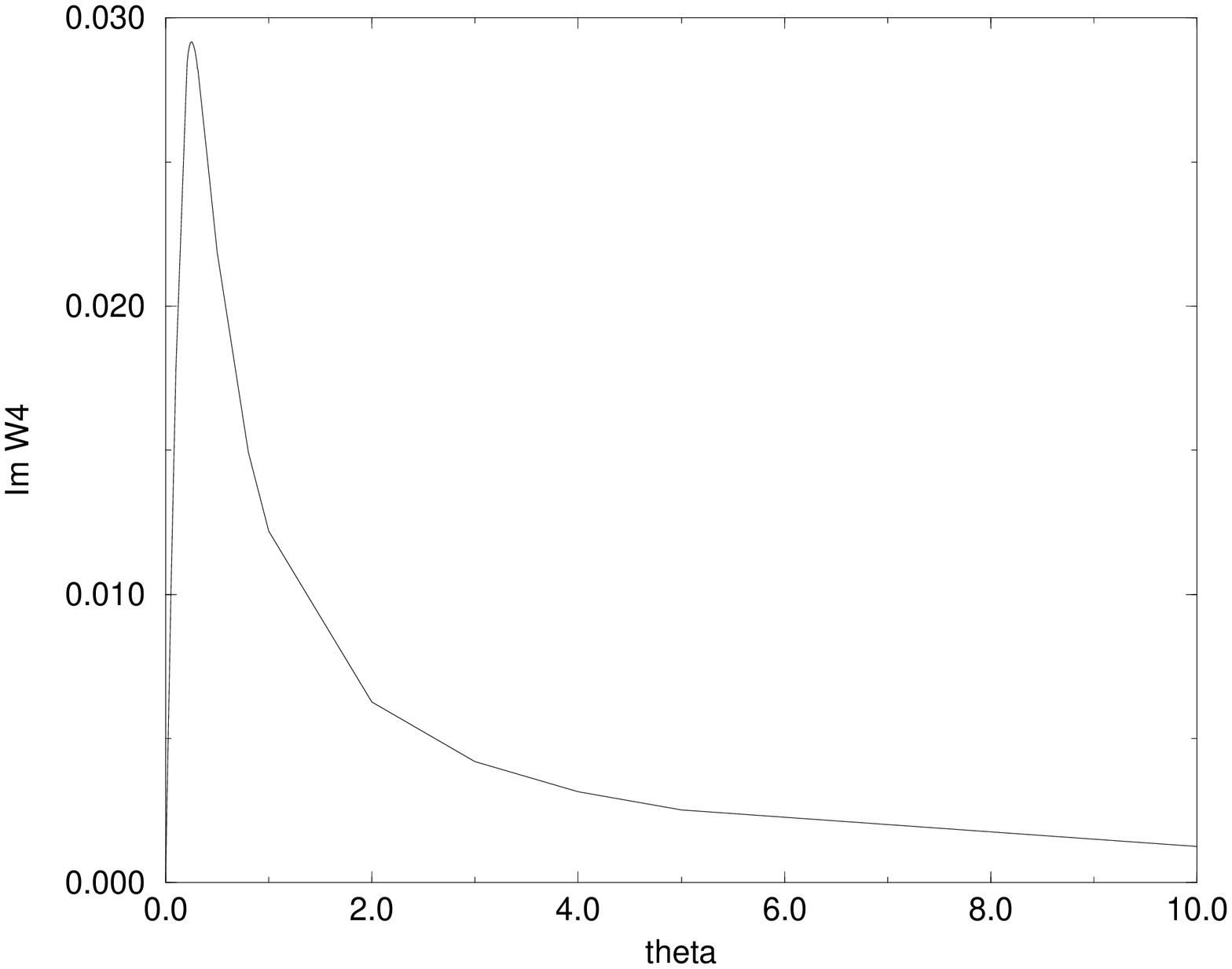}}

\centerline{\includegraphics[width=9.2cm,angle=270]{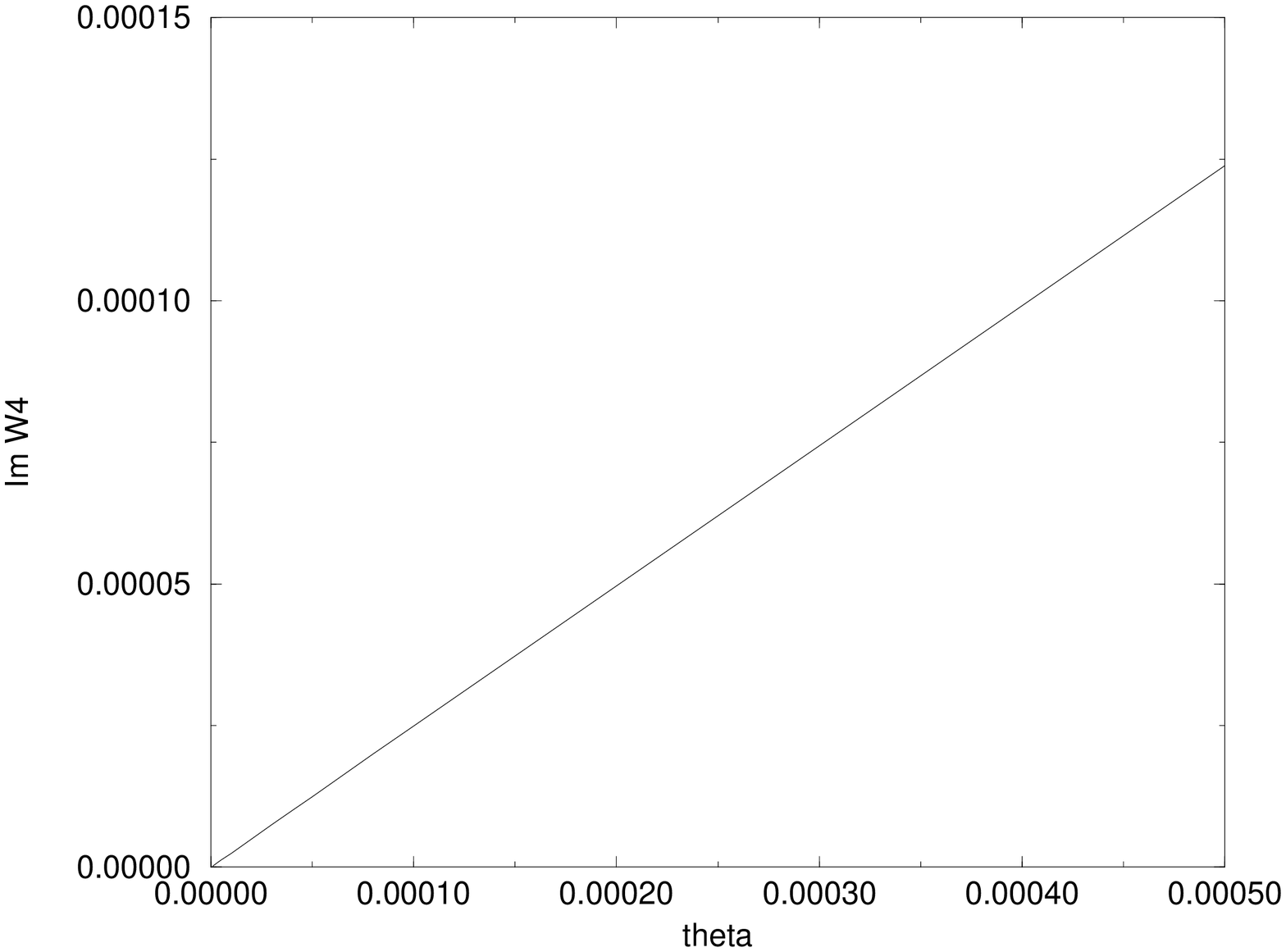}}

\bi

\bi

In the first plot we show the real part of $\WW_4^{(cr)}$ as a function of $\theta$, for the values of $n = 1$ and ${\cal A} = 1$. We see clearly that it approaches $- {{1}\over {8 {\pi}^2}}$ at infinity. The behaviour near the origin is best seen in logarithmic scale, see the second plot. The real part of the loop vanishes at a critical value $\theta \sim 0.41945$.  

In the second plot we show the real part of $\WW_4^{(cr)}$ as a function of $\log \theta$. We see that the limiting value $1 \over 24$ is reached quite soon and apparently smoothly when moving from real values of the noncommutativity parameter.

In the third plot we exhibit the imaginary part of $\WW_4^{(cr)}$ as a function of $\theta$, for the values of $n = 1$ and ${\cal A} = 1$. We see it goes to zero at both large and small $\theta$, in complete agreement with our asymptotic formulae. It has a maximum at $\theta \sim 0.25$. Near $\theta=\infty$ the behaviour ${{1}\over {8 {\pi}^2 \theta}}$ of Eq.(\ref{granteta}) is also manifest.   

In the fourth plot we have expanded the previous graphic near the origin, in order to see that it is in complete agreement with formula (\ref{respic}).

We notice, finally, that the terms of the expansion (\ref{granteta}) that are even in $\theta$ are also real, while the odd ones are imaginary. This is consistent with the observation we made at the beginning of Section 4.3.1, namely that the complex conjugate of $\WW_4^{(cr)}(\theta)$ is $\WW_4^{(cr)}(-\theta)$. One can notice as well that complex conjugation is equivalent to a change in the orientation of the windings $n \rightarrow -n$; at infinite $\theta$ the symmetry under this operation is restored.

\subsection{Scaling Laws}
Now we want to make some further remarks on the results we have obtained. First, we notice that we have recovered in the asymptotic regions exactly the Minkowski results, provided we rotate the area as usual: ${\cal A}$ is converted into $i {\cal A}$, as explained in Section 4.2. One can realize it comparing, for $n = 1$, Eq.(\ref{respic}) at $\theta =0$ and Eq.(\ref{tetazero}), and Eq.(\ref{granteta}) at $\theta=\infty$ with Eq.(\ref{finres}). In passing, we also notice that the Abelian-like exponentiation of the Wilson loop\cite{test} does not occur.

Second, we want to discuss the scaling properties of our result at ${\cal {O}}(g^4)$. Let us recall the commutative case. There, another remarkable difference appeared between 't Hooft's and WML formulations\cite{BGV}. When considering $n$ windings around the closed loop, a non-trivial symmetry concerning operations in the base manifold and over the fiber ($U(N)$) took place in the exact ('t Hooft's) solution, leading
to a peculiar scaling law intertwining the two integers $n$ and $N$
\begin{equation}
\label{scaling}
\WW_n({\cal A};N)=\WW_N(\frac{n}{N}{\cal A};n).
\end{equation}
When going around the loop the non-Abelian character of the
gauge group is felt.

The behaviour of the WML solution was instead fairly trivial (${\cal A}
\to n^2 {\cal A})$, as expected in
a genuinely perturbative treatment, and has just an abelian character, being substantially the same irrespectively of the rank of the gauge group. 

In the noncommutative context, this issue acquires new relevance: as we know, operations over the base manifold and over the fiber are intertwined and merged in a larger group; therefore this is really a favourite item in which to study the features and effects of the merging.  

From our formulae, we see that precisely at $\theta=0$ we recover the commutative result (Eq.(\ref{respic})), which has the trivial abelian scaling law ${\cal{A}} \rightarrow n^2 {\cal{A}}$. This scaling can be maintained at the level of the first correction in Eq.(\ref{respic}) provided one sends $\theta \rightarrow n\theta$; however, this is ruled out by the behaviour found in the other asymptotic region (large $\theta$). 

In fact, with great surprise, at $\theta=\infty$ the scaling is the symmetric one, characteristic of the exact commutative solution: $\WW_n({\cal A};N)=\WW_N(\frac{n}{N}{\cal A};n)$.

This law is destroyed by the next-to-leading orders, as one can see from Eq.(\ref{granteta}), and it is difficult to argue from these formulae if any simple scaling law can emerge at finite $\theta$, especially in view of the ${\theta}^{-2}$-term. Still, we find it very interesting to recover in the region of \it maximal \rm noncommutativity the symmetry of the exact commutative result, in spite of the fact that 't Hooft's prescription is now unacceptable. 

It is therefore crucial to investigate higher orders to see whether this finding is a general property of the noncommutative Wilson loop. This will be done in the next chapter.  

\chapter{Wilson Loop in 2D NCYM: Higher Orders} 

\section{Introduction}
Since we have seen the dramatic consequences to which 't Hooft's form of the free propagator gives rise when matched with noncommutativity, even in higher orders computations we will rely only on the WML prescription. Furthermore, we will continue to use the Euclidean formulation, which in particular, as we have already seen, can provide more compact and manageable formulae: this is of great importance in view of the complications we are going to face even in the immediately subsequent step of the perturbative expansion.

This research is reported in \cite{nostro2}.

\section{The ${\cal{O}}(g^6 )$ Calculation}
We organize the sixth order loop calculation according to the topologically different diagrams one can draw. If we order the
six vertices on the circle from 1 to 6, we denote by $\WW_{(ij)(kl)(mn)}$
the contribution of the graph corresponding to three propagators joining
the vertices $(ij),(kl),(mn)$, respectively. Thus  $\WW_{(14)(25)(36)}$
corresponds to the maximally crossed diagram 
(i.e. the one in which all propagators cross);
then we have three diagrams with double crossing, namely $\WW_{(14)(26)(35)},
\WW_{(13)(25)(46)}$ and $\WW_{(15)(24)(36)}$. Finally we have six diagrams
with a single crossing $\WW_{(12)(35)(46)}$, $\WW_{(16)(24)(35)}$, $\WW_{(15)(23)(46)}$,
$\WW_{(15)(26)(34)}$, $\WW_{(13)(26)(45)}$ and $\WW_{(13)(24)(56)}$.
Diagrams without any crossing are not interesting since they are not affected
by the Moyal phase: they indeed coincide with the corresponding ones in the
commutative case. All crossed diagrams, at this order in perturbation theory, have the same color factor (see Section 5.3). Just to summarize it graphically, we have the following situation:

\bi

\bi

\large ${\cal{W}}_6$\LARGE \ \ = \ \ \includegraphics[width=10cm]{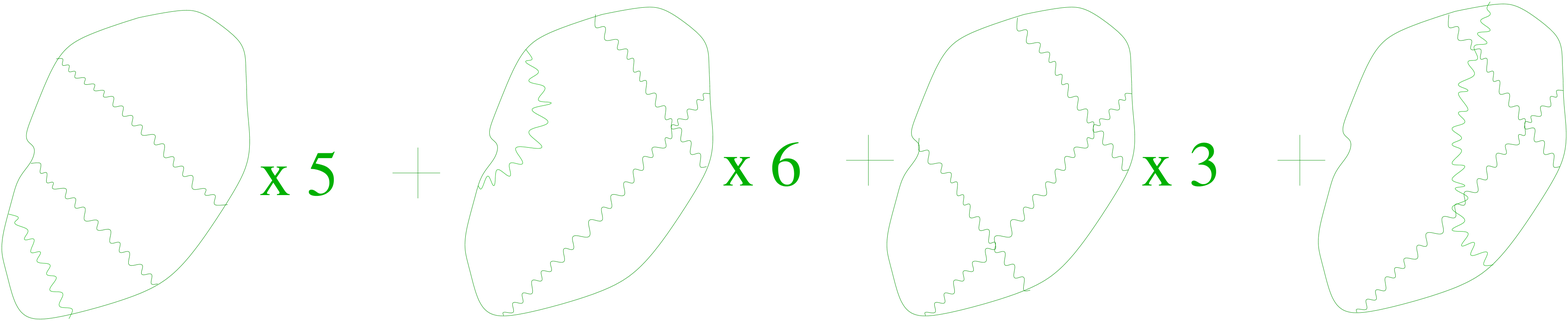}
\bi

\ \ \ \ \ \ =  \ \ \large${\cal{W}}_6^{pl}$\ \ \ \   $+$  \ \ \  \large${\cal{W}}_6^{1cr}(\theta )$  \ \   $+$ \, \large${\cal{W}}_6^{2cr}(\theta )$ \ $+$ \ \large${\cal{W}}_6^{3cr}(\theta ).$  
\normalsize 

\bi

\bi

We will analyse separately the three classes of diagrams, starting from the simplest singly-crossed ones.

\subsection{Singly-Crossed Diagrams}
We can choose one of them as an example, the other ones will be obtained simply by renaming the variables. We will consider $\WW_{(16)(24)(35)}$:
\begin{eqnarray}  
\label{scrossed}
&&\WW_{(16)(24)(35)}= \,- r^6 N n^6\, \int [ds]_6
\int_0^{\infty } \, {{dp}\over {p}} {{dq}\over {q}} {{dk}\over {k}} 
\, \int_0^{2 \pi } d\phi \, d\chi \, d\psi\no \\
&&\exp({- 2 i (\phi + \chi +\psi )}) \exp\Big( {2 i p \sin \phi \sin n\pi s^-_{16}}+{2 i q \sin \psi \sin n\pi s^-_{24}} +\no \\
&&{2 i k \sin \chi \sin n\pi s^-_{35} }\Big) \exp\left(i {{\theta }\over {r^2}} 
q\ k\ \sin [\psi - \chi + n\pi (s^+_{35} -  s^+_{24})] \right),
\end{eqnarray}
where 
\begin{eqnarray}
\label{short}
s^{\pm}_{ij}=s_i\pm s_j.
\end{eqnarray} 
We see that the integrations over $\phi$ and $p$ can be performed immediately putting $e^{i \phi} = z$, and then using formulae (\ref{terza}) and (\ref{quarta}) of Appendix A. We end up with an expression similar to (\ref{crociato}). This reflects the fact that this kind of diagrams are simply crossed ${\cal{O}}(g^4 )$ ones with added a ``non-interacting'' (i.e. non crossing) propagator. We thus simply recall our previous formulae, and integrate all residual phases and momenta as in Eq.(\ref{integrato}) and Eq.(\ref{reintegrato}). Our final expression is therefore 
\begin{eqnarray}  
\label{magic}
\WW_{(16)(24)(35)} \,
&=& -2 n^6 {\cal A}^3 N \int [ds]_6 \Big[ \, \frac{1}{2}  \no \\
&&+ \, \frac{2}{{\hat{\beta}}^2}
\Big( \exp[i{\hat\beta} \sin{\hat\alpha}]-1-i{\hat\beta} \sin {\hat\alpha}\Big) \Big],
\end{eqnarray}
where now 
\begin{eqnarray}
&&{\hat\alpha}=n\pi (s_2+s_4-s_3-s_5),\no \\
&&{\hat\beta}=\frac{4{\cal A}}{\pi \theta}\, 
\sin[n\pi (s_4-s_2)] \, \sin[n\pi(s_5-s_3)].\no
\end{eqnarray}  
As in Section 4.4, it is easy to find that this expression is continuous when $\theta$ goes to zero, where only the $1 / 2$ -term survives, and we recover the commutative result: $-2 n^6 {\cal A}^3 N \int [ds]_6 \, \frac{1}{2}$ $=$ $- n^6 {\cal A}^3 N \, / \, 6!$. The large-$\theta$ limit is again straightforwardly singled out:
\begin{eqnarray}  
\label{coseno}
\WW_{(16)(24)(35)} (\theta=\infty )= - n^6 {\cal A}^3 N \int [ds]_6 \, \cos 2 {\hat\alpha}.
\end{eqnarray}
To evaluate (\ref{coseno}) we have to perform two more integrations. The result is $- n^6 {\cal A}^3 N (9 - {\pi }^2 ) \, / \, (96 {\pi }^4 n^4) $. We are also able to find directly the values of the other five singly-crossed diagrams: it is sufficient to take Eq.(\ref{coseno}) with the suitable $\hat\alpha$, translated in terms of the variables of the two crossing propagators of each diagram. For example, for the diagram $(15)(26)(34)$ the corresponding $\hat\alpha$ is $n\pi (s_2+s_6-s_1-s_5)$. Once we have evaluated these other five contributions, we put all the six singly-crossed diagrams together and get
\begin{eqnarray}
\label{scd}
{\cal{W}}_6^{1cr}(\theta = \infty ) = - \, {\cal A}^3 \, N \, n^6  \, \Big[ {{1}\over {4 \pi^4 n^4 }} \, 
- \, {{1}\over {24 \pi^2 n^2}}\Big].  
\end{eqnarray}   

\subsection{Doubly-Crossed Diagrams}
 We study now the diagrams with two crossing propagators. Again, we compute one among all these diagrams, and obtain the full result just renaming the variables in the other ones. We do the job for $\WW_{(15)(24)(36)}$:
\begin{eqnarray}  
\label{dcrossed}
&&\WW_{(15)(24)(36)}= \,- r^6 N n^6\, \int [ds]_6
\int_0^{\infty } \, {{dp}\over {p}} {{dq}\over {q}} {{dk}\over {k}} 
\, \int_0^{2 \pi } d\phi \, d\chi \, d\psi\no \\
&&\exp({- 2 i (\phi + \chi +\psi )}) \exp({2 i p \sin \phi \sin n\pi s^-_{15} }) \exp({2 i q \sin \psi \sin n\pi s^-_{24}})\no \\
&&\exp(2 i k \sin \chi \sin n\pi s^-_{36}) \exp\Bigg(i {{\theta }\over {r^2}}\Big( 
p\ k\ \sin [\phi - \chi + n\pi (s^+_{36} -  s^+_{15}) ] \nonumber \\
&& \, \, \, \, \, \, \, \, \, \, \, \, \, \, \, \, \, \, \, \, \, \, \, \, \, \, \, \, \, \, \, \, \, \, \, \, \, \, \, \, \, \, \, \, \, \, \, \, \, \, \, \, \, \, \, \,  + \, \, q\ k\ \sin [\psi - \chi + n\pi (s^+_{36} -  s^+_{24})] \Big) \Bigg).
\end{eqnarray}
We recognize that the integral over the phase $\phi$ and its associate momentum $p$, and the one over $\psi$ and $q$, are of the same kind of the one we have encountered in passing from Eq.(\ref{crociato}) to Eq.(\ref{integrato}). We use the same result for both and get the expression
\begin{eqnarray}  
\label{duecast}
\WW_{(15)(24)(36)} \, &=& \, - r^6 N n^6 {\pi }^2 \int [ds]_6  \, 
\int_0^{\infty } {{dk} \over {k}} \, 
\ointop_{|z|=1}\, {{dz} \over {i {{z}^3}}} \, \, 
e^{-\ k  \sin [n \pi (s_6 - s_3 )] (z - {{1} \over {z}})} \nonumber \\
&& {{1 - {{\gamma }' \over {z}} e^{- i n \pi \sigma' }} 
\over {1 - \gamma' z e^{i n \pi \sigma' }}} \, \times \, {{1 - {{\gamma }'' \over {z}} e^{- i n \pi \sigma'' }} 
\over {1 - \gamma'' z e^{i n \pi \sigma'' }}},
\end{eqnarray}
where
$$
\sigma' = s_1 + s_5 - s_3 - s_6 \, \qquad \, \gamma' = \frac {\theta k }{2 r^2 \sin [n \pi (s_5 - s_1 )]} 
$$
$$
\sigma'' = s_2 + s_4 - s_3 - s_6 \, \qquad \, \gamma'' = \frac {\theta k }{2 r^2 \sin [n \pi (s_4 - s_2 )]}, 
$$
and now we have set $e^{i \chi}=z$. With this double integration we have also kept manifest the symmetry of (\ref{dcrossed}) under exchange of the two ``parallel'' propagators cut by the transversal $(36)$ one. It is a formidable task to evaluate this integral for a generic value of $\theta$. As before we are more interested in the extreme values, very small and very large $\theta$, so we try to find out these limits without performing the complete integration. The value at $\theta=0$ is easily extracted from (\ref{duecast}): the fractions become $1$ and we use formulae (\ref{terza}) and then (\ref{quarta}) , obtaining exactly the commutative result, which in particular is equal to the commutative singly-crossed diagram. Expression (\ref{duecast}) is very complicated but it exhibits all the features already emerging from the previous analysis: it is well defined for any $\theta\neq0$ and continuous in the commutative limit.

In order to extract the large-$\theta$ behaviour we adopt the following procedure, that will be justified in a moment: we write the exponential present in (\ref{duecast}) as $e^{-\ k  \sin [n \pi (s_6 - s_3 )] (z - {{1} \over {z}})}$ $=$ $[(e^{-\ k  \sin [n \pi (s_6 - s_3 )] (z - {{1} \over {z}})} - 1) + 1]$. Then (\ref{duecast}) becomes
\begin{eqnarray}  
\WW_{(15)(24)(36)}&=&-r^6 N n^6 {\pi }^2 \int [ds]_6 \int_0^{\infty } {{dk} \over {k}} \, 
\ointop_{|z|=1} {{dz} \over {i {{z}^3}}} 
(e^{-\ k  \sin [n \pi (s_6 - s_3 )] (z - {{1} \over {z}})}- 1) \nonumber \\
&& {{1 - {{\gamma }' \over {z}} e^{- i n \pi \sigma' }} 
\over {1 - \gamma' z e^{i n \pi \sigma' }}} \, \times \, {{1 - {{\gamma }'' \over {z}} e^{- i n \pi \sigma'' }} 
\over {1 - \gamma'' z e^{i n \pi \sigma'' }}}\nonumber \\
 &-& r^6 N n^6 {\pi }^2 \int [ds]_6  \, 
\int_0^{\infty } {{dk} \over {k}} \, 
\ointop_{|z|=1}\, {{dz} \over {i {{z}^3}}} \,  {{1 - {{\gamma }' \over {z}} e^{- i n \pi \sigma' }} 
\over {1 - \gamma' z e^{i n \pi \sigma' }}} \, \no \\
&&\qquad \qquad \qquad \qquad \times {{1 - {{\gamma }'' \over {z}} e^{- i n \pi \sigma'' }} 
\over {1 - \gamma'' z e^{i n \pi \sigma'' }}}\nonumber 
\end{eqnarray}
\begin{eqnarray}
\label{duecastsplit}
= {\WW^{(1)}}_{(15)(24)(36)} + {\WW^{(2)}}_{(15)(24)(36)}.      
\end{eqnarray}
Now, in the first term of the splitting (\ref{duecastsplit}) we can perform the limit $\theta \to \infty$ under the sign of integration, which was impossible directly from the (\ref{duecast}), because there large values of $\theta$ can be contrasted by small $k$, where the integrand is singular. Subtracting $1$ to the exponential suppresses this singularity, invalidating the obstruction to $\theta \to \infty$. This is a qualitative explanation, we proved analytically that one can exchange the limit with the integral in the first term of the splitting (\ref{duecastsplit}) (and not in (\ref{duecast})); the proof is reported in the Appendix B. This proof mainly relies on the fact that the rational functions present in the integrand are actually pure phase factors, a feature we have already observed elsewhere. 

The first contribution becomes
\begin{eqnarray}
\label{primspl}
{\WW^{(1)}}_{(15)(24)(36)}&&\rightarrow \, - r^6 N n^6 {\pi }^2 \int [ds]_6  \int_0^{\infty } {{dk} \over {k}} \, \times \nonumber \\
&&\ointop_{|z|=1}{{dz} \over {i {{z}^7}}} 
(e^{-\ k  \sin [n \pi (s_6 - s_3 )] (z - {{1} \over {z}})}- 1) e^{- 2 i n \pi(\sigma' + \sigma'')}. 
\end{eqnarray}
We use Eq.(\ref{terza}) and (\ref{quarta}), and obtain 
\begin{eqnarray}   
\label{conj}
{\WW^{(1)}}_{(15)(24)(36)}\rightarrow (- 1) {{r^6 \, N \, n^6 \, {\pi}^3}\over {3}} \, \int [ds]_6 \, e^{2 \pi i n (2 s_3 + 2 s_6 - s_1 - s_5 - s_2 - s_4 )}.
\end{eqnarray}
The calculation of ${\WW^{(2)}}_{(15)(24)(36)}$ is much more involved. First, we carry on the integration over $z$ by using standard complex integration. We remember once again that the integrand is a pure phase factor, and therefore no singularities are present on the path of integration, for any value of $\gamma $ and $\sigma $. The result is a function of $k$ which vanishes if $k$ is larger than a certain value, while, for k smaller, is actually a polynomial. Straightforward integration over k yields
\begin{eqnarray}
\label{easyhard}
{\WW^{(2)}}_{(15)(24)(36)}\, \rightarrow \, {\WW^{(2)}}_{(15)(24)(36)easy} \, + \, {\WW^{(2)}}_{(15)(24)(36)hard}, 
\end{eqnarray}
\begin{eqnarray}
\label{cos3}
{\WW^{(2)}}_{(15)(24)(36)easy}={{-r^6 N n^6 2 {\pi}^3}\over {3}} \int [ds]_6 \, e^{- i (\lambda + \omega )} {[\cos({\lambda - \omega })]^3},  
\end{eqnarray}
\begin{eqnarray}
\label{malefic}
{\WW^{(2)}}_{(15)(24)(36)hard}&=&{{i \, r^6 \, N \, n^6 \, 2 \, {\pi}^3}\over {3}} \, \int [ds]_6 \, \exp({- i (\lambda \, + \, \omega )}) \, [\sin({\lambda \, - \, \omega })]^3\nonumber \\
&\times&\, {{\Big( 1 + |{{d}\over {c}}| \exp({i({\lambda \, - \, \omega })})\Big)}\over {\Big( 1 - |{{d}\over {c}}| \exp({i({\lambda \, - \, \omega })})\Big)}},  
\end{eqnarray}
where
$$
c \, = \, - \sin [n \pi (s_1 - s_5 )] \, \exp({- i n \pi \sigma' }) \, = \, |c| \, \exp({i \omega }),
$$
$$
d \, = \, - \sin [n \pi (s_2 - s_4 )] \, \exp({- i n \pi \sigma'' }) \, = \, |d| \, \exp({i \lambda }).
$$
The names emphasize that expression (\ref{cos3}) is easily integrated, it just contains exponential factors of the same kind of the one present in Eq.(\ref{conj}). Expression (\ref{malefic}) is instead very difficult to deal with. Even if the variables $s_3$ and $s_6$ appear only in $\lambda + \omega$ and not in $\lambda - \omega$, and therefore they can be integrated over at once, generating a factor proportional to $n^{- 2}$, the remaining quadruple integral is extremely hard.  To deal with it, first of all we decide to sum up all the contributions from the three doubly-crossed diagrams, in order to cancel possible common terms and to reach a more symmetric form. Much of the (\ref{malefic}) in fact simplifies when put together with the analogous pieces from $\WW_{(14)(26)(35)}$ and $\WW_{(13)(25)(46)}$, after renaming variables in the way described under the (\ref{coseno}). The result is the following:
\begin{eqnarray}
\label{geometr}    
&&{\WW^{(2)}}_{(15)(24)(36)hard} + {\WW^{(2)}}_{(14)(26)(35)hard} + {\WW^{(2)}}_{(13)(25)(46)hard} \no \\
&&\, =\, - \, {{r^6 \, N \, n^4 \, 4 \, \pi }\over {3}} \, \int [ds*]_4  \, \, \cal{I},
\end{eqnarray}
where $\cal{I}$ is given by 
\begin{eqnarray}
\label{terrif}
{{{[\sin n\pi s_{23}^- ]}^2 {[\sin n\pi s_{14}^- ]}^2 {[\sin n\pi (s_{23}^+ - s_{14}^+ )]}^4}\over {{[\sin n\pi s_{23}^- ]}^2 + {[\sin n\pi s_{14}^- ]}^2 - 2 [\sin n\pi s_{23}^- ] [\sin n\pi s_{14}^- ] [\cos n\pi (s_{23}^+ - s_{14}^+ )]}}
\end{eqnarray}      
and
$$ 
\int [ds*]_4 = \int_{0}^{1} d{s_1} \int_{0}^{s_1} d{s_2} \int_{0}^{s_2} 
d{s_3 }  \Big( \int_{s_2}^{s_1} - \int_{0}^{s_3} \Big) d{s_4}. 
$$
This integral seems still refractory to further simplifications. Therefore we decided to evaluate it numerically. By using a FORTRAN program exploiting NAG subroutines we were able to find the following table of results:     
\begin{eqnarray}
\label{numer}
&&{\WW^{(2)}}_{(15)(24)(36)hard} + {\WW^{(2)}}_{(14)(26)(35)hard} + {\WW^{(2)}}_{(13)(25)(46)hard}\no \\
&&\, = \, {{4 \, \pi \, r^6 \, N \, n^4}\over {3}} \, \, 10^{- 3} \, \, J_{NUM}(n),
\end{eqnarray}
where
$$
\, \, \, \, J_{NUM} (n = 1) \, = \,  1.32236(80 \pm 37), \qquad \qquad \qquad \qquad \qquad \qquad \qquad \qquad \qquad \, \, \, \, \, \, \, \, \,    
$$
$$
J_{NUM} (n = 2) \, = \,  0.330(49 \pm 16), \qquad {{J_{NUM} (n = 1)}\over {4}} \, = \, 0.330592(01 \pm 93),      
$$
$$
J_{NUM} (n = 3) \, = \,  0.146(97 \pm 35), \qquad {{J_{NUM} (n = 1)}\over {9}} \, = \, 0.146929(80 \pm 41),     
$$
$$
J_{NUM} (n = 4) \, = \,  0.08(17 \pm 29), \qquad {{J_{NUM} (n = 1)}\over {16}} \, = \, 0.082648(00 \pm 23),      
$$
$$
J_{NUM} (n = 5) \, = \,  0.05(10 \pm 40), \qquad {{J_{NUM} (n = 1)}\over {25}} \, = \, 0.052894(72 \pm 15),     
$$
$$
J_{NUM} (n = 6) \, = \,  0.03(88 \pm 79), \qquad {{J_{NUM} (n = 1)}\over {36}} \, = \, 0.036732(45 \pm 10),      
$$
$$
J_{NUM} (n = 7) \, = \,  0.0(32 \pm 13), \qquad {{J_{NUM} (n = 1)}\over {49}} \, = \, 0.0269871(02 \pm 76).      
$$
All the errors are three standard deviations, and uncertainty is on the digits displayed between brackets. We have chosen to report our results in such a form for a precise reason: we see clearly from the comparison between the two columns that, within the numerical error, $J_{NUM}$ scales as ${{1} / {n^2}}$. The last value we have reported reveals that our computer capability was saturated for sufficiently large values of $n$, but we consider quite satisfactory that the scaling is confirmed within the error for $n=1-6(7)$ and we think that these results are quite sufficiently probing.   

We are finally able to collect all pieces to find the complete doubly-crossed contribution. We calculate easily (\ref{conj}) and (\ref{cos3}) and directly sum them up with the analogous easy parts of $\WW_{(14)(26)(35)}$ and $\WW_{(13)(25)(46)}$, getting ${{{\cal A}^3 \, N \, n^2 }\over {12 \, {\pi }^4 }}$; then we add to this value what comes out from the numerical evaluation of the hard parts, which amounts to ${{4 \, \pi \, r^6 \, N \, n^4}\over {3}} {{1.3224}\over {n^2}}$. The final answer is 
\begin{eqnarray}
\label{dcd}
{\cal{W}}_6^{2cr}(\theta = \infty ) = \frac{{\cal A}^3 N n^2}{12\pi^4}(1+
0.2088).
\end{eqnarray} 

\subsection{The Maximally-Crossed Diagram}
We are now able to consider the only remaining diagram, which is the one with three mutually crossing propagators. Its expression is the following:
\begin{eqnarray}
\label{mcrossed}
&&\WW_{(14)(25)(36)}= \,- r^6 N n^6\, \int [ds]_6
\int_0^{\infty } \, {{dp}\over {p}} {{dq}\over {q}} {{dk}\over {k}} 
\, \int_0^{2 \pi } d\phi \, d\chi \, d\psi \\
&&\exp[{- 2 i (\phi + \chi +\psi )}] \exp[{2 i p \sin \phi \sin n\pi s^-_{14} }]
\exp[{2 i q \sin \psi \sin n\pi s^-_{25} }] \nonumber \\ 
&&\exp({2 i k \sin \chi \sin n\pi s^-_{36}}) 
\exp\Bigg(i {{\theta }\over {r^2}}\Big( p\ q\ 
\sin [\phi - \psi + n\pi (s^+_{25} -  s^+_{14})]+ \nonumber \\
&& p\ k\ \sin [\phi - \chi + n\pi (s^+_{36} -  s^+_{14})]+
q\ k\ \sin [\psi - \chi + n\pi (s^+_{36} -  s^+_{25})] \Big) \Bigg) \nonumber.
\end{eqnarray} 
With the appropriate identifications we recognize once again that the integral over $\phi$ and $p$ has the same structure of the one solved in passing from Eq.(\ref{crociato}) to Eq.(\ref{integrato}). Therefore, we use the same formula once again, writing the result in the following slightly modified way 
\begin{eqnarray}
\label{intermedio}
&&\WW_{(14)(25)(36)}= \,- \pi \, r^6 N n^6\, \int [ds]_6 \int_0^{\infty } \, {{dq}\over {q}} {{dk}\over {k}} \int_0^{2 \pi } \, d\chi \, d\psi\, 
\no \\
&&\exp[{- 2 i (\chi +\psi )}] \exp[{2 i q \sin \psi \sin n\pi s^-_{25} }] \exp[{2 i k \sin \chi \sin n\pi s^-_{36}}]\nonumber \\  
&& \, \times \, \exp\Bigg[ i {{\theta }\over {r^2}} q\ k\ \sin [\psi - \chi + n\pi (s^+_{36} -  s^+_{25})] \Bigg] \nonumber \\
&& \times {{\sin[ \pi s^-_{14}]\, + \, {{\theta}\over {2 r^2}}\, \Big[ q \, e^{-i( \psi + \pi (s^+_{14} - s^+_{25}))} \, + \, k \, e^{-i( \chi + \pi (s^+_{14} - s^+_{36}))}\Big] }\over {\sin[ \pi s^-_{14}] + {{\theta}\over {2 r^2}}\Big[ \, q \, e^{i( \psi + \pi (s^+_{14} - s^+_{25}))} \, + \, k \, e^{i( \chi + \pi (s^+_{14} - s^+_{36}))}\Big] }}.  
\end{eqnarray}
Looking at Eq.(\ref{intermedio}), we see that the integral over $\chi$ and $k$ resembles the one encountered in passing from Eq.(\ref{integrato}) to Eq.(\ref{reintegrato}), with the only important distinction that the exponential has a more complicated exponent. With some efforts we are able to update our formula for this more general integral and the result is shown in the Appendix A in formula (\ref{quinta}). Using that relation we get
\begin{eqnarray} 
\label{storic}
&&\WW_{(14)(25)(36)} = \, - r^6 \, N \, n^6 \, 2 {\pi}^2 \, \int [ds]_6 \,\int_0^{\infty } \,  {{dq}\over {q}} \, \int_0^{2 \pi } \, d\psi \, \exp({- 2 i \psi })\nonumber  \\
&& \qquad \, \, \times \, \exp({4 i {{q r^2}\over {\theta}} { \sin \psi \sin [n \pi (s_2 - s_5 )]}}) \, \Bigg[ \, \, {{1}\over {2}} \, {{\bar{\alpha}}\over {\alpha}} \,{{\bar{\beta}}\over {\beta}} \, + \, \\
&& {{{\theta}^2}\over {8 r^4 {\alpha}^2 {\beta}^2}} \left( \exp\Big({{{4 i r^2 }\over {\theta}} \, {\rm Im} (e^{i n \pi {\sigma}'''} \bar{\alpha } \beta)}\Big) - {{4 i r^2 }\over {\theta}} {\rm Im} (e^{i n \pi {\sigma}'''} \bar{\alpha } \beta) - 1 \right)  \Bigg], \nonumber 
\end{eqnarray}     
where ${\sigma}''' = s_1 + s_4 - s_3 - s_6$, the bars denote complex conjugation
and
$$
\alpha \, = \, \sin [n \pi (s_1 - s_4 )] \, + \, q \, \exp\ i({\psi + n \pi (s_1 + s_4 - s_2 - s_5 )}), 
$$
$$
\beta \, = \, \sin [n \pi (s_3 - s_6 )] \, - \, q \, \exp\ i({\psi + n \pi (s_3 + s_6 - s_2 - s_5 )}) . 
$$
To deal with Eq.(\ref{storic}), first of all we parameterize $\alpha = |\alpha | \exp({i {\gamma }_{\alpha }})$ and $\beta = |\beta | \exp({i {\gamma }_{\beta }})$. Then we recall the following integral representation for the exponential function\cite{prudni3} 
$$
e^x \, - \sum_{k=0}^n \, {{x^k}\over {k!}} \, \, = \, \, {{1}\over{2 \pi i}} \int_{- \gamma - i \infty}^{- \gamma + i \infty} \Gamma (s) (- x)^{- s}\, \, , \, \, n+1>\gamma>n, 
$$
and we obtain
\begin{eqnarray} 
\label{drago}
&&\WW_{(14)(25)(36)} = - r^6 N n^6 2 {\pi}^2 \int [ds]_6 \int_0^{\infty } {{dq}\over {q}} \int_0^{2 \pi } d\psi \exp({- 2 i \psi }) \no \\
&& \times \exp({4 i {{q r^2}\over {\theta}} { \sin \psi \sin [n \pi (s_2 - s_5 )]}}) \, \exp({- 2 i ({\gamma }_{\alpha } + \, {\gamma }_{\beta })}) \no \\
&& \qquad \, \, \, \times \, \Bigg[ \, {{1}\over {2}} \, \cos({2 n \pi {\sigma }''' \, - \, 2 {\gamma }_{\alpha } \, + \, 2 {\gamma }_{\beta } }) \, + \\\
&& {{1}\over {\pi i}} \, \int_{\mu - i \infty }^{\mu + i \infty } \, ds \, \Gamma (- s) \, e^{- i {{\pi }\over {2}} s} \, {\Bigg[ {{4 |\alpha | |\beta | r^2 }\over {\theta }} \Bigg]}^{s - 2} \, {\big[ \sin({n \pi {\sigma }''' \, - \, {\gamma }_{\alpha } \, + \, {\gamma }_{\beta } }) \big]}^s \Bigg], \nonumber 
\end{eqnarray}     
with $2<\mu<3$. We succeeded in proving that the last integral goes to zero in the large-$\theta$ limit. The proof is based on a chain of majorizations of the modulus of the integral, and on the use of a Stirling formula for the majorization of the $\Gamma$-function. We will not report here all the details of this calculation, and only state the result: we proved that the modulus of this integral is smaller than a convergent integral independent of $\theta$, times a factor $\theta^{2 - \mu}$, which ends the proof when $\theta$ goes to zero. 

What remains to evaluate in this limit is therefore
\begin{eqnarray} 
\label{draghetto}
&&\WW_{(14)(25)(36)} = \, - r^6 \, N \, n^6 \, 2 {\pi}^2 \, \int [ds]_6 \, \int_0^{\infty } \,  {{dq}\over {q}} \, \int_0^{2 \pi } \, d\psi \, \exp({- 2 i \psi })\no \\
&&\, \times \, \exp ({ 4 i {{q r^2}\over {\theta}} { \sin \psi \sin [n \pi (s_2 - s_5 )]}}) \, \exp({- 2 i ({\gamma }_{\alpha } + \, {\gamma }_{\beta })}) \no \\
&&\, \times \, {{1}\over {2}} \, \cos(2 n \pi {\sigma }''' \, - \, 2 {\gamma }_{\alpha } \, + \, 2 {\gamma }_{\beta }).  
\end{eqnarray}
This is not a difficult task: we split formula (\ref{draghetto}) into two pieces, following the same philosophy described under Eq.(\ref{duecastsplit}):
\begin{eqnarray} 
\label{dragsplit}
&&\WW_{(14)(25)(36)} = \, - r^6 \, N \, n^6 \, 2 {\pi}^2 \, \int [ds]_6 \,\int_0^{\infty } \,  {{dq}\over {q}} \, \int_0^{2 \pi } \, d\psi \, \exp({- 2 i \psi })\times \no \\
&& \Big[ \exp[{4 i {{q r^2}\over {\theta}} { \sin \psi \sin n \pi (s_2 - s_5 )}}] - 1\Big] \, \exp({- 2 i ({\gamma }_{\alpha } + \, {\gamma }_{\beta })}) \times \no \\
&&{{1}\over {2}} \, \cos({2 n \pi {\sigma }''' \, - \, 2 {\gamma }_{\alpha } \, + \, 2 {\gamma }_{\beta } }) - r^6 \, N \, n^6 \, 2 {\pi}^2 \, \int [ds]_6 \,\int_0^{\infty } \,  {{dq}\over {q}} \, \int_0^{2 \pi } \, d\psi \no \\
&&\exp({- 2 i \psi + {\gamma }_{\alpha } + \, {\gamma }_{\beta }}) {{1}\over {2}} \cos({2 n \pi {\sigma }''' \, - \, 2 {\gamma }_{\alpha } \, + \, 2 {\gamma }_{\beta } }).  
\end{eqnarray}
In the first term of the splitting one can send $\theta \to \infty$ directly into the integrand, get a result similar to Eq.(\ref{primspl}), use again Eq.(\ref{terza}) and (\ref{quarta}) and obtain
\begin{eqnarray} 
\label{dragprimsplit} 
{{-r^6 N n^6 {\pi}^3}\over {3}} \int [ds]_6 \, e^{ 2 i n \pi (2 s_2 + 2 s_5 - s_1 - s_4 - s_3 - s_6 )}.
\end{eqnarray}
The second term of the splitting is directly evaluated setting $e^{i \psi } = z$ and using standard complex integration techniques, then integrating the result in the variable $q$:
\begin{eqnarray} 
\label{dragsecsplit}
&&{{-r^6 N n^6 {\pi}^3 }\over {3}}\int [ds]_6 \, [e^{ 2 i n \pi (2 s_1 + 2 s_4 - s_2 - s_5 - s_3 - s_6 )} \no \\
&& \qquad \qquad \qquad \qquad  \qquad \, \, \, \, + \, \, e^{ 2 i n \pi (2 s_3 + 2 s_6 - s_1 - s_4 - s_2 - s_5 )}].
\end{eqnarray}
We notice that the sum of (\ref{dragprimsplit}) and (\ref{dragsecsplit}) is completely symmetric in the three propagators (14)(25)(36), as it should (see for example (\ref{mcrossed})).  

The final result is easily evaluated to be
\begin{eqnarray} 
\label{symm}
{\cal{W}}_6^{3cr}(\theta = \infty ) = - \, {{{\cal A}^3 N n^2 }\over {64 {\pi }^4 }}.
\end{eqnarray}

\section{Scaling Properties}
We start collecting what we have obtained in the previous sections: we sum together all the contributions from crossed diagrams, Eq.(\ref{scd}), Eq.(\ref{dcd}) and Eq.(\ref{symm}), and get
\begin{equation}
\label{tot}
\WW_6^{(cr)}(\theta=\infty)=\ \frac{{\cal A}^3 N n^4}{24\pi^2}\left(1-\frac{1}{n^2\pi^2}
(\frac{35}{8}-0.4176)\right).
\end{equation}
The planar diagrams do not depend on $\theta$, so they are a sort of constant background uninteresting if we want to study the properties of noncommutativity. We decide therefore not to consider them anymore, and to focus our attention only on the crossed ones, just to catch the significant novel features of the noncommutative setup.

We realize that the symmetric scaling (\ref{scaling}), which was found to hold for the large-$\theta$ limit of ${\cal{O}}(g^4)$, does not hold for Eq.(\ref{tot}), frustrating the expectation that it could be a symmetry of the maximal noncommutativity regime \it tout court \rm. However, we have more freedom in playing with the parameters that enter the calculation: we realize that if we send simultaneously $\theta$ and $n$, the number of windings around the loop, to infinity, then the result exhibits again the symmetric scaling 
\begin{equation}
\label{scalingancora}
\WW_n({\cal A};N)=\WW_N(\frac{n}{N}{\cal A};n).
\end{equation}
Obviously, one can look at higher orders computations, in order to verify if this same situation persists, or if it is in turn destroyed at the next step. The difficulties in calculating higher order diagrams are really prohibitive, but we can start now constructing some general argument which can help in understanding what will happen going on in the perturbative series without performing the full computations. 

First, we see that building up a general diagram at a generic order and with a generic number of crossing propagators is a trivial matter: we compare Eq.(\ref{crociato}) with Eq.(\ref{scrossed}), Eq.(\ref{dcrossed}) and Eq.(\ref{mcrossed}), and easily understand how to generalize these structures to higher orders. The coefficient in front of the integral will contain a factor $n^{2m + 4}$ if we are at the $(2m + 4)$-th perturbative order. This factor always comes from the rescaling of the contour variables from the interval $[0,n]$ to $[0,1]$, as in what follows Eq.(\ref{crociato}). What is extremely difficult to prove, though highly plausible, is that the integral, which depends itself on $n$, will decrease as $n$ goes to infinity. We were unable to find any useful majorization, since we deal with entangled integrals of oscillating functions. Nevertheless, it is plausible that, as $n$ grows, the exponentials undergo larger and larger oscillations, cutting off the integral. What is possible to prove is that the value at $\theta = \infty$ of the integral depends only on the absolute value of $n$\footnote{To prove this fact it is sufficient to realize, as one can do with a suitable change of variables, that sending $n$ in $- n$ in the integrals is equivalent to sending $\theta$ in $- \theta$}. This means that, if we believe that the integral decreases with $n$ as a power law, it should decrease at least as $n^{-2}$ or faster. Then, taking into account the factor $n^{2m + 4}$ in front, a generic diagram will behave at least as $n^{2m + 2}$.     
 
Next, we consider the dependence on the quadratic Casimir operator of the gauge group, which appears in the coefficient in front of the integral. Inspired by the fact that we expect a symmetric scaling between $n$ and $N$, we also consider the large-$N$ regime. In this regime one can prove that the dominant class of diagrams at this order will behave as $N^m$ (recall the $N^{-1}$ in the definition of the Wilson loop). The case of three propagator is quite peculiar, since it provides Casimirs which are equal for single, double and triple crossing, while at higher orders the values of the Casimir operator start being quite different from diagram to diagram. 

Now we realize that at least a class of such diagrams, behaving like $n^{2m + 2} N^m$, always exists and provides a non-zero contribution. As we will show in a moment, the class of singly-crossed diagrams will always possess such a dependence, and will give a non-zero result at any perturbative order. We recognize that at ${\cal{O}}(g^6)$ this class of diagrams is dominant in the large-($\theta$,$n$,$N$) regime, as one can see comparing Eq.(\ref{scd}) with Eq.(\ref{dcd}) and Eq.(\ref{symm}). We could be sure they would furnish the leading behaviour, provided no cancellations occur from diagrams belonging to other classes, and with the same color factor at a given perturbative order. Therefore, we are led to propose a conjecture, namely, that only the singly-crossed diagrams will dominate, as far as the $n$-dependence is concerned, at any perturbative order in this particular asymptotic regime. Until now we have not been able to convert this conjecture in a theorem; the only evidence we have reached relies on the ${\cal{O}}(g^6)$ calculation.

Following our conjecture, first we prove that singly crossed diagrams behave in the large-$\theta$ limit
at least as 
${{1} \over {n^2}}$ or subleading in the limit of a large number of windings $n$. 
We start by realizing that we can always express the integral of 
a generic diagram with $m$ 
propagators and a single crossing, generalizing Eq.(\ref{reintegrato})\footnote{Recall that when $\beta$ in Eq.(\ref{reintegrato}) goes to zero, than the integrand becomes ${1\over 2} - \sin^2 \alpha$ = ${1\over 2} \cos 2\alpha$}, as follows:
\begin{eqnarray}
\label{ap1}
{\cal I}\equiv \int_{0}^{1}  dt \, \int_{0}^{t}  dz \, \int_{0}^{z}  dy \, \int_{0}^{y}  dx \, \int  [ds]_{2m - 4} \, \cos [2 \pi n (x + z - y - t)],
\end{eqnarray}
$[ds]_{2m-4}$ being a measure depending on $x,y,z,t$ only through the extremes 
of integration.
As a matter of fact, it is always possible to single out in this way the variables linked to
the propagators which cross, suitably rearranging the other kinematical integrations.
These integrations lead to polynomials:
\begin{eqnarray}
\label{ap2}
{\cal I}&=&\int_{0}^{1}  dt \, \int_{0}^{t}  dz \, \int_{0}^{z}  dy \, \int_{0}^{y}  dx \, \sum_{k_1 k_2 k_3 k_4} \, c_{k_1 k_2 k_3 k_4} \times \no \\
&&\qquad \qquad \qquad \, \, \, \, x^{k_1} y^{k_2} z^{k_3} t^{k_4} \, \cos [2 \pi n (x + z - y - t)].
\end{eqnarray}
Now we perform the change of variables $\alpha = y + x$, $\beta = y - x$, $\gamma= t +z$, $\delta = t - z$:
\begin{eqnarray}
\label{ap3}
{\cal I}&=&\int_0^1 d\delta \, \int_{\delta}^{2 - \delta} d\gamma \, 
\int_0^{{{\gamma - \delta }\over {2}}} d\beta \, 
\int_{\beta}^{\gamma - \delta - \beta } d\alpha \no \\ 
&&\, \, \, \, \, \, \sum_{q_1 q_2 q_3 q_4} \, {c'}_{q_1 q_2 q_3 q_4} \, {\alpha}^{q_1} {\beta}^{q_2} {\gamma}^{q_3} {\delta}^{q_4} \, \cos [2 \pi n (\beta + \delta )],
\end{eqnarray}
and then integrate over $\alpha$. Changing again variables to $\psi = \beta + \delta$, 
$\xi = \delta - \beta$, we end up with
\begin{eqnarray}
\label{ap4}
{\cal I}=\int_0^1 d\psi \, \int_{-\psi}^{\psi} d\xi \int_{\frac{3\psi-\xi}{2}}
^{2-\frac{\psi+\xi}{2}} d\gamma  \, \sum_{p_1 p_2 p_3} \, {C}_{p_1 p_2 p_3} \, {\psi}^{p_1} {\xi}^{p_2} {\gamma}^{p_3} \, \cos [2 \pi n \psi]. 
\end{eqnarray}     
The integrals over $\xi$ and $\gamma$ can be easily performed
giving, of course, a polynomial in $\psi$
\begin{eqnarray}
\label{ap5}
{\cal I}=\sum_r \, {C'}_r \, \int_0^1 \, {\psi}^r \, \cos [2 \pi n \psi]\, d\psi.        
\end{eqnarray}
Integrating by parts, we realize that only even inverse powers of $n$
are produced, starting from $n^{-2}$.

\smallskip

Now we turn our attention to the $U(N)$ factors. As we have already said, direct computation of the traces
involved in the diagrams with a single crossing at (${\cal O}(g^6)$) (and also a double or the triple crossing at this order) shows a factor $N^2$ .
As the Wilson loop is normalized with $N^{-1}$, at ${\cal O}(g^6)$ the single
factor $N$ ensues.
It is now trivial to realize that any insertion of $m-3$ lines no matter where in
such diagrams, provided that further crossings are avoided, produces the factor
$N^{m-3}$. 
       
Once we have shown that the $n$-dependence of singly 
crossed diagrams in the large-$\theta$ limit takes the form 
$\sum_{p=1}^{P} c_p \, n^{- 2 p}$, we have to evaluate $c_1$ and to show it does not vanish. This can be worked out as follows. At ${\cal O}(g^{2 m + 4})$, 
we start drawing a cross, and then add the remaining $m$ propagators in such a way 
they do not further cross. From Eqs.(\ref{ap2}-\ref{ap5}) one can realize that $c_1$ 
is different from zero only for a particular subset of these diagrams: if we label
the four sectors in which the cross divides the circular loop as North (the sector
containing the origin of the loop variables $s_i$), West, South and East, 
then only diagrams with $r$ propagator in the southern sector and $m - r$ in the 
northern one contribute to $c_1$; moreover, these contributions are all equal.
Therefore we can evaluate this integral once, and then multiply it by the number of
configurations in this subset.                  
 
We choose as representative the diagram with all the $m$ non-intersecting propagators 
in the northern sector, starting from the origin and connecting $s_1$ with $s_2$,
..., $s_{2m-1}$ with $s_{2m}$. In this way the crossed variables are the ones from $s_{2m+1}$ to $s_{2m+4}$. We obtain the integral
\begin{eqnarray}
\label{D1}
{\cal I} &=& {(- \pi )}^{m + 2} {(g r )}^{2 m + 4}  N^{m} n^{2 m + 4}  
\int_{0}^{1}  dt \, \int_{0}^{t}  dz \, \int_{0}^{z}  dy \, \int_{0}^{y}  dx \, \no \\ 
&&\qquad \qquad \, \times \, {{x^{2 m}}\over {(2 m)!}} 
\cos [2 \pi n (x + z - y - t)].
\end{eqnarray}
Following the procedure just described to reach Eq.(\ref{ap5}), we get
\begin{eqnarray}
\label{D2}
{\cal I} &=& {{{(- \pi )}^{m + 2} {(g r )}^{2 m + 4}}\over {{(2 m)!}}}  N^{m}
n^{2 m + 4}   \int_{0}^{1}  d\psi {{1}\over {(2 m + 1) (2 m + 2)}}\no \\ 
&&\qquad \qquad \, \times \, \psi {(1 - \psi )}^{2 m + 2} \cos [2 \pi n \psi],
\end{eqnarray}
and finally
\begin{eqnarray}
\label{D3}
{\cal I} = - N^{m} {{{(- g^2 {\cal A} \, n^2 ) }^{m + 2}}\over {(2 m + 2)!}}  
\Bigg( {{1}\over {4 {\pi}^2 n^2}} + {\cal{O}}({{1}\over {n^4}})\Bigg). 
\end{eqnarray}
Now we have to count. We denote by $S_{2r}$ the ways in which the $r$ propagators 
in the southern sector can be arranged without crossing. A little thought provides the recursive relation\footnote{It is sufficient to draw the loop as a line with the two extremes identified, then draw $2(n-1)$ vertices: now draw a single vertex, say, at the extreme right, with attached a ``floating'' propagator, and look for the possibilities of attaching the other vertex of this floating propagator among the others $2(n-1)$, such to build a non-mutually intersecting configuration. Each time one has to overcomes two vertices, and gets an $S_{2k-2} S_{2r-2k}$}
\begin{eqnarray}
\label{D4}
S_0=1, S_{2r}=\sum_{k=1}^{r} S_{2k-2} S_{2r-2k},
\end{eqnarray}
which can easily be solved using the method of the generating function\footnote{If one defines $S=\sum_{n=1}^{\infty} t^{n} S_{2n}$, then from (\ref{D4}) one finds in particular that $S = t {(1 + S)}^2$}
\begin{eqnarray}
\label{D5}
S_{2r} = {{{2^{2r}} \Gamma (r + {1\over 2})}\over {{\Gamma ({1\over 2})} 
\Gamma (r + 2)}}.
\end{eqnarray}
The $m - r$ propagators in the northern sector lead to the weight $S_{2(m - r)}$ 
times the number of possible insertions of the origin, namely $[2(m - r) + 1]$.
The number of relevant diagrams is therefore
\begin{eqnarray}
\label{D6}
{\cal N}_m = \sum_{r=0}^m S_{2r}\, S_{2(m - r)} \, [2(m - r) + 1] \, =
 \, {{{2^{2 m + 2}} (m + 1) \Gamma (m + {3\over 2}) }\over {\Gamma ({1\over 2})
\Gamma (m + 3)}}.
\end{eqnarray}
Multiplying Eqs.(\ref{D3}) and (\ref{D6}) we are led to 
\begin{eqnarray}
\label{conject}
g^{2m+4}\WW_{2m+4}^{(cr)}(\theta=\infty)
\simeq -\frac{(g^2{\cal A}n)^2}{4\pi ^2}\,\frac{1}{m!}
\frac{(-g^2{\cal A} N n^2)^m}{(m+2)!}.
\end{eqnarray}
The related perturbative series can be easily resummed
\begin{eqnarray}
\label{resu}
\WW^{(cr)}(\theta=\infty)=-\frac{g^2{\cal A}}{4\pi^2N}\, J_2(2\sqrt{g^2{\cal A}
n^2N}).
\end{eqnarray}
First, we notice that, if we compare Eq.(\ref{conject}) with the corresponding term due to planar diagrams\footnote{This term can be derived, for instance, as the leading term in $N$ of Eq.(\ref{wml}), since planar diagrams are the same as in the commutative case. Looking at the following Eq.(\ref{wuplanar}), one becomes aware of the fact we mentioned, namely that the commutative WML solution does not confine in the 't Hooft limit of an infinite number of colors and 't Hooft coupling $g^2 N$ finite. Eq.(\ref{wuplanar}) is the 't Hooft limit of Eq.(\ref{wml}) (the theory becomes planar in this limit), and no exponential suppression is found\cite{stau}} 
\begin{eqnarray}
\label{wuplanar}
\WW^{(pl)}_{WML}=\sum_{m=0}^{\infty}\frac{(-g^2{\cal A}n^2N)^m}{m!(m+1)!}=
\frac{1}{\sqrt{g^2{\cal A}n^2N}}J_1(2\sqrt{g^2{\cal A}n^2N}),
\end{eqnarray}
then in the 't Hooft limit $N\to \infty$ with
fixed $g^2N$ we see the planar diagrams dominate by a factor $n^2 N^2$, as expected.

Second, we see that the leading term Eq.(\ref{conject}) exhibits the symmetric scaling 
\begin{equation}
\label{persemprescaling}
\WW_n({\cal A};N)=\WW_N(\frac{n}{N}{\cal A};n).
\end{equation}
It is finally interesting to see the way in which this scaling is realized: in the planar contribution (\ref{wuplanar}), it is encoded in the invariant combination $g^2{\cal A}n^2N$, which we could call the ``planar-block''. Now we rewrite the formula (\ref{resu}) as follows:
\begin{eqnarray}
\label{resu2}
\WW^{(cr)}(\theta=\infty)=-\frac{g^2{\cal A}n^2N}{4\pi^2 {(Nn)}^2}\, J_2(2\sqrt{g^2{\cal A}
n^2N}).
\end{eqnarray}
In this way, the scaling is obtained through the dependence on the planar-block, which encodes the ladder of uncrossed propagators that constitutes a singly-crossed diagram, and the symmetric combination $(Nn)^2$, which seems to encode the single crossing, that is the part which feels the noncommutativity.  

\chapter{Open String Amplitudes}

\section{Introduction}
We now return to the second problem we mentioned, namely the problem of explaining the different behaviour of magnetic and electric-type noncommutative field theories in the light of their stringy derivation. In particular, as we saw, the problem of the perturbative unitarity of noncommutative quantum field theories was encountered since the appearance of these theories as an effective description of open string theory in an antisymmetric constant background, and was one of the first concrete evidence of this difference. The absence of a straightforward  Hamiltonian formulation for the case of an electric-type noncommutativity (in which the time variable is involved) and the bizarre dynamical features of its scattering amplitudes soon casted doubts on the consistency of this kind of theories. We have reported in Section 3.4 the results obtained in \cite{gm} for a noncommutative scalar theory: Cutkoski's rules were found to hold only in the magnetic case; in the electric case an additional tachyonic branch cut is present \footnote{The case of a light-like noncommutativity is peculiar and was discussed in \cite{agm}}. We reported also the analogous result that was found in \cite{bgnv} for noncommutative gauge theories: here again the cutting rules hold if the noncommutativity does not involve the time variable, otherwise new intermediate tachyonic states call for being inserted in order to explain the analytic structure of the vacuum polarization tensor \footnote{For a discussion on the role of UV/IR mixing in relation to unitarity in NC field theories see \cite{clz}}. These new possible states were considered in \cite{abz}. This phenomenon is related to the fact that in the electric case the Seiberg-Witten limit does not succeed in a complete decoupling of the string states. The best one can do in general in this situation is to send the electric field at its critical value, ending with a theory of only open strings (NCOS), namely decoupling all closed strings but keeping the open massive tower of states (\cite{ssto,gmms}, see also \cite{kar,kp}). 

From the point of view of the behaviour of open strings in electric backgrounds, the situation is well understood \cite{ft,acny,clny,b,n,bp,ba}: there is a critical value of the field beyond which the string becomes unstable. In particular, an electric field along the string can stretch it and balance its energy, reducing its effective tension to zero, making pair production possible from the vacuum without energy loss\footnote{One way to see how this phenomenon manifests itself is to look, for example, at the effect of the electric field on the open string metric $G_{\mu \nu}$: basically, its components along the directions of the electric field become zero when this last approaches the critical value (see in the next, under formulae (\ref{9}), (\ref{16})), which corresponds to an infinite stretching of the distance in these directions}. The limit of Seiberg and Witten precisely forces the electric field to overcome this critical value, even if the full string theory from which one starts is taken in the region of stability. 

Our aim is to explore this situation starting from the analytical structure of the full string theory amplitude, namely we want to reproduce the results obtained in a noncommutative field theory from the study of the analytical structure of the string two-point function, precisely looking at what happens to its branch cuts when one performs the Seiberg-Witten limit. We see that the string theory two-point amplitude, when viewed as a function of the squared momentum of the external leg, has two positive branch cuts in the complex plane below the critical value of the electric field. We find that one of this branch cuts is parameterized by a quantity that changes sign when the electric field overcomes its critical value: the branch cut becomes tachyonic in this situation, and, consequently, the field theory amplitude one gets as a limit, which coincides precisely with the noncommutative electric-type amplitude, loses unitarity. 

This matter is reported in \cite{mio}.

\section{The Full String Theory Amplitude}
Let us begin considering one-loop tachyon amplitudes in the bosonic open string theory in the presence of a constant $B$-field living on a D$p$-brane, as anticipated in Section 2.2\footnote{For the derivation of amplitudes for photon-vertex operators the reader is referred to \cite{bcr,gkmrs}}. These amplitudes have been studied in \cite{ad,kl,bcr,l,crs}; in particular in \cite{ad} it was shown that the Seiberg-Witten zero-slope limit gives precisely the amplitudes of the NC ${\phi}^3$ theory. The NC ${\phi}^3$ theory is also one of the cases examined in \cite{gm}, where the breakdown of cutting rules was found; we are going to analyze it in detail. 

We therefore start writing the string amplitude for a generic dimension $p$ of the brane, governed by the sigma model action (\ref{sigma}) (see Chapter 2), 
\begin{eqnarray}
\label{1}
S={{1}\over{4\pi {\alpha}'}}\int_{C_2}d^2 z \, (g_{ij} {\partial}_a X^i {\partial}^a X^j - 2 i \pi {\alpha}' B_{ij} {\epsilon}^{ab}{\partial}_a X^i {\partial}_b X^j).     
\end{eqnarray}
At one loop the string world-sheet is conformally mapped on the cylinder $C_2 = \{0\leq \Re w\leq 1, w = w + 2 i \tau$\}; $\tau$ is the modulus of the cilinder, and we recall that the metric $g$ and the antisymmetric tensor $B$ are constant closed-string sector backgrounds. The indices $i$ and $j$ live on the brane (see section 2.2.1). The one loop propagator we adopt in this situation, with the new boundary conditions imposed by the $B$-term in (\ref{1}), has been constructed in \cite{acny},\cite{clny},\cite{ad}. If one sets $w=x+iy$, the relevant propagator on the boundary of the cilinder ($x=0,1$) can be written as \cite{ad}
\begin{eqnarray}
\label{2}
G(y,y')={{1}\over{2}}{\alpha}' g^{-1} \log q - 2 {\alpha}' G^{-1} \log \Big[{{q^{{1}\over {4}}}\over {D(\tau)}} \, {\vartheta_4}({{|y-y'|}\over {2 \tau}}, {{i}\over {\tau}})\Big],\, \, \, x\neq x', 
\end{eqnarray}
\begin{eqnarray}
\label{3}
G(y,y')={{\pm i \theta }\over{2}} {\epsilon}_{\perp} (y-y') - 2{\alpha}' G^{-1} \log \Big[{{1}\over {D(\tau)}} {\vartheta_1}({{|y-y'|}\over {2 \tau}}, {{i}\over {\tau}})\Big],\, \, \, x=x',
\end{eqnarray}  
where $q=e^{-{{\pi}\over{\tau}}}$, $\pm$ correspond to $x=1$ and $x=0$ respectively, and ${\epsilon}_{\perp} (y) = sign(y) - {{y}\over{\tau}}$.
The open string parameters are as in \cite{sw}:
\begin{eqnarray}
\label{4}
G=(g-2\pi {\alpha}' B)g^{-1} (g+2\pi {\alpha}' B)
\end{eqnarray}
is the open string metric, and
\begin{eqnarray}
\label{5}
\theta=-{(2\pi{\alpha}' )}^2 {(g+2\pi {\alpha}' B)}^{-1}B {(g-2\pi {\alpha}' B)}^{-1}
\end{eqnarray}
is the noncommutativity parameter. ${\vartheta_{1,4}}(\nu, \tau )$ are Jacobi theta functions, while $D(\tau)={\tau}^{-1} {[\eta({i\over \tau})]}^3$, where $\eta$ is the Dedekind eta function\cite{p}. One can notice that, while at tree level the boundary propagator is entirely written in terms of the open string parameters (see Eq.(\ref{propbound}), now there is a piece in Eq.(\ref{2}) in which the closed string metric is explicitly present. This non decoupled closed string part will have a crucial role in the amplitude, originating in the Seiberg-Witten limit the noncommutative Moyal phase of non-planar graphs.

With this propagator and the suitable modular measure, the amplitude for the insertion of $N$ tachyonic vertex operators at $x=1$ and $M-N$ at $x=0$ is easily calculated: 
\begin{eqnarray}
\label{6}
A_{N.M}&=&{\cal{N}}{(2\pi)}^{d} {({\alpha}')}^{\Delta} {G_s}^M \int_0^{\infty} {{d\tau}\over{\tau}} {\tau}^{-{{d}\over{2}}} {[\eta(i\tau)]}^{2-d} q^{{{1}\over{2}}{\alpha}' Kg^{-1} K}  \nonumber \\
&&\times \Big(\prod_{a=1}^M \int_0^{y_{a-1}} dy_a \Big) \prod_{i=1}^N \prod_{j=N+1}^M {\Bigg[ q^{{1}\over {4}} \, {\vartheta_4}({{|y_i - y_j|}\over {2 \tau}}, {{i}\over {\tau}}) / D(\tau)\Bigg]}^{2 {\alpha}' k_i G^{-1} k_j } \no \\
&&\times \prod_{i<j=1}^N e^{- {{1}\over{2}}i {\epsilon}_{\perp} (y_i - y_j)k_i \theta k_j} {\Bigg[ {\vartheta_1}({{|y_i -y_j|}\over {2 \tau}}, {{i}\over {\tau}}) / D(\tau)\ \Bigg]}^{2 {\alpha}' k_i G^{-1} k_j }\nonumber \\
&&\times \prod_{i<j=N+1}^M e^{{{1}\over{2}}i {\epsilon}_{\perp} (y_i - y_j)k_i \theta k_j} {\Bigg[ {\vartheta_1}({{|y_i -y_j|}\over {2 \tau}}, {{i}\over {\tau}}) / D(\tau)\ \Bigg]}^{2 {\alpha}' k_i G^{-1} k_j } \no \\
&&\, \, + \, \, \, non-cyclic \, \, \, \, permutations.
\end{eqnarray}
Constants are as in \cite{ad}: ${\cal{N}}$ is the normalization, $\Delta = M {{d-2}\over{4}} - {{d}\over {2}}$, $G_s$ is the open string coupling. Moreover $d=p+1$, $K=\sum_{i=1}^N k_i$ is the sum of all momenta associated with the vertex operators inserted on the $x=1$ boundary, and ${y_0} = 2\tau$. Whenever $N,M\neq0$, this amplitude corresponds to nonplanar graphs: this is most easily realized when mapping the cilinder into an annulus whose internal boundary corresponds, say, to $x=0$ and the external one to $x=1$. We have also omitted a global delta function due to momentum conservation, and the traces of the Chan-Paton matrices.

\section{Calculation of the Branch Cuts}
We want to discuss the case of $N=1$, $M=2$, that is a diagram with two vertex inserted on opposite boundaries, which, in the field theory limit, will become the nonplanar contribution to the two-point function. We recall once again that nonplanar diagrams are the ones in which the dependence on noncommutativity does not factor out of the loop integral, and therefore they are substantially different from their commutative counterpart (see Section 3.2); they are indeed the diagrams we are interested in. 

We rescale $t=2\pi {\alpha}' \tau$ and $\nu_{1,2} ={{y_{1,2}}/ {2 \tau }}$. Then, we set $\nu_{2} =0$ to fix the residual invariance. The result we obtain from Eq.(\ref{6}) is 
\begin{eqnarray}
\label{7}
A_{1.2}&=&{\cal{N}} {G_s}^2 \, 2^{{3 d} \over 2}{\pi}^{{{3 d}\over{2}}-2} {\alpha '}^{{{d}\over{2}}-3} \int_0^{\infty} dt \, t^{1-{{d}\over{2}}} \, {\Big[ \eta({{i t}\over{2\pi {\alpha '}}})\Big] }^{2-d}\times \nonumber \\
&&\, e^{- {{{\pi}^2}{\alpha '}^2 \over {t}} kg^{-1} k} 
\int_0^1 d\nu {\Bigg[{{e^{-{{{\pi}^2}{\alpha '}\over{2t}}} {\vartheta_4}(\nu, {{2 \pi i {\alpha '}}\over {t}})} \over { {{2 \pi{\alpha '}}\over{t}} {[\eta({2\pi i {\alpha '}\over {t}})]}^3 }}\Bigg]}^{- 2 {\alpha '} kG^{-1} k},
\end{eqnarray}
$k$ being the external momentum. In deriving this expression we have taken into account momentum conservation $k_1 = - k_2 = k$; moreover, with two momenta, there are no non-cyclic permutations. 

Now we perform the zero-slope limit of Seiberg-Witten we have discussed in Section 2.2.2. We recall it consists in sending ${\alpha}' \to 0$ keeping $\theta$ and $G$ fixed (in this case we will keep $t$ and $\nu$ fixed as well). This can be done as indicated in (\ref{SWlimit}), setting ${\alpha}' \sim {\epsilon}^{1\over 2}$ and the closed string metric $g\sim \epsilon$, and then sending $\epsilon \to 0$ \cite{sw}. The formulae for the asymptotic values of the functions in the integrand are as follows: from the expression $\eta (s) = x^{1\over 24}\prod_{m=1}^{\infty} (1 - x^m)$ where $x=\exp [2 \pi i s]$, we deduce the behaviour $$\eta ({{i t}\over{2\pi {\alpha '}}}) \sim \exp [-{t\over{24 \alpha '}}];$$ by using the property $\eta (s) = {(- i s)}^{- {{1}\over {2}}} \eta (- {{1}\over {s}})$ we get $$\eta({2\pi i {\alpha '}\over {t}})\sim {({2\pi {\alpha '}\over {t}})}^{-{{1}\over {2}}} \exp [-{t\over{24 \alpha '}}],$$ and finally, since one has that ${\vartheta_4}(\nu, \tau ) = {(- i \tau)}^{- {{1}\over {2}}} \exp [- {{i \pi {\nu }^2 }\over {\tau}}] \, {\vartheta_2}({{\nu }\over {\tau}},- {{1}\over {\tau}})\, $, and $\, {\vartheta_2}(\rho, \sigma)=\sum_{n=-\infty}^{\infty} e^{\pi i \sigma {(n + {1\over 2})}^2 + 2 \pi i \rho (n + {1\over 2})}$, we obtain $${\vartheta_4}(\nu, {{2 \pi i {\alpha '}}\over {t}})\sim {({2\pi {\alpha '}\over {t}})}^{-{{1}\over {2}}} e^{{t\over {2 \pi {\alpha '}}}[- {{\pi}\over {4}} + \pi (\nu - {\nu}^2)]}.$$

We relate the field theory coupling constant to the open string parameters as $g_f = G_s \, {{\alpha}'}^{{{d-6}\over {4}}}$: in this way, all powers of $\alpha'$ in front of the integral are absorbed in the field theory coupling. Then, we recall that the open string spectrum we consider here is determined by $m^2 = {{N - a}\over {\alpha '}}$, where $a = {{d-2}\over {24}}$ and $d = p + 1$ is the space-time dimension of the brane over which the indices $i,j$ in Eq.(\ref{1}) run. The tachyon mass is therefore ${m^2} = {{2-d}\over {24 {\alpha}'}}$. Furthermore, $g^{-1}=- {{1}\over {(2 \pi {\alpha}')}^2} \theta G \theta$ in the Seiberg-Witten limit. Putting everything together, one obtains the following\cite{ad}:
\begin{eqnarray}
\label{8}
A_{1.2}^{lim} ={\cal{N}} 2^{{3 d} \over 2}{\pi}^{{{3 d}\over{2}}-2} {g_f}^2 \int_0^{\infty} dt t^{1-{{d}\over{2}}} e^{- m^2 t \, + \, k\theta G\theta k/4 t}\int_0^1 d\nu e^{- t\, \nu (1 - \nu )\, k G^{-1} k},
\end{eqnarray}
This reproduces the expression for the two-point function in the noncommutative $\phi^3$ theory. An important point is that now we choose the brane to be actually a string, therefore $d=2$. This case is peculiar and can present a series of pathologies; it can be considered as a tool in order to simply understand the mechanism we consider to be at the origin of the phenomenon under scrutiny even in more general situations. The tachyon mass squared goes to zero from below (if $d>2$ the effective theory is naturally tachyonic, ${m^2} = {{2-d}\over {24 {\alpha}'}}$, we recall that this represents a two-tachyon amplitude in string theory). 

In this situation, the effective field theory limit (\ref{8}) is the nonplanar amplitude of a massless scalar in two-dimensional NC $\phi^3$ theory
\begin{eqnarray}
\label{10}
{A_{1.2}^{lim}}_{d=2} ={\cal{N}} 8 \pi \, {g_f}^2 \int_0^{\infty} dt \int_0^1 d\nu \, e^{- t \, \nu \, (1 - \nu )\, kG^{-1} k\,  + \, {1\over {4t}}k\theta G\theta k}.
\end{eqnarray}
In two dimensions the noncommutativity parameter $\theta$ is necessarily proportional to the antisymmetric tensor. We are once again in the right situation to study the effects of electric-type backgrounds \footnote{We also recall that the noncommutative nonplanar diagram equals the planar one if we set $\theta=0$}. We can also take the open string metric to be proportional to the flat Minkowski metric ${\eta}_{\mu \nu }$, and we define a constant G such that $G^{-1} k^2 = k_{\mu} {[G^{-1}]}^{\mu \nu} k_{\nu}$, where $k^2$ is the usual Minkowski invariant. The field theory amplitude can then be written as 
\begin{eqnarray}
\label{11}
{A_{1.2}^{lim}}_{\, d=2} ={\cal{N}} 8 \pi \, {g_f}^2 \int_0^{\infty} dt \int_0^1 d\nu \, e^{- k^2 [G^{-1} t \, \nu (1 - \nu ) - {{ \theta}^2 \over {4 G^{-1} t}}]}.
\end{eqnarray}
We add a small regulator ${m_0}^2$ , and get 
\begin{eqnarray}
\label{12}
{A_{1.2}^{lim}}_{\, d=2} ={\cal{N}} 8 \pi \, {g_f}^2 \int_0^{\infty} dt \int_0^1 d\nu \, e^{- {m_0}^2 G^{-1} t - k^2 [G^{-1} t \, \nu \, (1 - \nu ) - {{ \theta}^2 \over {4 G^{-1} t}}]}.
\end{eqnarray}
We assume that $G$ is positive, as well as ${m_0}^2$. We see that the integral in (\ref{12}) is convergent in the strip $\{- 4 {m_0}^2 <\Re k^2<0\}$:   
\begin{eqnarray}
\label{13}
{A_{1.2}^{lim}}_{\, d=2} ={{-{\cal{N}} 8 \pi {g_f}^2 {\theta}^2 k^2}\over {G^{-1}}} \int_0^1 d\nu \Bigg[{{K_1\Big( \sqrt{- {\theta}^2 k^2 ({m_0}^2 + k^2 \nu (1 - \nu))}\Big)}\over {\sqrt{- {\theta}^2 k^2 ({m_0}^2 + k^2 \nu (1 - \nu))}}}\Bigg].
\end{eqnarray}
After analytic continuation, it defines an analytic functions with two branch cuts: $\{\Re k^2 <-4 {m_0}^2\}$ and $\{\Re k^2>0\}$\footnote{As a side remark, we notice that the amplitude (\ref{13}) is continuous in the limit $\theta \to 0$, thanks to the lack of UV/IR mixing in two dimensions (see Section 4.4.1)}. One of these is necessarily tachyonic, and cutting rules are invalidated. If we send the regulator to zero, the two branch points get closer and closer and eventually coalesce.

Let us analyse now the full string theory diagram. The complete string amplitude, before performing the field theory limit, in the case of $d=2$, is easily found from (\ref{7}) to be
\begin{eqnarray}
\label{9}
A_{1.2}^{\, d=2} ={{{\cal{N}} {G_s}^2 8\pi}\over {{{\alpha}'}^2}}\int_0^{\infty} dt \int_0^1 d\nu \, e^{- {{{\pi}^2}{\alpha '}^2 \over {t}} kg^{-1} k} {\Bigg[{{e^{-{{{\pi}^2}{\alpha '}\over{2t}}} {\vartheta_4}(\nu, {{2 \pi i {\alpha '}}\over {t}})} \over { {{2 \pi{\alpha '}}\over{t}} {[\eta({2\pi i {\alpha '}\over {t}})]}^3 }}\Bigg].}^{- 2 {\alpha '} kG^{-1} k}
\end{eqnarray}
We set consistently $g_{\mu \nu } = g {\eta }_{\mu \nu }$ and $B_{\mu \nu } = B {\epsilon }_{\mu \nu }$. A simple calculation shows that $G_{\mu \nu }={{1}\over {g}}[g^2 - {(2 \pi {{\alpha '} B)}^2}] {\eta }_{\mu \nu }$. We can therefore rewrite the string amplitude as   
\begin{eqnarray}
\label{14}
A_{1.2}^{\, d=2} &=&{{{\cal{N}} {G_s}^2 8\pi}\over {{{\alpha}'}^2}}\int_0^{\infty} dt \int_0^1 d\nu \, \exp [-k^2 {{{\pi}^2 {{\alpha '}^2 }}\over {g \, t}}] \nonumber \\
&&\times \exp \Bigg[ {{- 2{\alpha '} g k^2 }\over {[g^2 - {(2 \pi {\alpha '} B)}^2]}}\log [{{e^{-{{{\pi}^2}{\alpha '}\over{2t}}} {\vartheta_4}(\nu, {{2 \pi i {\alpha '}}\over {t}})} \over { {{2 \pi{\alpha '}}\over{t}} {[\eta({2\pi i {\alpha '}\over {t}})]}^3 }}] \Bigg].  
\end{eqnarray}
In order to evaluate the convergence of the integral over $t$ in (\ref{14}), we study the regions close to the two extremes. When regarded as a function of $t$, the integrand of the (\ref{14}) is a regular function in between. At small $t$, the integrand behaves as 
\begin{eqnarray}
\label{15}
{\Bigg( {{t}\over {2 \pi \alpha '}}\Bigg) }^{- 2 \alpha ' k^2 {{g}\over {[g^2 - {(2 \pi {\alpha '} B)}^2}]}} \exp \Big( - k^2 {{{\pi}^2 {{\alpha '}^2 }}\over {g \, t}}\Big).
\end{eqnarray}
In fact, ${\theta_4}(\nu, \tau)=\sum_{n=-\infty}^{\infty} {(-1)}^n \exp [\pi i \tau n^2 ] \exp [2 \pi i n \nu ]$, therefore at small t (and fixed $\alpha'$)
\begin{eqnarray}
\label{vaauno}
{\vartheta_4}(\nu, {{2 \pi i {\alpha '}}\over {t}}) \, \sim \, 1; \nonumber
\end{eqnarray}
then again $\eta (s) = x^{1\over 24}\prod_{m=1}^{\infty} (1 - x^m)$ where $x=\exp [2 \pi i s]$, from which directly we get 
\begin{eqnarray}
\label{laeta}
{\eta}^3 ({{2 \pi i {\alpha '}}\over {t}}) \, \sim \, \exp [- {{{\pi}^2 \alpha'}\over {2 t}}]. 
\end{eqnarray}
Therefore, looking at Eq.(\ref{15}), we have convergence for $g \Re k^2 >0$ and a branch cut along the opposite axis. At large $t$ instead, the first exponential becomes $\sim 1$, while for the argument of the $\log$ we can equivalently use the asymptotic formulae found for small $\alpha'$ : the argument of the log behaves like $e^{{t \nu (1 - \nu )}\over {2 {\alpha '}}}$, therefore the $\log$ actually produces a linear term, with a net result
\begin{eqnarray}
\label{16}
\exp \Big(  - k^2 t \, {{g}\over {[g^2 - {(2 \pi {{\alpha '} B)}^2}]}}\, \nu (1 - \nu ) \Big).
\end{eqnarray}
We see that the branch cut\footnote{Subleading terms coming from $\exp( - k^2 {{{\pi}^2 {{\alpha '}^2 }}\over {g \, t}})$ cooperate to make it a cut} is parameterized by the quantity ${{g}\over {[g^2 - {(2 \pi {{\alpha '} B)}^2}]}} = {{1}\over {g}} {{1}\over {1 - {\tilde E}^2}}$, where we have defined the effective electric field ${\tilde E} = {2 \pi {{\alpha '} B }\over g}$. This is exactly the ratio $E/E_{cr}$ of \cite{ssto} which discriminates the stability of the string in this background\footnote{$E_{cr}={{g}\over {2 \pi \alpha'}}$ is proportional to the inverse of the fundamental string scale $\alpha'$}: $1 - {\tilde E}^2$ must be positive in order to avoid tachyonic instability. This condition is precisely what we find from cutting rules: if this parameter is positive, the region of convergence is $g \Re k^2 >0$, and the branch cut is superimposed to the one coming from the small $t$ analysis; therefore the total amplitude exhibits a single ``physical'' branch cut. We choose $g$ negative, in order to have the cut on the positive real axis. If, on the contrary, this parameter changes sign and becomes negative, as it does in the Seiberg-Witten limit because $g^2 \sim {\epsilon}^2$ and ${\alpha '}^2 \sim\epsilon$, a new branch cut appears on the opposite axis, which is just the unphysical one we find in the effective field theory\footnote{Another way to see it  is that the Seiberg-Witten limit is of a strong electric field, such that it overcomes its finite critical value. Analogously, one says that the critical value goes to zero in their limit}. The key seems to be from here the instability of the string, rather than (or mixed with) the absence of decoupling. 

\section{Cutting Rules}
In this section we perform some explicit calculations of the discontinuities across the branch cuts we have found in the previous analysis. When the electric field overcomes its critical value, we have seen that already at the level of the full open string amplitude a tachyonic branch cut appears. Therefore, we will restrict our subsequent analysis to the case of $|\tilde{E}| < 1$. 

When the electric field is smaller than the critical value, we found a single physical branch cut for the string. The field theory amplitude in the Seiberg-Witten limit manifests however an unphysical branch cut, because this limit forces $|\tilde{E}|$ to become larger than $1$. Again in the limit there is no hope to have an optical theorem. This would not be the case, could we start with a magnetic-like theory (which is impossible in $d=2$). The noncommutative field theory analysis of cutting rules was performed in \cite{gm}. Focusing on the full string theory result, an analysis of factorization of the loop amplitude for a general $M$-point function in the presence of the $B$-field was carried out in \cite{ad} \footnote{The reader is also referred to the fundamental treatment of \cite{gsw}}. Their analysis showed the appearance of the closed string channel in the usual way when ${{t}\over {2 \pi \alpha '}} = \tau \to 0$. This was obtained  in 26 dimensions, with the standard criterion of putting all external momenta on shell and using momentum conservation, such that $\sum_{i<j=1}^{M} 2 \alpha ' k_i G^{-1} k_j = - M$. The analysis of singularities is then performed in the Mandelstam variables, and it revealed the expected appearance of the closed tachyon pole. 

The two-point function is quite peculiar in this respect, since it depends only on one invariant, $k^2$. As in the previous section, we will perform an off-shell analysis in $k^2$\cite{div1,div2,div3} (see also \cite{bern1,bern2,bern3} and \cite{bcr}).  

We choose to continue the variable $d$ to complex values, and go to a strip in the $d$ complex plane in which the tachyon mass is positive, as a mathematical tool in order to recover a better specification of the mechanism. The amplitude we study now is therefore Eq.(\ref{7}), and the same procedure we used to single out the asymptotic behaviours (\ref{15}) and (\ref{16}) leads us to the following results: at small $t$, 
\begin{eqnarray}
\label{17}
{\Bigg( {{t}\over {2 \pi \alpha '}}\Bigg) }^{2 \alpha ' k^2 {{|g|}\over {[g^2 - {(2 \pi {\alpha '} B)}^2}]}} \exp \Big( k^2 {{{\pi}^2 {{\alpha '}^2 }}\over {|g| \, t}} - {{{\pi}^2 {{\alpha '}} (2 - d)}\over {6 t}}\Big),
\end{eqnarray}
where we are assuming $g$ negative. To get Eq.(\ref{17}), we have also used the already reported relation $\eta (s) = {(- i s)}^{- {{1}\over {2}}} \eta (- {{1}\over {s}})$ and obtained
\begin{eqnarray}
\label{ancoraeta}
\eta ({{i t }\over {2 \pi \alpha'}}) \, \sim \, {({{t}\over {2 \pi \alpha'}})}^{- 1/2} \, \exp[- {{\pi^2 \alpha'}\over {6 t}}].
\end{eqnarray} 
We select the strip $(2 - d)>0$, which corresponds to $m^2={{(2-d)}\over {24\alpha '}}>0$. The branch cut is for $\Re k^2 > 4 m^2 |g|$. At large $t$ we have the behaviour 
\begin{eqnarray}
\label{18}     
t^{1-{{d}\over {2}}} \, \, \exp \Big( k^2 t \, {{|g|}\over {[g^2 - {(2 \pi {{\alpha '} B)}^2}]}}\, \nu (1 - \nu )  - m^2 t \Big).
\end{eqnarray}
The branch cut is for $\Re k^2 > 4 m^2 [g^2 - {(2 \pi {{\alpha '} B)}^2}] / |g|$ where we recall that we restrict the analysis to $g^2 - {(2 \pi {\alpha '} B)}^2 > 0$. As long as $B\neq 0$, this branch cut starts below the previous one, being $[g^2 - {(2 \pi {{\alpha '} B)}^2}] / {|g|^2} < 1$. We remark that, going to complex dimensions $2+(d-2)$, we have chosen to keep the antisymmetric field still of electric type, and, together with the momentum, with strictly two components. 

Our results show, therefore, that the two branch cuts of the string amplitude, which are superimposed in $d=2$, are separated in the strip $d<2$. They have, of course, different nature, which reveals their different origin: the first one, coming from the small $t$ corner when the internal radius of the annulus becomes negligible, is driven by the closed string metric $g$, while the second one, coming from the corner in which the modulus $t$ of the annulus is very large, is driven by the open string metric $G$. It starts well below the first one. 

Evaluating the discontinuities of the full analytic formulae is quite a formidable task, therefore we choose to perform the calculation in the approximation in which the momentum of the incoming external tachyon is just above the starting point of the lower cut of a very small amount. In this situation the higher cut is not yet active, as well as the higher string levels, which start\footnote{We recall that the open string spectrum we consider is determined by $m^2 = {{N - a}\over {\alpha'}}$, where $a = {{d-2}\over {24}}$ and $d=p+1$ is the space-time dimensions of the brane-worldvolume} from $m^2 = {{1 - {{d-2}\over {24}}}\over {\alpha '}}$: we split the integration region in order to separate small $t$ values from large ones, then only this second term develops a discontinuity, the first one resulting in an entire function of the momentum. Here $\alpha '$ is finite, but the momentum is at the threshold, therefore only very high values of $t$ are important, and the discontinuity is the one of the field theory-like expression 
\begin{eqnarray}
\label{19}
C \int_0^1 d\nu \, \int^{\infty} dt \, t^{1-{{d}\over{2}}} \, \exp \Big( k^2 t \, {{|g|}\over {[g^2 - {(2 \pi {{\alpha '} B)}^2}]}}\, \nu (1 - \nu )  - m^2 t \Big),  
\end{eqnarray}
with $C = {\cal{N}} {G_s}^2 \, 2^{{3 d} \over 2}{\pi}^{{{3 d}\over{2}}-2} {\alpha '}^{{{d}\over{2}}-3}$. Integrating over $t$ we exploit the freedom we have in choosing the lower extreme, which we fix at ${{1}\over {\gamma}}$, $\gamma = |g| / [g^2 - {(2 \pi {{\alpha '} B)}^2}]$, and get\footnote{We use that $\int_1^{\infty} t^a \, \exp [- b\, t] \, = \, b^{- 1- a} \, \Gamma [a + 1, b]$ if $\Re b >0$}
\begin{eqnarray}
\label{20}
{\cal{N}} {G_s}^2 \, 2^{{3 d} \over 2}{\pi}^{{{3 d}\over{2}}-2} {\alpha '}^{{{d}\over{2}}-3} {\gamma}^{{{d}\over {2}} - 2} \int_0^1 d\nu \, {{\Gamma [\omega, {\mu}^2 - k^2 \nu (1 - \nu )]}\over {{[{\mu}^2 - k^2 \nu (1 - \nu )]}^{\omega }}}, 
\end{eqnarray}
$\Gamma$ being the incomplete gamma function, $\omega = 2 - (d/2)$, and $\mu^2 = m^2 / \gamma$. The discontinuity is found to be of a logarithmic type, and, using transformation properties of $\Gamma$, it can be explicitly evaluated near the threshold
\begin{eqnarray}
\label{21}
Disc A_{1.2} &\propto& i{\cal{N}} {G_s}^2 \, 2^{{3 d} \over 2}{\pi}^{{{3 d}\over{2}}-1} {\alpha '}^{{{d}\over{2}}-3} {\gamma}^{{{d}\over {2}} - 2} {{\Gamma [1/2]}\over {\Gamma [(3/2) - \omega ]}} \no \\
&&{\left( {{k^2} \over {4}}\right) }^{- \omega } {\left( 1 - {{4 \mu^2}\over {k^2}}\right) }^{(1/2) - \omega} \Theta (k^2 - 4 \mu^2 ).
\end{eqnarray}
The proof is reported in the Appendix C. 

Expression (\ref{21}) equals the phase space of one scalar particle going into two with a $\Phi^3$ dynamics in $d$ dimensions and with a mass $\mu$:
\begin{eqnarray}
\label{phasespace}
phase \, space \, \propto \, \int {{d{\vec {p_1}} d{\vec {p_2}}}\over {2 E_1 2 E_2}} \, \delta ({\vec {p_1}} +{\vec {p_2}} -{\vec k}) \, \delta (E_1 + E_2 - E),
\end{eqnarray}     
where $E_i = \sqrt{{\vec {p_i}}^2 + \mu^2}$ and $E = \sqrt{{\vec k}^2 + \mu^2}$. We see therefore that close to the threshold $k^2 \sim 4 \mu^2$ the cutting rules are satisfied with a $\Phi^3$ dynamics: higher string levels cannot contribute since they are too high in energy, as well as the second cut, and only two scalar particles can be exchanged, with a field theory-like vertex.

We will try now to isolate also the discontinuity across the second cut. We will use the same trick as before, focusing this time on the small $t$ region. We study the leading behaviour at small $t$:
\begin{eqnarray}
\label{22}
\int_0^1 d\nu \, \int_0 dt \, {\left( {{t}\over {2 \pi \alpha '}}\right) }^{2 \alpha ' k^2 {{|g|}\over {[g^2 - {(2 \pi {{\alpha '} B)}^2}]}}} e^{k^2 {{{\pi}^2 {{\alpha '}^2 }}\over {|g| \, t}} - {{{\pi}^2 {{\alpha '}} (2 - d)}\over {6 t}}}. 
\end{eqnarray} 
We choose as the upper extreme of integration $t = 2 \pi \alpha '$ and make the change of variable $t \rightarrow {{1}\over {t}}$ to reduce the integral to the already analysed form (\ref{19}). In so doing we also freeze $k$ in the exponent of ${\left( {{t}\over {2 \pi \alpha '}}\right) }^{- 2 \alpha ' k^2 {{|g|}\over {[g^2 - {(2 \pi {{\alpha '} B)}^2}]}}}$ to be at the actual threshold: this combination cannot interfere with the singularity coming from the exponential. After redoing an analysis similar to the previous one, but without the integration over $\nu$, we find that the discontinuity is proportional to
\begin{eqnarray}
\label{23}
{\left( 1 - {{4 |g| m^2}\over {k^2}}\right) }^{1 + {|g|}^2 {{2 - d} \over {3 [g^2 - {(2 \pi {{\alpha '} B)}^2}]}}} \, \Theta (k^2 - 4 m^2 |g|).
\end{eqnarray} 

We finally notice also that for $B=0$ the two branch cuts are superimposed even for $d<2$, and the open string metric coincide with the closed one. In this situation one should instead perform the effective field theory limit keeping this metric fixed, and one would obtain the commutative $\Phi^3$ theory.  

\chapter{Conclusions and Perspectives}

\section{Moral}
In this thesis we have summarized the work we have performed during three years of investigation on noncommutative field theories. Our analysis has been twofold: first, we have investigated noncommutative gauge theories in their own, trying to understand their features, and to clarify, through particular examples, the main effects that noncommutativity, especially of electric type, produces in a quantum field theory. Then, we have turned our attention to the relation with string theory, and analysed the link between string amplitudes and their noncommutative zero-slope limits. Before summarizing the main achievements, and describing the open problems as well as the perspectives, we want to draw a conclusion of our work. Noncommutative theories are a wonderful tool in order to investigate, through the instruments of QFT, energy regimes in which the space-time has lost its usual properties. The mixture between the standard framework of quantum field theory and the ingredient of noncommutative geometry seems to have the power to reveal a lot of features of the space-time at the scale of the unification of gravity with the other forces. At these scales, string theory is likely to have the ultimate answer. Under certain constraints noncommutative field theories turn out to be an effective description for the ground-state of the string spectrum, in the presence of an antisymmetric constant background. They behave similarly to string theory, and their ability to reproduce some very peculiar features of it is at the basis of their importance. On their own, they are however often affected by some problems. We think one should learn to use them as effective tools for analyzing the complexity of string theory, in particular mathematical difficulties and geometrical consequences, from a privileged point of view, without trying to make them candidates for a final model. It is following these ideas that we have turned our attention to some particular examples, which could clarify the mathematical properties of the perturbative expansion in such a landscape, in the light of string theory.   

\section{Main Achievements}
In usual commutative two-dimensional Yang-Mills theories ($YM_{1+1}$), closed
Wilson loops can be exactly computed, either using geometrical techniques
or by summing a perturbative series in light-cone gauge, in which the
propagator is handled with a Cauchy principal value prescription.
As a matter of fact in this case contributions from crossed diagrams vanish
thanks to peculiar support conditions. Only planar diagrams survive, enforcing 
the picture of the loop as an exchange potential between two (static)
colour sources. A simple area exponentiation is the final outcome.

If the propagator is instead prescribed in a causal way, crossed diagrams no 
longer vanish; the result one obtains by summing the series no longer 
exhibits Abelian-like exponentiation and only captures the zero instanton
sector of the theory\cite{anlu}.

When considering non-commutative $YM_{1+1}$, one may wonder whether such a theory 
should make sense at all, owing to the pathologies occurring when non-commutativity
involves the time variable. Indeed even simple models exhibit either trivial
or inconsistent solutions in such a case \cite{seme}.

Therefore, we found quite remarkable that a sensible expression can be obtained
for a closed Wilson loop as a fourth-order
perturbative calculation using the WML propagator. 
It exhibits a smooth limit when the non-commutative
parameter $\theta$ tends to zero, thereby recovering the usual commutative result;
in the large non-commutative limit $\theta \to \infty$, the crossed diagrams
contribution changes sign,
and reveals not to be depressed with respect to the planar part. 
The loop does not obey the Abelian-like
exponentiation constraint, but this is not surprising as it happens also
in the usual commutative case.

More dramatic is the situation when considering  't Hooft's form of
the free propagator, which does not allow a smooth transition to Euclidean
variables. Here the presence of the Moyal phase produces singularities
even at ${\cal O}(g^4)$, which cannot be cured. Strictly speaking, the
Wilson loop diverges (at least perturbatively). Nevertheless we find
extremely interesting the circumstance that the difference between its Moyal
dependent term and the analogous one in the WML case turns out to be 
confined only in two singular terms, precisely ascribed to the delta-like contribution present in the PV propagator. This signal the incompatibility of a point-like interaction with noncommutative geometry.

Turning back to WML, and considering the next order and higher ones, we can say that, when winding
$n$-times around the Wilson loop, the non-Abelian nature of the gauge group
in the non-commutative
case is felt, even in a perturbative calculation making use of the WML
prescription for the vector propagator. This is due to the merging of 
space-time properties with
``internal'' symmetries in a large invariance unitary group, which can be identified with $U_{cpt}({\cal H})$\cite{har,sza}: this is the group of operators of the form $1 + K$, with $K$ compact.
One gets the clear impression that in a non-commutative formulation
what is really relevant are not separately the space-time properties
of the ``base'' manifold and of the ``fiber'' $U(N)$, but rather
the overall algebraic structure of the resulting invariance group. To properly understand its topological features is
certainly beyond any perturbative approach. Rather, one should possibly resort
to suitable ${\cal N}$-truncations of the Hilbert space in the form 
of matrix models leading to the invariance groups $U({\cal N})$.

It is not clear how many perturbative features might eventually be
embodied in those contexts, especially in view of the difficulty in performing the
inductive limit ${\cal N}\to \infty$. 

For this reason we think
that our perturbative results are challenging.
They indicate that the intertwining between $n$, controlling the
space-time geometry, and $N$, related to the gauge group, is far
from trivial. The presence of corrections to the scaling laws
occurring at $\theta=0$ and at $\theta=\infty$, while frustrating
at a first sight in view of a generalization to all values
of $\theta$ and, possibly, to an all-order 
resummation of the perturbative series, might be taken
instead as a serious indication
that $n$ and $N$ separately are not perhaps the best parameters to be chosen
unless large values for both (and for $\theta$!) are considered. In such a
situation, perhaps surprisingly, the relation (\ref{scaling}) is recovered.

Eqs.(\ref{wuplanar},\ref{conject}) 
are concrete realizations of the more general structure
\begin{equation}
\label{gener}
\WW_{2m+4}=({\cal A}n^2 N)^{m+2}\, f_m(n,N),
\end{equation}
$f_m$ being a symmetric function of its arguments. We stress that Eq.(\ref{wuplanar}) 
concerns only {\it planar} diagrams (which we recall coincide with the ones in the commutative case); crossed graphs contributions
in the commutative case cannot be put in the form (\ref{gener}) and violate the relation
(\ref{scaling}) \footnote{The structure (\ref{gener}) is shared
also by the exact geometrical solution of the commutative case.}. 
In the non-commutative case, for large $n,N$ and maximal
non-commutativity ($\theta=\infty$), the structure (\ref{gener})
is instead restored for the leading contribution
of {\it crossed} diagrams. The presence of the function $f_m$ in the $WML$ context
might be thought as a sign of the merging of space-time
and internal symmetries. 

All these difficult, but intriguing questions are worthy in our opinion
of thorough investigations and promise further exciting, unexpected developments. We refer, for instance, to some nonperturbative attempts made in this direction by the authors of \cite{lu,bieten,sza}.

We then turn to open string amplitudes and their noncommutative zero-slope limits. We have shown that the breakdown of perturbative unitarity in noncommutative electric-type field theories can be related, from the point of view of cutting rules, to the appearance of a tachyonic branch cut in the corresponding string theory amplitude when the electric field overcomes its critical value. We have analyzed the simple case of a two dimensional brane-worldvolume, which leads to a massless scalar noncommutative $\phi^3$ theory. The string amplitude below the critical field has two branch cuts, both positive, but the zero-slope limit forces the electric field to overcome its critical value. At the same time the quantity that parameterizes one of the branch cuts becomes negative, and the amplitude enters the region of instability. The corresponding noncommutative field theory violates unitarity in this situation. 

We have studied the discontinuities of the full string theory amplitude across the branch cuts, keeping the electric field below the critical value in order to have a stable string. These two branch cuts have different nature, and, in dimensions of the brane-worldvolume different from two, start from different points. This separation, furnished by the $B$-field, allows us to put ourselves in the approximation of an incoming momentum just above the lower threshold, the upper one, as well as the higher string levels, being not still active. In this situation we have recovered, for a small region above the lower branch point, the cutting rules for simple $\phi^3$ phase space. We have also discussed the features of the upper threshold: in the case of $B=0$ it would be superimposed to the lower one, and is driven by the closed string metric. 

\section{Perspectives}

The researches that now call for being performed, along the lines indicated by the calculations we have presented here, have, as their main points, on one hand, the analysis of possible nonperturbative derivations of our formulae for the two-dimensional Wilson loop, and the connection of these relations with a geometrical study of the noncommutative larger gauge group, on the other hand, the investigation of the quantum spectrum of noncommutative field theories in relation to the string theory spectrum, extending our results to higher open string levels, as well as to the truly clarifying case of Superstrings. 

In order to really understand what happens at the level of the unifying larger gauge group, it is necessary to be able to interpret the meaning of our scaling results for the Wilson loop. They are open to non-perturbative tests, and constitutes the natural \it paradigma \rm for all the attempts to a geometrical analysis of the structure of noncommutative gauge theories, which have to reproduce them from a consistent mathematical framework.  

From the string theory point of view, our one-loop analysis can add some clarifications about
the origin of the breaking of perturbative unitarity when time is involved in noncommutativity. We have to overcome the problems of our simplified analysis. First of all we have to get rid of the tachyon: one should first consider higher bosonic string levels, subtracting the tachyon by hand, but our project is to reach a fully clarifying analysis, finally passing to Superstrings, and consistently including supersymmetry in our formulae.

\chapter{Acknowledgements}
I have not enough words for thanking Professor Antonio Bassetto, for having been, on one hand, such an enlightened, wise, indefatigable, helpful and kind an advisor, and, on the other hand, such an expert, stimulating, untiring, friendly, unexcelled collaborator. I thank him for teaching me an infinite number of notions, which I am still trying to fix in mind, for showing me the best way at every cross-roads, for discussing with me and listening to my poor ideas, and correct them, and for never stopping unravelling Physics. 

I wish to thank heartily Giuseppe Nardelli, for his invaluable collaboration, help, disposal and enthusiasm, his friendly discussions, and for being able, along with Antonio, to drive me poor through some such intricate calculations as I never performed in my life before. 

I wish to thank Federica Vian, for many useful discussions and extremely fruitful collaboration at various stages of this work.

I wish to thank Luca Griguolo, for very interesting discussions and comments.

I wish to thank Professor Loriano Bonora, for having read the paper on which Chapter 6 of this thesis is based, and for his kind and useful suggestions.

I wish to thank Chong-Sun Chu and Jaume Gomis for very useful suggestions. I wish to thank Rodolfo Russo for very useful suggestions and extremely interesting discussion. 

I wish to thank Roberto Valandro for his help, for stimulating and fruitful collaboration, and for very deep, useful and beautiful discussions.

I wish to thank Supriya Kumar Kar and Mofazzal Azam for useful suggestions.

I wish to thank Professor Mario Tonin, for his lessons and teachings in string theory.

I wish to thank Matteo Viel, for having faced the computer for us, writing the program that has been used for the numerical calculations we have performed in Section 5.2.2, and for being, together with Nicola Bartolo, my inseparable fellows.

I wish to thank Lorenzo Fortunato, for teaching me so many subtleties of quantum mechanics, and for some nights spent on a pizza calculating atomic crazy matrix elements.

I wish to thank Carla Biggio, for innumerable extremely stimulating discussions and invaluable suggestions.

I wish to thank all my friends, my PhD colleagues with whom I spent all of my week, for their inestimable friendship, for their infinite help and for their being so great.        

Every drop of my sweat, every thought of my mind, every second of my life, every type of this work is dedicated to my family.

\appendix
\chapter{Formulae}
\bf 1) Formula 1 \rm
 
\bigskip
This is the formula used in general as the first step in the integration of the Euclidean diagrams:
\begin{eqnarray}
\label{prima}
&&\int {{dp}\over {p}} \int d\varphi \, \exp [- 2 i \varphi] \, \exp [i p \Sigma  \sin \varphi ] \exp [i p (a \sin (\varphi - \alpha ) + b \sin (\varphi - \beta ))] \nonumber \\
&& \, \, = \, \, \pi {{\Sigma + a \exp [- i \alpha ] + b \exp [- i \beta ]} \over {\Sigma + a \exp [i \alpha ] + b \exp [i \beta ]}},
\end{eqnarray}
where all quantities $\Sigma$, $a$, $\alpha$, $b$ and $\beta$ are considered here as real ones. 

\bigskip
\bf 2) Formula 2\rm

\bigskip
In crossed diagrams, one needs this formula to carry on the second significant step, after having used the formula 1:

\begin{eqnarray}
\label{seconda}
\int {{dp}\over {p}} \int d\varphi \, \exp [- 2 i \varphi] \, \exp [i p \Sigma  \sin \varphi ] \, {{1 - B p \exp [- i \varphi ]}\over {1 - \bar{B} p \exp [i \varphi ]}} \nonumber \\
= 2\pi \Bigg( {{2}\over {z^2}} \exp [- i z \sin \tau ] \, + \, {{2 i}\over {z}} \sin \tau \, + \, - {{2}\over {z^2}} + {{1}\over {2}}\Bigg),
\end{eqnarray}

where $z = \Sigma / |B|$, $\exp [i \tau ] = \bar B / |B|$, and $\Sigma$ is real. 

To derive this formula, which is not listed in tables to our knowledge, and is far from trivial, one needs a lot of intermediate steps, which we will only sketch very briefly here. First, one has to define $\zeta = \exp [i \varphi ]$, then expand the remaining exponential in a double Taylor series, one for its part in $\zeta $ and one for ${{1}\over {\zeta}}$. One wants to use the geometric series for the piece $1 / (1 - \bar{B} p \zeta )$, so one has to split the $p$-integration into two domains, according to the regions of convergence of the geometric series. We recall that the rational factor is a pure phase really, it is never singular, nor the rest of the integrand in the domain of integration, which is the unit circle for $\zeta$. We are allowed to integrate over $\zeta$ by series, and use the Cauchy theorem. Our result is an integral in $p$ split into two regions, for each of them the integrand being a triple series in powers of $p$. One of the features of this kind of integrals is that the integration on the angular variables, which has to be performed before the one over the momenta \it by prescription \rm (see Section 4.4.1), results in a function which behaves in an extreme regular way both at small and at large values of the momenta. This happens sistematically in our calculations, and in particular, in the present case, this triple series results in a combination of indices that precisely prevents any singularity in the respective regions of $p$-integration. We can furthermore integrate by series and then try to resum. We succeed in resum all the three series, using many transformations and redefinition of the indices themselves: first two resummations lead to Bessel functions, which still have to be summed over their indices. All pieces fit together to give (${{1}\over {2}}$ +) a single sum, which coincides with the ``Jacobi-Anger expansion'' of Bessel functions: it gives the simple exponential of formula (\ref{seconda}), subtracted of its pieces that are singular at small $\zeta$\cite{prudni}.

\bigskip
\bf 3) Formula 3\rm
 
\bigskip
We report here a simple formula useful for single decoupled propagators:
\begin{eqnarray}
\label{terza}
\oint_{|z|=1} \, {{dz}\over {i z^n}} \, \exp [a z - {{b}/{z}}] \, = \, 2 \pi \, J_{n-1} (2 \sqrt{a b}) \, \, {({{a}/{b}})}^{{{n-1}\over {2}}}, 
\end{eqnarray}  
where $J$ is a Bessel function of the first kind.

\bigskip
\bf 4) Formula 4\rm 

\bigskip
Directly connected with the previous one, the following formula is often used:
\begin{eqnarray}
\label{quarta}
\int_0^{\infty} {{dp}\over {p}} \, J_{\nu }(p) \, = \, {{1}\over {\nu}},\qquad \Re{\nu} > 0. 
\end{eqnarray}

\bigskip
\bf 5) Formula 5\rm 

\bigskip
Following basically the same procedure used to get Eq.(\ref{seconda}), we have proved also the following relation
\begin{eqnarray}
\label{quinta}
&&\int_0^{\infty} {{dk}\over {k}} \oint_{|z|=1} {{dz}\over {z^3}} \, \exp [k (a_1 z - a_2 / z)] \, \, {{1 - ({{B k}/ {z}})}\over {1 - \bar{B} k z}} \nonumber \\
&&= \, {{2 \pi i}\over {D^2}} \Bigg[ {{2}\over {z^2}} \exp [{{z}\over {2}} (t - {{1}\over {t}})] \, - \, {{1}\over {z}} (t - {{1}\over {t}}) \, - \, {{2}\over {z^2}} \, + \, {{1}\over {2}} \Bigg], 
\end{eqnarray}
where $D = {(a_2 / a_1 )}^{{{1}\over {2}}}$, $t = \exp [- i \tau ] / D$, $\exp [i \tau ] = \bar B / |B|$ and $z = {{2 {(a_1 a_2 )}^{{{1}\over {2}}}}\over {|B|}}$.

\chapter{Proof of Eq.(\ref{primspl})}
We will sketch here only the main points of the proof for a simpler paradigmatic case; the proof can be generalized to similar situations, in particular to Eq.(\ref{primspl}). Let us prove that, in the limit $\gamma' \to \infty$,  
\begin{eqnarray}
\label{magior}
&&\int_0^{\infty } {{dk} \over {k}}\ointop_{|z|=1}\, {{dz} \over {i {{z}^3}}}(e^{-\ k  \sin [n \pi s_{ij}] (z - {{1} \over {z}})}- 1){{1 - {{\gamma }' \over {z}} e^{- i n \pi \sigma }} \over {1 - \gamma' z e^{i n \pi \sigma }}} \nonumber \\
&&\longrightarrow \int_0^{\infty } {{dk} \over {k}}\ointop_{|z|=1}\, {{dz} \over {i {{z}^3}}}(e^{-\ k  \sin [n \pi s_{ij}] (z - {{1} \over {z}})}- 1) {{e^{- 2 i \pi n \sigma} }\over {z^2}}.
\end{eqnarray}
We consider their difference and show that it goes to zero. Noting that the fraction is a pure phase, we will write it as $e^{i \Theta}$. Passing again from $z$ to $e^{i \chi}$, and using some properties of exponentials, one finds that their difference can be casted in a form proportional to 
\begin{eqnarray}
\label{sinusoidi}
&&4 \int_0^{\infty } {{dk} \over {k}}\int_0^{2 \pi} d\chi \exp [- 2 i \chi ] \exp [- i k \sin \chi \sin \pi s_{ij}] \sin [k \sin \chi \sin \pi s_{ij}] \nonumber \\
&& \times \sin [{{\Theta }\over {2}} + \chi + n \pi \sigma ] \exp [i {{\Theta }\over {2}} - \chi - n \pi \sigma ].  
\end{eqnarray}
Let us now split the integral over $k$ into two pieces, from $0$ to $\eta$ and from $\eta$ to $\infty$: the modulus of the expression (\ref{sinusoidi}) is certainly smaller than
\begin{eqnarray}
\label{sinusoidi2}
4 \int_0^{\eta } {{dk} \over {k}} \int_0^{2 \pi} d\chi \, \,  k \, + \, 4 \int_{\eta }^{\infty} {{dk} \over {k}} \int_0^{2 \pi} d\chi \, \, |{{\Theta }\over {2}} - \chi - n \pi \sigma |.
\end{eqnarray}
Through a careful study of the function $\Theta$ (which is asymptotically close to $2 \chi + 2 n \pi \sigma$) one can show that, for sufficiently large noncommutativity parameter $\theta>{{1}\over {\eta^2}}$, both the first and the second piece are of order ${\cal{O}}(\eta )$; therefore, the whole expression goes to zero with $\eta \to 0$ ($\theta \to \infty$). 

\chapter{Proof of Eq.(\ref{21})}
We can rewrite the integrand in (\ref{20}) as a sum of two pieces:
\begin{eqnarray}
\label{lephi}
&&\int_0^1 d\nu \, {{e^{- [{\mu}^2 - k^2 \nu (1 - \nu )]}}\over {{[{\mu}^2 - k^2 \nu (1 - \nu )]}^{\omega}}} \Bigg[ \Gamma (\omega ) \, \phi (1-\omega,1-\omega,{\mu}^2 - k^2 \nu (1 - \nu ))\nonumber \\
&&+ {{\Gamma (-\omega )}\over {\Gamma (1-\omega )}} \, {[{\mu}^2 - k^2 \nu (1 - \nu )]}^{\omega} \, \phi (1,1+\omega,{\mu}^2 - k^2 \nu (1 - \nu ))\Bigg],   
\end{eqnarray}
in terms of confluent hypergeometric functions $\phi$, which are entire functions of their arguments. Taking the common factor into account, we see that the second piece is an entire function in $k^2$, while the first piece exhibits a logarithmic discontinuity: we write ${[{\mu}^2 - k^2 \nu (1 - \nu )]}^{-\omega} = \exp [-\omega \log [{\mu}^2 - k^2 \nu (1 - \nu )]$ and obtain the discontinuity as 
\begin{eqnarray}
\label{discontinuita}
\Delta \Big( \exp [-\omega \log [{\mu}^2 - k^2 \nu (1 - \nu )] \Big) &\propto&  - 2 i \sin (\pi \omega ) \, {[- {\mu}^2 + k^2 \nu (1 - \nu )]}^{-\omega} \nonumber \\
&& \, \times \, \Theta (- {\mu}^2 + k^2 \nu (1 - \nu )).
\end{eqnarray}
Inserting (\ref{discontinuita}) in the integral in (\ref{20}), considering that $\phi (d,d,x) = e^x$, and that $\sin (\pi \omega ) = {{\pi}\over {\Gamma (\omega ) \Gamma (1-\omega )}}$, the discontinuity of the integral turns out to be proportional to
\begin{eqnarray}
\label{intecsi}
- 2 i \int_0^1 {{\pi}\over {\Gamma (1-\omega )}} \, {[- {\mu}^2 + k^2 \nu (1 - \nu )]}^{-\omega} \, \Theta (- {\mu}^2 + k^2 \nu (1 - \nu )).
\end{eqnarray}
One uses now the standard trick of defining a new variable such that $\nu = {{1+\lambda}\over {2}}$, $1-\nu ={{1-\lambda}\over {2}}$; then the integral is rewritten as
\begin{eqnarray}
\label{trick}
- 2 \pi i \int_0^{\sqrt{1- {{4 \mu^2 }\over {k^2}}}} d\lambda \, {{1}\over {\Gamma (1-\omega )}}\, {\Big( {{k^2}\over {4}}\Big) }^{-\omega} \, {\Big[ 1-{{4 \mu^2 }\over {k^2}} - \lambda^2 \Big] }^{-\omega}. 
\end{eqnarray}
The upper limit comes from the $\Theta$ function which was present. After a straightforward computation equation (\ref{21}) is recovered.

\end{document}

%% file: colors.tex
\definecolor{snow}{rgb}{1.00,0.98,0.98}
\definecolor{ghostwhite}{rgb}{0.97,0.97,1.00}
\definecolor{GhostWhite}{rgb}{0.97,0.97,1.00}
\definecolor{whitesmoke}{rgb}{0.96,0.96,0.96}
\definecolor{WhiteSmoke}{rgb}{0.96,0.96,0.96}
\definecolor{gainsboro}{rgb}{0.86,0.86,0.86}
\definecolor{floralwhite}{rgb}{1.00,0.98,0.94}
\definecolor{FloralWhite}{rgb}{1.00,0.98,0.94}
\definecolor{oldlace}{rgb}{0.99,0.96,0.90}
\definecolor{OldLace}{rgb}{0.99,0.96,0.90}
\definecolor{linen}{rgb}{0.98,0.94,0.90}
\definecolor{antiquewhite}{rgb}{0.98,0.92,0.84}
\definecolor{AntiqueWhite}{rgb}{0.98,0.92,0.84}
\definecolor{papayawhip}{rgb}{1.00,0.93,0.83}
\definecolor{PapayaWhip}{rgb}{1.00,0.93,0.83}
\definecolor{blanchedalmond}{rgb}{1.00,0.92,0.80}
\definecolor{BlanchedAlmond}{rgb}{1.00,0.92,0.80}
\definecolor{bisque}{rgb}{1.00,0.89,0.77}
\definecolor{peachpuff}{rgb}{1.00,0.85,0.72}
\definecolor{PeachPuff}{rgb}{1.00,0.85,0.72}
\definecolor{navajowhite}{rgb}{1.00,0.87,0.68}
\definecolor{NavajoWhite}{rgb}{1.00,0.87,0.68}
\definecolor{moccasin}{rgb}{1.00,0.89,0.71}
\definecolor{cornsilk}{rgb}{1.00,0.97,0.86}
\definecolor{ivory}{rgb}{1.00,1.00,0.94}
\definecolor{lemonchiffon}{rgb}{1.00,0.98,0.80}
\definecolor{LemonChiffon}{rgb}{1.00,0.98,0.80}
\definecolor{seashell}{rgb}{1.00,0.96,0.93}
\definecolor{honeydew}{rgb}{0.94,1.00,0.94}
\definecolor{mintcream}{rgb}{0.96,1.00,0.98}
\definecolor{MintCream}{rgb}{0.96,1.00,0.98}
\definecolor{azure}{rgb}{0.94,1.00,1.00}
\definecolor{aliceblue}{rgb}{0.94,0.97,1.00}
\definecolor{AliceBlue}{rgb}{0.94,0.97,1.00}
\definecolor{lavender}{rgb}{0.90,0.90,0.98}
\definecolor{lavenderblush}{rgb}{1.00,0.94,0.96}
\definecolor{LavenderBlush}{rgb}{1.00,0.94,0.96}
\definecolor{mistyrose}{rgb}{1.00,0.89,0.88}
\definecolor{MistyRose}{rgb}{1.00,0.89,0.88}
\definecolor{white}{rgb}{1.00,1.00,1.00}
\definecolor{black}{rgb}{0.00,0.00,0.00}
\definecolor{darkslategray}{rgb}{0.18,0.31,0.31}
\definecolor{DarkSlateGray}{rgb}{0.18,0.31,0.31}
\definecolor{darkslategrey}{rgb}{0.18,0.31,0.31}
\definecolor{DarkSlateGrey}{rgb}{0.18,0.31,0.31}
\definecolor{dimgray}{rgb}{0.41,0.41,0.41}
\definecolor{DimGray}{rgb}{0.41,0.41,0.41}
\definecolor{dimgrey}{rgb}{0.41,0.41,0.41}
\definecolor{DimGrey}{rgb}{0.41,0.41,0.41}
\definecolor{slategray}{rgb}{0.44,0.50,0.56}
\definecolor{SlateGray}{rgb}{0.44,0.50,0.56}
\definecolor{slategrey}{rgb}{0.44,0.50,0.56}
\definecolor{SlateGrey}{rgb}{0.44,0.50,0.56}
\definecolor{lightslategray}{rgb}{0.46,0.53,0.60}
\definecolor{LightSlateGray}{rgb}{0.46,0.53,0.60}
\definecolor{lightslategrey}{rgb}{0.46,0.53,0.60}
\definecolor{LightSlateGrey}{rgb}{0.46,0.53,0.60}
\definecolor{gray}{rgb}{0.74,0.74,0.74}
\definecolor{grey}{rgb}{0.74,0.74,0.74}
\definecolor{lightgrey}{rgb}{0.82,0.82,0.82}
\definecolor{LightGrey}{rgb}{0.82,0.82,0.82}
\definecolor{lightgray}{rgb}{0.82,0.82,0.82}
\definecolor{LightGray}{rgb}{0.82,0.82,0.82}
\definecolor{midnightblue}{rgb}{0.10,0.10,0.44}
\definecolor{MidnightBlue}{rgb}{0.10,0.10,0.44}
\definecolor{navy}{rgb}{0.00,0.00,0.50}
\definecolor{navyblue}{rgb}{0.00,0.00,0.50}
\definecolor{NavyBlue}{rgb}{0.00,0.00,0.50}
\definecolor{cornflowerblue}{rgb}{0.39,0.58,0.93}
\definecolor{CornflowerBlue}{rgb}{0.39,0.58,0.93}
\definecolor{darkslateblue}{rgb}{0.28,0.24,0.54}
\definecolor{DarkSlateBlue}{rgb}{0.28,0.24,0.54}
\definecolor{slateblue}{rgb}{0.41,0.35,0.80}
\definecolor{SlateBlue}{rgb}{0.41,0.35,0.80}
\definecolor{mediumslateblue}{rgb}{0.48,0.41,0.93}
\definecolor{MediumSlateBlue}{rgb}{0.48,0.41,0.93}
\definecolor{lightslateblue}{rgb}{0.52,0.44,1.00}
\definecolor{LightSlateBlue}{rgb}{0.52,0.44,1.00}
\definecolor{mediumblue}{rgb}{0.00,0.00,0.80}
\definecolor{MediumBlue}{rgb}{0.00,0.00,0.80}
\definecolor{royalblue}{rgb}{0.25,0.41,0.88}
\definecolor{RoyalBlue}{rgb}{0.25,0.41,0.88}
\definecolor{blue}{rgb}{0.00,0.00,1.00}
\definecolor{dodgerblue}{rgb}{0.12,0.56,1.00}
\definecolor{DodgerBlue}{rgb}{0.12,0.56,1.00}
\definecolor{deepskyblue}{rgb}{0.00,0.75,1.00}
\definecolor{DeepSkyBlue}{rgb}{0.00,0.75,1.00}
\definecolor{skyblue}{rgb}{0.53,0.80,0.92}
\definecolor{SkyBlue}{rgb}{0.53,0.80,0.92}
\definecolor{lightskyblue}{rgb}{0.53,0.80,0.98}
\definecolor{LightSkyBlue}{rgb}{0.53,0.80,0.98}
\definecolor{steelblue}{rgb}{0.27,0.51,0.70}
\definecolor{SteelBlue}{rgb}{0.27,0.51,0.70}
\definecolor{lightsteelblue}{rgb}{0.69,0.77,0.87}
\definecolor{LightSteelBlue}{rgb}{0.69,0.77,0.87}
\definecolor{lightblue}{rgb}{0.68,0.84,0.90}
\definecolor{LightBlue}{rgb}{0.68,0.84,0.90}
\definecolor{powderblue}{rgb}{0.69,0.88,0.90}
\definecolor{PowderBlue}{rgb}{0.69,0.88,0.90}
\definecolor{paleturquoise}{rgb}{0.68,0.93,0.93}
\definecolor{PaleTurquoise}{rgb}{0.68,0.93,0.93}
\definecolor{darkturquoise}{rgb}{0.00,0.80,0.82}
\definecolor{DarkTurquoise}{rgb}{0.00,0.80,0.82}
\definecolor{mediumturquoise}{rgb}{0.28,0.82,0.80}
\definecolor{MediumTurquoise}{rgb}{0.28,0.82,0.80}
\definecolor{turquoise}{rgb}{0.25,0.88,0.81}
\definecolor{cyan}{rgb}{0.00,1.00,1.00}
\definecolor{lightcyan}{rgb}{0.88,1.00,1.00}
\definecolor{LightCyan}{rgb}{0.88,1.00,1.00}
\definecolor{cadetblue}{rgb}{0.37,0.62,0.63}
\definecolor{CadetBlue}{rgb}{0.37,0.62,0.63}
\definecolor{mediumaquamarine}{rgb}{0.40,0.80,0.66}
\definecolor{MediumAquamarine}{rgb}{0.40,0.80,0.66}
\definecolor{aquamarine}{rgb}{0.50,1.00,0.83}
\definecolor{darkgreen}{rgb}{0.00,0.39,0.00}
\definecolor{DarkGreen}{rgb}{0.00,0.39,0.00}
\definecolor{darkolivegreen}{rgb}{0.33,0.42,0.18}
\definecolor{DarkOliveGreen}{rgb}{0.33,0.42,0.18}
\definecolor{darkseagreen}{rgb}{0.56,0.73,0.56}
\definecolor{DarkSeaGreen}{rgb}{0.56,0.73,0.56}
\definecolor{seagreen}{rgb}{0.18,0.54,0.34}
\definecolor{SeaGreen}{rgb}{0.18,0.54,0.34}
\definecolor{mediumseagreen}{rgb}{0.23,0.70,0.44}
\definecolor{MediumSeaGreen}{rgb}{0.23,0.70,0.44}
\definecolor{lightseagreen}{rgb}{0.13,0.70,0.66}
\definecolor{LightSeaGreen}{rgb}{0.13,0.70,0.66}
\definecolor{palegreen}{rgb}{0.59,0.98,0.59}
\definecolor{PaleGreen}{rgb}{0.59,0.98,0.59}
\definecolor{springgreen}{rgb}{0.00,1.00,0.50}
\definecolor{SpringGreen}{rgb}{0.00,1.00,0.50}
\definecolor{lawngreen}{rgb}{0.48,0.98,0.00}
\definecolor{LawnGreen}{rgb}{0.48,0.98,0.00}
\definecolor{green}{rgb}{0.00,1.00,0.00}
\definecolor{chartreuse}{rgb}{0.50,1.00,0.00}
\definecolor{mediumspringgreen}{rgb}{0.00,0.98,0.60}
\definecolor{MediumSpringGreen}{rgb}{0.00,0.98,0.60}
\definecolor{greenyellow}{rgb}{0.68,1.00,0.18}
\definecolor{GreenYellow}{rgb}{0.68,1.00,0.18}
\definecolor{limegreen}{rgb}{0.20,0.80,0.20}
\definecolor{LimeGreen}{rgb}{0.20,0.80,0.20}
\definecolor{yellowgreen}{rgb}{0.60,0.80,0.20}
\definecolor{YellowGreen}{rgb}{0.60,0.80,0.20}
\definecolor{forestgreen}{rgb}{0.13,0.54,0.13}
\definecolor{ForestGreen}{rgb}{0.13,0.54,0.13}
\definecolor{olivedrab}{rgb}{0.42,0.55,0.14}
\definecolor{OliveDrab}{rgb}{0.42,0.55,0.14}
\definecolor{darkkhaki}{rgb}{0.74,0.71,0.42}
\definecolor{DarkKhaki}{rgb}{0.74,0.71,0.42}
\definecolor{khaki}{rgb}{0.94,0.90,0.55}
\definecolor{palegoldenrod}{rgb}{0.93,0.91,0.66}
\definecolor{PaleGoldenrod}{rgb}{0.93,0.91,0.66}
\definecolor{lightgoldenrodyellow}{rgb}{0.98,0.98,0.82}
\definecolor{LightGoldenrodYellow}{rgb}{0.98,0.98,0.82}
\definecolor{lightyellow}{rgb}{1.00,1.00,0.88}
\definecolor{LightYellow}{rgb}{1.00,1.00,0.88}
\definecolor{yellow}{rgb}{1.00,1.00,0.00}
\definecolor{gold}{rgb}{1.00,0.84,0.00}
\definecolor{lightgoldenrod}{rgb}{0.93,0.86,0.51}
\definecolor{LightGoldenrod}{rgb}{0.93,0.86,0.51}
\definecolor{goldenrod}{rgb}{0.85,0.64,0.13}
\definecolor{darkgoldenrod}{rgb}{0.72,0.52,0.04}
\definecolor{DarkGoldenrod}{rgb}{0.72,0.52,0.04}
\definecolor{rosybrown}{rgb}{0.73,0.56,0.56}
\definecolor{RosyBrown}{rgb}{0.73,0.56,0.56}
\definecolor{indianred}{rgb}{0.80,0.36,0.36}
\definecolor{IndianRed}{rgb}{0.80,0.36,0.36}
\definecolor{saddlebrown}{rgb}{0.54,0.27,0.07}
\definecolor{SaddleBrown}{rgb}{0.54,0.27,0.07}
\definecolor{sienna}{rgb}{0.63,0.32,0.18}
\definecolor{peru}{rgb}{0.80,0.52,0.25}
\definecolor{burlywood}{rgb}{0.87,0.72,0.53}
\definecolor{beige}{rgb}{0.96,0.96,0.86}
\definecolor{wheat}{rgb}{0.96,0.87,0.70}
\definecolor{sandybrown}{rgb}{0.95,0.64,0.38}
\definecolor{SandyBrown}{rgb}{0.95,0.64,0.38}
\definecolor{tan}{rgb}{0.82,0.70,0.55}
\definecolor{chocolate}{rgb}{0.82,0.41,0.12}
\definecolor{firebrick}{rgb}{0.70,0.13,0.13}
\definecolor{brown}{rgb}{0.64,0.16,0.16}
\definecolor{darksalmon}{rgb}{0.91,0.59,0.48}
\definecolor{DarkSalmon}{rgb}{0.91,0.59,0.48}
\definecolor{salmon}{rgb}{0.98,0.50,0.45}
\definecolor{lightsalmon}{rgb}{1.00,0.63,0.48}
\definecolor{LightSalmon}{rgb}{1.00,0.63,0.48}
\definecolor{orange}{rgb}{1.00,0.64,0.00}
\definecolor{darkorange}{rgb}{1.00,0.55,0.00}
\definecolor{DarkOrange}{rgb}{1.00,0.55,0.00}
\definecolor{coral}{rgb}{1.00,0.50,0.31}
\definecolor{lightcoral}{rgb}{0.94,0.50,0.50}
\definecolor{LightCoral}{rgb}{0.94,0.50,0.50}
\definecolor{tomato}{rgb}{1.00,0.39,0.28}
\definecolor{orangered}{rgb}{1.00,0.27,0.00}
\definecolor{OrangeRed}{rgb}{1.00,0.27,0.00}
\definecolor{red}{rgb}{1.00,0.00,0.00}
\definecolor{hotpink}{rgb}{1.00,0.41,0.70}
\definecolor{HotPink}{rgb}{1.00,0.41,0.70}
\definecolor{deeppink}{rgb}{1.00,0.08,0.57}
\definecolor{DeepPink}{rgb}{1.00,0.08,0.57}
\definecolor{pink}{rgb}{1.00,0.75,0.79}
\definecolor{lightpink}{rgb}{1.00,0.71,0.75}
\definecolor{LightPink}{rgb}{1.00,0.71,0.75}
\definecolor{palevioletred}{rgb}{0.86,0.44,0.57}
\definecolor{PaleVioletRed}{rgb}{0.86,0.44,0.57}
\definecolor{maroon}{rgb}{0.69,0.19,0.38}
\definecolor{mediumvioletred}{rgb}{0.78,0.08,0.52}
\definecolor{MediumVioletRed}{rgb}{0.78,0.08,0.52}
\definecolor{violetred}{rgb}{0.81,0.13,0.56}
\definecolor{VioletRed}{rgb}{0.81,0.13,0.56}
\definecolor{magenta}{rgb}{1.00,0.00,1.00}
\definecolor{violet}{rgb}{0.93,0.51,0.93}
\definecolor{plum}{rgb}{0.86,0.63,0.86}
\definecolor{orchid}{rgb}{0.85,0.44,0.84}
\definecolor{mediumorchid}{rgb}{0.73,0.33,0.82}
\definecolor{MediumOrchid}{rgb}{0.73,0.33,0.82}
\definecolor{darkorchid}{rgb}{0.60,0.20,0.80}
\definecolor{DarkOrchid}{rgb}{0.60,0.20,0.80}
\definecolor{darkviolet}{rgb}{0.58,0.00,0.82}
\definecolor{DarkViolet}{rgb}{0.58,0.00,0.82}
\definecolor{blueviolet}{rgb}{0.54,0.17,0.88}
\definecolor{BlueViolet}{rgb}{0.54,0.17,0.88}
\definecolor{purple}{rgb}{0.63,0.13,0.94}
\definecolor{mediumpurple}{rgb}{0.57,0.44,0.86}
\definecolor{MediumPurple}{rgb}{0.57,0.44,0.86}
\definecolor{thistle}{rgb}{0.84,0.75,0.84}
\definecolor{snow1}{rgb}{1.00,0.98,0.98}
\definecolor{snow2}{rgb}{0.93,0.91,0.91}
\definecolor{snow3}{rgb}{0.80,0.79,0.79}
\definecolor{snow4}{rgb}{0.54,0.54,0.54}
\definecolor{seashell1}{rgb}{1.00,0.96,0.93}
\definecolor{seashell2}{rgb}{0.93,0.89,0.87}
\definecolor{seashell3}{rgb}{0.80,0.77,0.75}
\definecolor{seashell4}{rgb}{0.54,0.52,0.51}
\definecolor{AntiqueWhite1}{rgb}{1.00,0.93,0.86}
\definecolor{AntiqueWhite2}{rgb}{0.93,0.87,0.80}
\definecolor{AntiqueWhite3}{rgb}{0.80,0.75,0.69}
\definecolor{AntiqueWhite4}{rgb}{0.54,0.51,0.47}
\definecolor{bisque1}{rgb}{1.00,0.89,0.77}
\definecolor{bisque2}{rgb}{0.93,0.83,0.71}
\definecolor{bisque3}{rgb}{0.80,0.71,0.62}
\definecolor{bisque4}{rgb}{0.54,0.49,0.42}
\definecolor{PeachPuff1}{rgb}{1.00,0.85,0.72}
\definecolor{PeachPuff2}{rgb}{0.93,0.79,0.68}
\definecolor{PeachPuff3}{rgb}{0.80,0.68,0.58}
\definecolor{PeachPuff4}{rgb}{0.54,0.46,0.39}
\definecolor{NavajoWhite1}{rgb}{1.00,0.87,0.68}
\definecolor{NavajoWhite2}{rgb}{0.93,0.81,0.63}
\definecolor{NavajoWhite3}{rgb}{0.80,0.70,0.54}
\definecolor{NavajoWhite4}{rgb}{0.54,0.47,0.37}
\definecolor{LemonChiffon1}{rgb}{1.00,0.98,0.80}
\definecolor{LemonChiffon2}{rgb}{0.93,0.91,0.75}
\definecolor{LemonChiffon3}{rgb}{0.80,0.79,0.64}
\definecolor{LemonChiffon4}{rgb}{0.54,0.54,0.44}
\definecolor{cornsilk1}{rgb}{1.00,0.97,0.86}
\definecolor{cornsilk2}{rgb}{0.93,0.91,0.80}
\definecolor{cornsilk3}{rgb}{0.80,0.78,0.69}
\definecolor{cornsilk4}{rgb}{0.54,0.53,0.47}
\definecolor{ivory1}{rgb}{1.00,1.00,0.94}
\definecolor{ivory2}{rgb}{0.93,0.93,0.88}
\definecolor{ivory3}{rgb}{0.80,0.80,0.75}
\definecolor{ivory4}{rgb}{0.54,0.54,0.51}
\definecolor{honeydew1}{rgb}{0.94,1.00,0.94}
\definecolor{honeydew2}{rgb}{0.88,0.93,0.88}
\definecolor{honeydew3}{rgb}{0.75,0.80,0.75}
\definecolor{honeydew4}{rgb}{0.51,0.54,0.51}
\definecolor{LavenderBlush1}{rgb}{1.00,0.94,0.96}
\definecolor{LavenderBlush2}{rgb}{0.93,0.88,0.89}
\definecolor{LavenderBlush3}{rgb}{0.80,0.75,0.77}
\definecolor{LavenderBlush4}{rgb}{0.54,0.51,0.52}
\definecolor{MistyRose1}{rgb}{1.00,0.89,0.88}
\definecolor{MistyRose2}{rgb}{0.93,0.83,0.82}
\definecolor{MistyRose3}{rgb}{0.80,0.71,0.71}
\definecolor{MistyRose4}{rgb}{0.54,0.49,0.48}
\definecolor{azure1}{rgb}{0.94,1.00,1.00}
\definecolor{azure2}{rgb}{0.88,0.93,0.93}
\definecolor{azure3}{rgb}{0.75,0.80,0.80}
\definecolor{azure4}{rgb}{0.51,0.54,0.54}
\definecolor{SlateBlue1}{rgb}{0.51,0.43,1.00}
\definecolor{SlateBlue2}{rgb}{0.48,0.40,0.93}
\definecolor{SlateBlue3}{rgb}{0.41,0.35,0.80}
\definecolor{SlateBlue4}{rgb}{0.28,0.23,0.54}
\definecolor{RoyalBlue1}{rgb}{0.28,0.46,1.00}
\definecolor{RoyalBlue2}{rgb}{0.26,0.43,0.93}
\definecolor{RoyalBlue3}{rgb}{0.23,0.37,0.80}
\definecolor{RoyalBlue4}{rgb}{0.15,0.25,0.54}
\definecolor{blue1}{rgb}{0.00,0.00,1.00}
\definecolor{blue2}{rgb}{0.00,0.00,0.93}
\definecolor{blue3}{rgb}{0.00,0.00,0.80}
\definecolor{blue4}{rgb}{0.00,0.00,0.54}
\definecolor{DodgerBlue1}{rgb}{0.12,0.56,1.00}
\definecolor{DodgerBlue2}{rgb}{0.11,0.52,0.93}
\definecolor{DodgerBlue3}{rgb}{0.09,0.45,0.80}
\definecolor{DodgerBlue4}{rgb}{0.06,0.30,0.54}
\definecolor{SteelBlue1}{rgb}{0.39,0.72,1.00}
\definecolor{SteelBlue2}{rgb}{0.36,0.67,0.93}
\definecolor{SteelBlue3}{rgb}{0.31,0.58,0.80}
\definecolor{SteelBlue4}{rgb}{0.21,0.39,0.54}
\definecolor{DeepSkyBlue1}{rgb}{0.00,0.75,1.00}
\definecolor{DeepSkyBlue2}{rgb}{0.00,0.70,0.93}
\definecolor{DeepSkyBlue3}{rgb}{0.00,0.60,0.80}
\definecolor{DeepSkyBlue4}{rgb}{0.00,0.41,0.54}
\definecolor{SkyBlue1}{rgb}{0.53,0.80,1.00}
\definecolor{SkyBlue2}{rgb}{0.49,0.75,0.93}
\definecolor{SkyBlue3}{rgb}{0.42,0.65,0.80}
\definecolor{SkyBlue4}{rgb}{0.29,0.44,0.54}
\definecolor{LightSkyBlue1}{rgb}{0.69,0.88,1.00}
\definecolor{LightSkyBlue2}{rgb}{0.64,0.82,0.93}
\definecolor{LightSkyBlue3}{rgb}{0.55,0.71,0.80}
\definecolor{LightSkyBlue4}{rgb}{0.38,0.48,0.54}
\definecolor{SlateGray1}{rgb}{0.77,0.88,1.00}
\definecolor{SlateGray2}{rgb}{0.72,0.82,0.93}
\definecolor{SlateGray3}{rgb}{0.62,0.71,0.80}
\definecolor{SlateGray4}{rgb}{0.42,0.48,0.54}
\definecolor{LightSteelBlue1}{rgb}{0.79,0.88,1.00}
\definecolor{LightSteelBlue2}{rgb}{0.73,0.82,0.93}
\definecolor{LightSteelBlue3}{rgb}{0.63,0.71,0.80}
\definecolor{LightSteelBlue4}{rgb}{0.43,0.48,0.54}
\definecolor{LightBlue1}{rgb}{0.75,0.93,1.00}
\definecolor{LightBlue2}{rgb}{0.70,0.87,0.93}
\definecolor{LightBlue3}{rgb}{0.60,0.75,0.80}
\definecolor{LightBlue4}{rgb}{0.41,0.51,0.54}
\definecolor{LightCyan1}{rgb}{0.88,1.00,1.00}
\definecolor{LightCyan2}{rgb}{0.82,0.93,0.93}
\definecolor{LightCyan3}{rgb}{0.70,0.80,0.80}
\definecolor{LightCyan4}{rgb}{0.48,0.54,0.54}
\definecolor{PaleTurquoise1}{rgb}{0.73,1.00,1.00}
\definecolor{PaleTurquoise2}{rgb}{0.68,0.93,0.93}
\definecolor{PaleTurquoise3}{rgb}{0.59,0.80,0.80}
\definecolor{PaleTurquoise4}{rgb}{0.40,0.54,0.54}
\definecolor{CadetBlue1}{rgb}{0.59,0.96,1.00}
\definecolor{CadetBlue2}{rgb}{0.55,0.89,0.93}
\definecolor{CadetBlue3}{rgb}{0.48,0.77,0.80}
\definecolor{CadetBlue4}{rgb}{0.32,0.52,0.54}
\definecolor{turquoise1}{rgb}{0.00,0.96,1.00}
\definecolor{turquoise2}{rgb}{0.00,0.89,0.93}
\definecolor{turquoise3}{rgb}{0.00,0.77,0.80}
\definecolor{turquoise4}{rgb}{0.00,0.52,0.54}
\definecolor{cyan1}{rgb}{0.00,1.00,1.00}
\definecolor{cyan2}{rgb}{0.00,0.93,0.93}
\definecolor{cyan3}{rgb}{0.00,0.80,0.80}
\definecolor{cyan4}{rgb}{0.00,0.54,0.54}
\definecolor{DarkSlateGray1}{rgb}{0.59,1.00,1.00}
\definecolor{DarkSlateGray2}{rgb}{0.55,0.93,0.93}
\definecolor{DarkSlateGray3}{rgb}{0.47,0.80,0.80}
\definecolor{DarkSlateGray4}{rgb}{0.32,0.54,0.54}
\definecolor{aquamarine1}{rgb}{0.50,1.00,0.83}
\definecolor{aquamarine2}{rgb}{0.46,0.93,0.77}
\definecolor{aquamarine3}{rgb}{0.40,0.80,0.66}
\definecolor{aquamarine4}{rgb}{0.27,0.54,0.45}
\definecolor{DarkSeaGreen1}{rgb}{0.75,1.00,0.75}
\definecolor{DarkSeaGreen2}{rgb}{0.70,0.93,0.70}
\definecolor{DarkSeaGreen3}{rgb}{0.61,0.80,0.61}
\definecolor{DarkSeaGreen4}{rgb}{0.41,0.54,0.41}
\definecolor{SeaGreen1}{rgb}{0.33,1.00,0.62}
\definecolor{SeaGreen2}{rgb}{0.30,0.93,0.58}
\definecolor{SeaGreen3}{rgb}{0.26,0.80,0.50}
\definecolor{SeaGreen4}{rgb}{0.18,0.54,0.34}
\definecolor{PaleGreen1}{rgb}{0.60,1.00,0.60}
\definecolor{PaleGreen2}{rgb}{0.56,0.93,0.56}
\definecolor{PaleGreen3}{rgb}{0.48,0.80,0.48}
\definecolor{PaleGreen4}{rgb}{0.33,0.54,0.33}
\definecolor{SpringGreen1}{rgb}{0.00,1.00,0.50}
\definecolor{SpringGreen2}{rgb}{0.00,0.93,0.46}
\definecolor{SpringGreen3}{rgb}{0.00,0.80,0.40}
\definecolor{SpringGreen4}{rgb}{0.00,0.54,0.27}
\definecolor{green1}{rgb}{0.00,1.00,0.00}
\definecolor{green2}{rgb}{0.00,0.93,0.00}
\definecolor{green3}{rgb}{0.00,0.80,0.00}
\definecolor{green4}{rgb}{0.00,0.54,0.00}
\definecolor{chartreuse1}{rgb}{0.50,1.00,0.00}
\definecolor{chartreuse2}{rgb}{0.46,0.93,0.00}
\definecolor{chartreuse3}{rgb}{0.40,0.80,0.00}
\definecolor{chartreuse4}{rgb}{0.27,0.54,0.00}
\definecolor{OliveDrab1}{rgb}{0.75,1.00,0.24}
\definecolor{OliveDrab2}{rgb}{0.70,0.93,0.23}
\definecolor{OliveDrab3}{rgb}{0.60,0.80,0.20}
\definecolor{OliveDrab4}{rgb}{0.41,0.54,0.13}
\definecolor{DarkOliveGreen1}{rgb}{0.79,1.00,0.44}
\definecolor{DarkOliveGreen2}{rgb}{0.73,0.93,0.41}
\definecolor{DarkOliveGreen3}{rgb}{0.63,0.80,0.35}
\definecolor{DarkOliveGreen4}{rgb}{0.43,0.54,0.24}
\definecolor{khaki1}{rgb}{1.00,0.96,0.56}
\definecolor{khaki2}{rgb}{0.93,0.90,0.52}
\definecolor{khaki3}{rgb}{0.80,0.77,0.45}
\definecolor{khaki4}{rgb}{0.54,0.52,0.30}
\definecolor{LightGoldenrod1}{rgb}{1.00,0.92,0.54}
\definecolor{LightGoldenrod2}{rgb}{0.93,0.86,0.51}
\definecolor{LightGoldenrod3}{rgb}{0.80,0.74,0.44}
\definecolor{LightGoldenrod4}{rgb}{0.54,0.50,0.30}
\definecolor{LightYellow1}{rgb}{1.00,1.00,0.88}
\definecolor{LightYellow2}{rgb}{0.93,0.93,0.82}
\definecolor{LightYellow3}{rgb}{0.80,0.80,0.70}
\definecolor{LightYellow4}{rgb}{0.54,0.54,0.48}
\definecolor{yellow1}{rgb}{1.00,1.00,0.00}
\definecolor{yellow2}{rgb}{0.93,0.93,0.00}
\definecolor{yellow3}{rgb}{0.80,0.80,0.00}
\definecolor{yellow4}{rgb}{0.54,0.54,0.00}
\definecolor{gold1}{rgb}{1.00,0.84,0.00}
\definecolor{gold2}{rgb}{0.93,0.79,0.00}
\definecolor{gold3}{rgb}{0.80,0.68,0.00}
\definecolor{gold4}{rgb}{0.54,0.46,0.00}
\definecolor{goldenrod1}{rgb}{1.00,0.75,0.14}
\definecolor{goldenrod2}{rgb}{0.93,0.70,0.13}
\definecolor{goldenrod3}{rgb}{0.80,0.61,0.11}
\definecolor{goldenrod4}{rgb}{0.54,0.41,0.08}
\definecolor{DarkGoldenrod1}{rgb}{1.00,0.72,0.06}
\definecolor{DarkGoldenrod2}{rgb}{0.93,0.68,0.05}
\definecolor{DarkGoldenrod3}{rgb}{0.80,0.58,0.05}
\definecolor{DarkGoldenrod4}{rgb}{0.54,0.39,0.03}
\definecolor{RosyBrown1}{rgb}{1.00,0.75,0.75}
\definecolor{RosyBrown2}{rgb}{0.93,0.70,0.70}
\definecolor{RosyBrown3}{rgb}{0.80,0.61,0.61}
\definecolor{RosyBrown4}{rgb}{0.54,0.41,0.41}
\definecolor{IndianRed1}{rgb}{1.00,0.41,0.41}
\definecolor{IndianRed2}{rgb}{0.93,0.39,0.39}
\definecolor{IndianRed3}{rgb}{0.80,0.33,0.33}
\definecolor{IndianRed4}{rgb}{0.54,0.23,0.23}
\definecolor{sienna1}{rgb}{1.00,0.51,0.28}
\definecolor{sienna2}{rgb}{0.93,0.47,0.26}
\definecolor{sienna3}{rgb}{0.80,0.41,0.22}
\definecolor{sienna4}{rgb}{0.54,0.28,0.15}
\definecolor{burlywood1}{rgb}{1.00,0.82,0.61}
\definecolor{burlywood2}{rgb}{0.93,0.77,0.57}
\definecolor{burlywood3}{rgb}{0.80,0.66,0.49}
\definecolor{burlywood4}{rgb}{0.54,0.45,0.33}
\definecolor{wheat1}{rgb}{1.00,0.90,0.73}
\definecolor{wheat2}{rgb}{0.93,0.84,0.68}
\definecolor{wheat3}{rgb}{0.80,0.73,0.59}
\definecolor{wheat4}{rgb}{0.54,0.49,0.40}
\definecolor{tan1}{rgb}{1.00,0.64,0.31}
\definecolor{tan2}{rgb}{0.93,0.60,0.29}
\definecolor{tan3}{rgb}{0.80,0.52,0.25}
\definecolor{tan4}{rgb}{0.54,0.35,0.17}
\definecolor{chocolate1}{rgb}{1.00,0.50,0.14}
\definecolor{chocolate2}{rgb}{0.93,0.46,0.13}
\definecolor{chocolate3}{rgb}{0.80,0.40,0.11}
\definecolor{chocolate4}{rgb}{0.54,0.27,0.07}
\definecolor{firebrick1}{rgb}{1.00,0.19,0.19}
\definecolor{firebrick2}{rgb}{0.93,0.17,0.17}
\definecolor{firebrick3}{rgb}{0.80,0.15,0.15}
\definecolor{firebrick4}{rgb}{0.54,0.10,0.10}
\definecolor{brown1}{rgb}{1.00,0.25,0.25}
\definecolor{brown2}{rgb}{0.93,0.23,0.23}
\definecolor{brown3}{rgb}{0.80,0.20,0.20}
\definecolor{brown4}{rgb}{0.54,0.14,0.14}
\definecolor{salmon1}{rgb}{1.00,0.55,0.41}
\definecolor{salmon2}{rgb}{0.93,0.51,0.38}
\definecolor{salmon3}{rgb}{0.80,0.44,0.33}
\definecolor{salmon4}{rgb}{0.54,0.30,0.22}
\definecolor{LightSalmon1}{rgb}{1.00,0.63,0.48}
\definecolor{LightSalmon2}{rgb}{0.93,0.58,0.45}
\definecolor{LightSalmon3}{rgb}{0.80,0.50,0.38}
\definecolor{LightSalmon4}{rgb}{0.54,0.34,0.26}
\definecolor{orange1}{rgb}{1.00,0.64,0.00}
\definecolor{orange2}{rgb}{0.93,0.60,0.00}
\definecolor{orange3}{rgb}{0.80,0.52,0.00}
\definecolor{orange4}{rgb}{0.54,0.35,0.00}
\definecolor{DarkOrange1}{rgb}{1.00,0.50,0.00}
\definecolor{DarkOrange2}{rgb}{0.93,0.46,0.00}
\definecolor{DarkOrange3}{rgb}{0.80,0.40,0.00}
\definecolor{DarkOrange4}{rgb}{0.54,0.27,0.00}
\definecolor{coral1}{rgb}{1.00,0.45,0.34}
\definecolor{coral2}{rgb}{0.93,0.41,0.31}
\definecolor{coral3}{rgb}{0.80,0.36,0.27}
\definecolor{coral4}{rgb}{0.54,0.24,0.18}
\definecolor{tomato1}{rgb}{1.00,0.39,0.28}
\definecolor{tomato2}{rgb}{0.93,0.36,0.26}
\definecolor{tomato3}{rgb}{0.80,0.31,0.22}
\definecolor{tomato4}{rgb}{0.54,0.21,0.15}
\definecolor{OrangeRed1}{rgb}{1.00,0.27,0.00}
\definecolor{OrangeRed2}{rgb}{0.93,0.25,0.00}
\definecolor{OrangeRed3}{rgb}{0.80,0.21,0.00}
\definecolor{OrangeRed4}{rgb}{0.54,0.14,0.00}
\definecolor{red1}{rgb}{1.00,0.00,0.00}
\definecolor{red2}{rgb}{0.93,0.00,0.00}
\definecolor{red3}{rgb}{0.80,0.00,0.00}
\definecolor{red4}{rgb}{0.54,0.00,0.00}
\definecolor{DeepPink1}{rgb}{1.00,0.08,0.57}
\definecolor{DeepPink2}{rgb}{0.93,0.07,0.54}
\definecolor{DeepPink3}{rgb}{0.80,0.06,0.46}
\definecolor{DeepPink4}{rgb}{0.54,0.04,0.31}
\definecolor{HotPink1}{rgb}{1.00,0.43,0.70}
\definecolor{HotPink2}{rgb}{0.93,0.41,0.65}
\definecolor{HotPink3}{rgb}{0.80,0.38,0.56}
\definecolor{HotPink4}{rgb}{0.54,0.23,0.38}
\definecolor{pink1}{rgb}{1.00,0.71,0.77}
\definecolor{pink2}{rgb}{0.93,0.66,0.72}
\definecolor{pink3}{rgb}{0.80,0.57,0.62}
\definecolor{pink4}{rgb}{0.54,0.39,0.42}
\definecolor{LightPink1}{rgb}{1.00,0.68,0.72}
\definecolor{LightPink2}{rgb}{0.93,0.63,0.68}
\definecolor{LightPink3}{rgb}{0.80,0.55,0.58}
\definecolor{LightPink4}{rgb}{0.54,0.37,0.39}
\definecolor{PaleVioletRed1}{rgb}{1.00,0.51,0.67}
\definecolor{PaleVioletRed2}{rgb}{0.93,0.47,0.62}
\definecolor{PaleVioletRed3}{rgb}{0.80,0.41,0.54}
\definecolor{PaleVioletRed4}{rgb}{0.54,0.28,0.36}
\definecolor{maroon1}{rgb}{1.00,0.20,0.70}
\definecolor{maroon2}{rgb}{0.93,0.19,0.65}
\definecolor{maroon3}{rgb}{0.80,0.16,0.56}
\definecolor{maroon4}{rgb}{0.54,0.11,0.38}
\definecolor{VioletRed1}{rgb}{1.00,0.24,0.59}
\definecolor{VioletRed2}{rgb}{0.93,0.23,0.55}
\definecolor{VioletRed3}{rgb}{0.80,0.20,0.47}
\definecolor{VioletRed4}{rgb}{0.54,0.13,0.32}
\definecolor{magenta1}{rgb}{1.00,0.00,1.00}
\definecolor{magenta2}{rgb}{0.93,0.00,0.93}
\definecolor{magenta3}{rgb}{0.80,0.00,0.80}
\definecolor{magenta4}{rgb}{0.54,0.00,0.54}
\definecolor{orchid1}{rgb}{1.00,0.51,0.98}
\definecolor{orchid2}{rgb}{0.93,0.48,0.91}
\definecolor{orchid3}{rgb}{0.80,0.41,0.79}
\definecolor{orchid4}{rgb}{0.54,0.28,0.54}
\definecolor{plum1}{rgb}{1.00,0.73,1.00}
\definecolor{plum2}{rgb}{0.93,0.68,0.93}
\definecolor{plum3}{rgb}{0.80,0.59,0.80}
\definecolor{plum4}{rgb}{0.54,0.40,0.54}
\definecolor{MediumOrchid1}{rgb}{0.88,0.40,1.00}
\definecolor{MediumOrchid2}{rgb}{0.82,0.37,0.93}
\definecolor{MediumOrchid3}{rgb}{0.70,0.32,0.80}
\definecolor{MediumOrchid4}{rgb}{0.48,0.21,0.54}
\definecolor{DarkOrchid1}{rgb}{0.75,0.24,1.00}
\definecolor{DarkOrchid2}{rgb}{0.70,0.23,0.93}
\definecolor{DarkOrchid3}{rgb}{0.60,0.20,0.80}
\definecolor{DarkOrchid4}{rgb}{0.41,0.13,0.54}
\definecolor{purple1}{rgb}{0.61,0.19,1.00}
\definecolor{purple2}{rgb}{0.57,0.17,0.93}
\definecolor{purple3}{rgb}{0.49,0.15,0.80}
\definecolor{purple4}{rgb}{0.33,0.10,0.54}
\definecolor{MediumPurple1}{rgb}{0.67,0.51,1.00}
\definecolor{MediumPurple2}{rgb}{0.62,0.47,0.93}
\definecolor{MediumPurple3}{rgb}{0.54,0.41,0.80}
\definecolor{MediumPurple4}{rgb}{0.36,0.28,0.54}
\definecolor{thistle1}{rgb}{1.00,0.88,1.00}
\definecolor{thistle2}{rgb}{0.93,0.82,0.93}
\definecolor{thistle3}{rgb}{0.80,0.71,0.80}
\definecolor{thistle4}{rgb}{0.54,0.48,0.54}
\definecolor{gray0}{rgb}{0.00,0.00,0.00}
\definecolor{grey0}{rgb}{0.00,0.00,0.00}
\definecolor{gray1}{rgb}{0.01,0.01,0.01}
\definecolor{grey1}{rgb}{0.01,0.01,0.01}
\definecolor{gray2}{rgb}{0.02,0.02,0.02}
\definecolor{grey2}{rgb}{0.02,0.02,0.02}
\definecolor{gray3}{rgb}{0.03,0.03,0.03}
\definecolor{grey3}{rgb}{0.03,0.03,0.03}
\definecolor{gray4}{rgb}{0.04,0.04,0.04}
\definecolor{grey4}{rgb}{0.04,0.04,0.04}
\definecolor{gray5}{rgb}{0.05,0.05,0.05}
\definecolor{grey5}{rgb}{0.05,0.05,0.05}
\definecolor{gray6}{rgb}{0.06,0.06,0.06}
\definecolor{grey6}{rgb}{0.06,0.06,0.06}
\definecolor{gray7}{rgb}{0.07,0.07,0.07}
\definecolor{grey7}{rgb}{0.07,0.07,0.07}
\definecolor{gray8}{rgb}{0.08,0.08,0.08}
\definecolor{grey8}{rgb}{0.08,0.08,0.08}
\definecolor{gray9}{rgb}{0.09,0.09,0.09}
\definecolor{grey9}{rgb}{0.09,0.09,0.09}
\definecolor{gray10}{rgb}{0.10,0.10,0.10}
\definecolor{grey10}{rgb}{0.10,0.10,0.10}
\definecolor{gray11}{rgb}{0.11,0.11,0.11}
\definecolor{grey11}{rgb}{0.11,0.11,0.11}
\definecolor{gray12}{rgb}{0.12,0.12,0.12}
\definecolor{grey12}{rgb}{0.12,0.12,0.12}
\definecolor{gray13}{rgb}{0.13,0.13,0.13}
\definecolor{grey13}{rgb}{0.13,0.13,0.13}
\definecolor{gray14}{rgb}{0.14,0.14,0.14}
\definecolor{grey14}{rgb}{0.14,0.14,0.14}
\definecolor{gray15}{rgb}{0.15,0.15,0.15}
\definecolor{grey15}{rgb}{0.15,0.15,0.15}
\definecolor{gray16}{rgb}{0.16,0.16,0.16}
\definecolor{grey16}{rgb}{0.16,0.16,0.16}
\definecolor{gray17}{rgb}{0.17,0.17,0.17}
\definecolor{grey17}{rgb}{0.17,0.17,0.17}
\definecolor{gray18}{rgb}{0.18,0.18,0.18}
\definecolor{grey18}{rgb}{0.18,0.18,0.18}
\definecolor{gray19}{rgb}{0.19,0.19,0.19}
\definecolor{grey19}{rgb}{0.19,0.19,0.19}
\definecolor{gray20}{rgb}{0.20,0.20,0.20}
\definecolor{grey20}{rgb}{0.20,0.20,0.20}
\definecolor{gray21}{rgb}{0.21,0.21,0.21}
\definecolor{grey21}{rgb}{0.21,0.21,0.21}
\definecolor{gray22}{rgb}{0.22,0.22,0.22}
\definecolor{grey22}{rgb}{0.22,0.22,0.22}
\definecolor{gray23}{rgb}{0.23,0.23,0.23}
\definecolor{grey23}{rgb}{0.23,0.23,0.23}
\definecolor{gray24}{rgb}{0.24,0.24,0.24}
\definecolor{grey24}{rgb}{0.24,0.24,0.24}
\definecolor{gray25}{rgb}{0.25,0.25,0.25}
\definecolor{grey25}{rgb}{0.25,0.25,0.25}
\definecolor{gray26}{rgb}{0.26,0.26,0.26}
\definecolor{grey26}{rgb}{0.26,0.26,0.26}
\definecolor{gray27}{rgb}{0.27,0.27,0.27}
\definecolor{grey27}{rgb}{0.27,0.27,0.27}
\definecolor{gray28}{rgb}{0.28,0.28,0.28}
\definecolor{grey28}{rgb}{0.28,0.28,0.28}
\definecolor{gray29}{rgb}{0.29,0.29,0.29}
\definecolor{grey29}{rgb}{0.29,0.29,0.29}
\definecolor{gray30}{rgb}{0.30,0.30,0.30}
\definecolor{grey30}{rgb}{0.30,0.30,0.30}
\definecolor{gray31}{rgb}{0.31,0.31,0.31}
\definecolor{grey31}{rgb}{0.31,0.31,0.31}
\definecolor{gray32}{rgb}{0.32,0.32,0.32}
\definecolor{grey32}{rgb}{0.32,0.32,0.32}
\definecolor{gray33}{rgb}{0.33,0.33,0.33}
\definecolor{grey33}{rgb}{0.33,0.33,0.33}
\definecolor{gray34}{rgb}{0.34,0.34,0.34}
\definecolor{grey34}{rgb}{0.34,0.34,0.34}
\definecolor{gray35}{rgb}{0.35,0.35,0.35}
\definecolor{grey35}{rgb}{0.35,0.35,0.35}
\definecolor{gray36}{rgb}{0.36,0.36,0.36}
\definecolor{grey36}{rgb}{0.36,0.36,0.36}
\definecolor{gray37}{rgb}{0.37,0.37,0.37}
\definecolor{grey37}{rgb}{0.37,0.37,0.37}
\definecolor{gray38}{rgb}{0.38,0.38,0.38}
\definecolor{grey38}{rgb}{0.38,0.38,0.38}
\definecolor{gray39}{rgb}{0.39,0.39,0.39}
\definecolor{grey39}{rgb}{0.39,0.39,0.39}
\definecolor{gray40}{rgb}{0.40,0.40,0.40}
\definecolor{grey40}{rgb}{0.40,0.40,0.40}
\definecolor{gray41}{rgb}{0.41,0.41,0.41}
\definecolor{grey41}{rgb}{0.41,0.41,0.41}
\definecolor{gray42}{rgb}{0.42,0.42,0.42}
\definecolor{grey42}{rgb}{0.42,0.42,0.42}
\definecolor{gray43}{rgb}{0.43,0.43,0.43}
\definecolor{grey43}{rgb}{0.43,0.43,0.43}
\definecolor{gray44}{rgb}{0.44,0.44,0.44}
\definecolor{grey44}{rgb}{0.44,0.44,0.44}
\definecolor{gray45}{rgb}{0.45,0.45,0.45}
\definecolor{grey45}{rgb}{0.45,0.45,0.45}
\definecolor{gray46}{rgb}{0.46,0.46,0.46}
\definecolor{grey46}{rgb}{0.46,0.46,0.46}
\definecolor{gray47}{rgb}{0.47,0.47,0.47}
\definecolor{grey47}{rgb}{0.47,0.47,0.47}
\definecolor{gray48}{rgb}{0.48,0.48,0.48}
\definecolor{grey48}{rgb}{0.48,0.48,0.48}
\definecolor{gray49}{rgb}{0.49,0.49,0.49}
\definecolor{grey49}{rgb}{0.49,0.49,0.49}
\definecolor{gray50}{rgb}{0.50,0.50,0.50}
\definecolor{grey50}{rgb}{0.50,0.50,0.50}
\definecolor{gray51}{rgb}{0.51,0.51,0.51}
\definecolor{grey51}{rgb}{0.51,0.51,0.51}
\definecolor{gray52}{rgb}{0.52,0.52,0.52}
\definecolor{grey52}{rgb}{0.52,0.52,0.52}
\definecolor{gray53}{rgb}{0.53,0.53,0.53}
\definecolor{grey53}{rgb}{0.53,0.53,0.53}
\definecolor{gray54}{rgb}{0.54,0.54,0.54}
\definecolor{grey54}{rgb}{0.54,0.54,0.54}
\definecolor{gray55}{rgb}{0.55,0.55,0.55}
\definecolor{grey55}{rgb}{0.55,0.55,0.55}
\definecolor{gray56}{rgb}{0.56,0.56,0.56}
\definecolor{grey56}{rgb}{0.56,0.56,0.56}
\definecolor{gray57}{rgb}{0.57,0.57,0.57}
\definecolor{grey57}{rgb}{0.57,0.57,0.57}
\definecolor{gray58}{rgb}{0.58,0.58,0.58}
\definecolor{grey58}{rgb}{0.58,0.58,0.58}
\definecolor{gray59}{rgb}{0.59,0.59,0.59}
\definecolor{grey59}{rgb}{0.59,0.59,0.59}
\definecolor{gray60}{rgb}{0.60,0.60,0.60}
\definecolor{grey60}{rgb}{0.60,0.60,0.60}
\definecolor{gray61}{rgb}{0.61,0.61,0.61}
\definecolor{grey61}{rgb}{0.61,0.61,0.61}
\definecolor{gray62}{rgb}{0.62,0.62,0.62}
\definecolor{grey62}{rgb}{0.62,0.62,0.62}
\definecolor{gray63}{rgb}{0.63,0.63,0.63}
\definecolor{grey63}{rgb}{0.63,0.63,0.63}
\definecolor{gray64}{rgb}{0.64,0.64,0.64}
\definecolor{grey64}{rgb}{0.64,0.64,0.64}
\definecolor{gray65}{rgb}{0.65,0.65,0.65}
\definecolor{grey65}{rgb}{0.65,0.65,0.65}
\definecolor{gray66}{rgb}{0.66,0.66,0.66}
\definecolor{grey66}{rgb}{0.66,0.66,0.66}
\definecolor{gray67}{rgb}{0.67,0.67,0.67}
\definecolor{grey67}{rgb}{0.67,0.67,0.67}
\definecolor{gray68}{rgb}{0.68,0.68,0.68}
\definecolor{grey68}{rgb}{0.68,0.68,0.68}
\definecolor{gray69}{rgb}{0.69,0.69,0.69}
\definecolor{grey69}{rgb}{0.69,0.69,0.69}
\definecolor{gray70}{rgb}{0.70,0.70,0.70}
\definecolor{grey70}{rgb}{0.70,0.70,0.70}
\definecolor{gray71}{rgb}{0.71,0.71,0.71}
\definecolor{grey71}{rgb}{0.71,0.71,0.71}
\definecolor{gray72}{rgb}{0.72,0.72,0.72}
\definecolor{grey72}{rgb}{0.72,0.72,0.72}
\definecolor{gray73}{rgb}{0.73,0.73,0.73}
\definecolor{grey73}{rgb}{0.73,0.73,0.73}
\definecolor{gray74}{rgb}{0.74,0.74,0.74}
\definecolor{grey74}{rgb}{0.74,0.74,0.74}
\definecolor{gray75}{rgb}{0.75,0.75,0.75}
\definecolor{grey75}{rgb}{0.75,0.75,0.75}
\definecolor{gray76}{rgb}{0.76,0.76,0.76}
\definecolor{grey76}{rgb}{0.76,0.76,0.76}
\definecolor{gray77}{rgb}{0.77,0.77,0.77}
\definecolor{grey77}{rgb}{0.77,0.77,0.77}
\definecolor{gray78}{rgb}{0.78,0.78,0.78}
\definecolor{grey78}{rgb}{0.78,0.78,0.78}
\definecolor{gray79}{rgb}{0.79,0.79,0.79}
\definecolor{grey79}{rgb}{0.79,0.79,0.79}
\definecolor{gray80}{rgb}{0.80,0.80,0.80}
\definecolor{grey80}{rgb}{0.80,0.80,0.80}
\definecolor{gray81}{rgb}{0.81,0.81,0.81}
\definecolor{grey81}{rgb}{0.81,0.81,0.81}
\definecolor{gray82}{rgb}{0.82,0.82,0.82}
\definecolor{grey82}{rgb}{0.82,0.82,0.82}
\definecolor{gray83}{rgb}{0.83,0.83,0.83}
\definecolor{grey83}{rgb}{0.83,0.83,0.83}
\definecolor{gray84}{rgb}{0.84,0.84,0.84}
\definecolor{grey84}{rgb}{0.84,0.84,0.84}
\definecolor{gray85}{rgb}{0.85,0.85,0.85}
\definecolor{grey85}{rgb}{0.85,0.85,0.85}
\definecolor{gray86}{rgb}{0.86,0.86,0.86}
\definecolor{grey86}{rgb}{0.86,0.86,0.86}
\definecolor{gray87}{rgb}{0.87,0.87,0.87}
\definecolor{grey87}{rgb}{0.87,0.87,0.87}
\definecolor{gray88}{rgb}{0.88,0.88,0.88}
\definecolor{grey88}{rgb}{0.88,0.88,0.88}
\definecolor{gray89}{rgb}{0.89,0.89,0.89}
\definecolor{grey89}{rgb}{0.89,0.89,0.89}
\definecolor{gray90}{rgb}{0.89,0.89,0.89}
\definecolor{grey90}{rgb}{0.89,0.89,0.89}
\definecolor{gray91}{rgb}{0.91,0.91,0.91}
\definecolor{grey91}{rgb}{0.91,0.91,0.91}
\definecolor{gray92}{rgb}{0.92,0.92,0.92}
\definecolor{grey92}{rgb}{0.92,0.92,0.92}
\definecolor{gray93}{rgb}{0.93,0.93,0.93}
\definecolor{grey93}{rgb}{0.93,0.93,0.93}
\definecolor{gray94}{rgb}{0.94,0.94,0.94}
\definecolor{grey94}{rgb}{0.94,0.94,0.94}
\definecolor{gray95}{rgb}{0.95,0.95,0.95}
\definecolor{grey95}{rgb}{0.95,0.95,0.95}
\definecolor{gray96}{rgb}{0.96,0.96,0.96}
\definecolor{grey96}{rgb}{0.96,0.96,0.96}
\definecolor{gray97}{rgb}{0.96,0.96,0.96}
\definecolor{grey97}{rgb}{0.96,0.96,0.96}
\definecolor{gray98}{rgb}{0.98,0.98,0.98}
\definecolor{grey98}{rgb}{0.98,0.98,0.98}
\definecolor{gray99}{rgb}{0.98,0.98,0.98}
\definecolor{grey99}{rgb}{0.98,0.98,0.98}
\definecolor{gray100}{rgb}{1.00,1.00,1.00}
\definecolor{grey100}{rgb}{1.00,1.00,1.00}

%% file: copiathesis.bbl
\begin{thebibliography}{99}

\bibitem{Connes}A.~Connes,
\it ``Noncommutative Geometry''. \rm San Diego: Acad. Pr. (1994).

\bibitem{dopli}
S.~Doplicher, K.~Fredenhagen and J.~E.~Roberts,
Commun.\ Math.\ Phys.\  {\bf 172} (1995) 187.

\bibitem{cds}
A.~Connes, M.~R.~Douglas and A.~Schwarz,
JHEP {\bf 9802} (1998) 003
[arXiv:hep-th/9711162].

\bibitem{dh}
M.~R.~Douglas and C.~M.~Hull,
JHEP {\bf 9802} (1998) 008
[arXiv:hep-th/9711165].

\bibitem{sw}
N.~Seiberg and E.~Witten,
JHEP {\bf 9909} (1999) 032
[arXiv:hep-th/9908142].

\bibitem{tasi}
J.~Polchinski,
\it ``TASI Lectures on D-branes'',\rm
arXiv:hep-th/9611050.

\bibitem{Harveyrev}
J.~A.~Harvey,
\it ``Komaba lectures on noncommutative solitons and D-branes,'' \rm
arXiv:hep-th/0102076.

\bibitem{Douglasrev}
M.~R.~Douglas and N.~A.~Nekrasov,
Rev.\ Mod.\ Phys.\  {\bf 73} (2001) 977
[arXiv:hep-th/0106048].

\bibitem{Szaborev}
R.~J.~Szabo,
\it ``Quantum field theory on noncommutative spaces,'' \rm
arXiv:hep-th/0109162.

\bibitem{Witten}
E.~Witten,
Nucl.\ Phys.\ B {\bf 460} (1996) 335
[arXiv:hep-th/9510135].

\bibitem{ft}
E.~S.~Fradkin and A.~A.~Tseytlin,
Phys.\ Lett.\ B {\bf 163} (1985) 123.

\bibitem{acny}
A.~Abouelsaood, C.~G.~Callan, C.~R.~Nappi and S.~A.~Yost,
Nucl.\ Phys.\ B {\bf 280} (1987) 599.

\bibitem{clny}
C.~G.~Callan, C.~Lovelace, C.~R.~Nappi and S.~A.~Yost,
Nucl.\ Phys.\ B {\bf 288} (1987) 525.

\bibitem{Schom}
V.~Schomerus,
JHEP {\bf 9906} (1999) 030
[arXiv:hep-th/9903205].

\bibitem{G}
H.~J.~Groenewold,
Physica {\bf 12} (1946) 405.

\bibitem{M}
J.~E.~Moyal,
Proc.\ Cambridge Phil.\ Soc.\  {\bf 45} (1949) 99.

\bibitem{ch}
C.~S.~Chu and P.~M.~Ho,
Nucl.\ Phys.\ B {\bf 550} (1999) 151
[arXiv:hep-th/9812219].

\bibitem{chka}
C.~S.~Chu, P.~M.~Ho and Y.~C.~Kao,
Phys.\ Rev.\ D {\bf 60} (1999) 126003
[arXiv:hep-th/9904133].

\bibitem{aas}
F.~Ardalan, H.~Arfaei and M.~M.~Sheikh-Jabbari,
Nucl.\ Phys.\ B {\bf 576} (2000) 578
[arXiv:hep-th/9906161].

\bibitem{ch2}
C.~S.~Chu and P.~M.~Ho,
Nucl.\ Phys.\ B {\bf 568} (2000) 447
[arXiv:hep-th/9906192].

\bibitem{c}
C.~S.~Chu,
\it ``Noncommutative open string: Neutral and charged,'' \rm
arXiv:hep-th/0001144.

\bibitem{dir}
P.A.M.~Dirac
``Lectures on Quantum Mechanics''
Yeshiva University, 1964

\bibitem{nakoji}
N.~Nakanishi and I.~Ojima,
\it``Covariant Operator Formalism of Gauge Theories And Quantum Gravity''. \rm
 World Sci.\ Lect.\ Notes Phys.\  {\bf 27} (1990).

\bibitem{kar}
S.~Kar,
\it ``D-branes, cyclic symmetry and noncommutative geometry,'' \rm
arXiv:hep-th/0006073.

\bibitem{witt}
E.~Witten,     
Int.\ J.\ Mod.\ Phys.\ A {\bf 16} (2001) 693
[arXiv:hep-th/0007175].

\bibitem{mvs}
S.~Minwalla, M.~Van Raamsdonk and N.~Seiberg,
JHEP {\bf 0002} (2000) 020
[arXiv:hep-th/9912072].

\bibitem{alvawadia}
L.~Alvarez-Gaume and S.~R.~Wadia,
Phys.\ Lett.\ B {\bf 501} (2001) 319
[arXiv:hep-th/0006219].

\bibitem{gronekra}
D.~J.~Gross and N.~A.~Nekrasov,
JHEP {\bf 0103} (2001) 044
[arXiv:hep-th/0010090].

\bibitem{Filk}
T.~Filk,
Phys.\ Lett.\ B {\bf 376} (1996) 53.

\bibitem{mst}
A.~Matusis, L.~Susskind and N.~Toumbas,
JHEP {\bf 0012} (2000) 002
[arXiv:hep-th/0002075].

\bibitem{armoni}
A.~Armoni,
Nucl.\ Phys.\ B {\bf 593} (2001) 229
[arXiv:hep-th/0005208].

\bibitem{Sheikh}
M.~M.~Sheikh-Jabbari,
JHEP {\bf 9906} (1999) 015
[arXiv:hep-th/9903107].

\bibitem{Chepelev}
I.~Chepelev and R.~Roiban,
JHEP {\bf 0005} (2000) 037
[arXiv:hep-th/9911098].

\bibitem{shmic}
A.~Micu and M.~M.~Sheikh Jabbari,
JHEP {\bf 0101} (2001) 025
[arXiv:hep-th/0008057].

\bibitem{Chepelev2}
I.~Chepelev and R.~Roiban,
JHEP {\bf 0103} (2001) 001
[arXiv:hep-th/0008090].

\bibitem{bonorasali}
L.~Bonora and M.~Salizzoni,
Phys.\ Lett.\ B {\bf 504} (2001) 80
[arXiv:hep-th/0011088].

\bibitem{martin}
C.~P.~Martin and D.~Sanchez-Ruiz,
Nucl.\ Phys.\ B {\bf 598} (2001) 348
[arXiv:hep-th/0012024].

\bibitem{lucamass}
L.~Griguolo and M.~Pietroni,
Phys.\ Rev.\ Lett.\  {\bf 88} (2002) 071601
[arXiv:hep-th/0102070].

\bibitem{lucamass2}
L.~Griguolo and M.~Pietroni,
JHEP {\bf 0105} (2001) 032
[arXiv:hep-th/0104217].

\bibitem{Becchi}
C.~Becchi, S.~Giusto and C.~Imbimbo,
Nucl.\ Phys.\ B {\bf 633} (2002) 250
[arXiv:hep-th/0202155].

\bibitem{gsw}
M.~B.~Green, J.~H.~Schwarz and E.~Witten,
\it ``Superstring Theory''. \rm Vol.1 and Vol.2 Cambridge, Uk: Univ. Pr. ( 1987). 

\bibitem{sst}
N.~Seiberg, L.~Susskind and N.~Toumbas,
JHEP {\bf 0006} (2000) 044
[arXiv:hep-th/0005015].

\bibitem{rey}
S.~J.~Rey,
\it ``Exact answers to approximate questions: Noncommutative dipoles, open  Wilson lines, and UV-IR duality,'' \rm
arXiv:hep-th/0207108.

\bibitem{rey1}
Y.~Kiem, S.~J.~Rey, H.~T.~Sato and J.~T.~Yee,
Eur.\ Phys.\ J.\ C {\bf 22} (2002) 757
[arXiv:hep-th/0107106].

\bibitem{rey2}
Y.~j.~Kiem, S.~S.~Kim, S.~J.~Rey and H.~T.~Sato,
Nucl.\ Phys.\ B {\bf 641} (2002) 256
[arXiv:hep-th/0110066].

\bibitem{rey3}
Y.~Kiem, S.~Lee, S.~J.~Rey and H.~T.~Sato,
Phys.\ Rev.\ D {\bf 65} (2002) 046003
[arXiv:hep-th/0110215].

\bibitem{sibold}
Y.~Liao and K.~Sibold,
Eur.\ Phys.\ J.\ C {\bf 25} (2002) 469
[arXiv:hep-th/0205269].

\bibitem{gm}
J.~Gomis and T.~Mehen,
Nucl.\ Phys.\ B {\bf 591} (2000) 265
[arXiv:hep-th/0005129].

\bibitem{bgnv}
A.~Bassetto, L.~Griguolo, G.~Nardelli and F.~Vian,
JHEP {\bf 0107} (2001) 008
[arXiv:hep-th/0105257].

\bibitem{rr}
F.~R.~Ruiz,
Phys.\ Lett.\ B {\bf 502} (2001) 274
[arXiv:hep-th/0012171].

\bibitem{Matsubara}
K.~Matsubara,
Phys.\ Lett.\ B {\bf 482} (2000) 417
[arXiv:hep-th/0003294].

\bibitem{BonoraSchnabl}
L.~Bonora, M.~Schnabl, M.~M.~Sheikh-Jabbari and A.~Tomasiello,
Nucl.\ Phys.\ B {\bf 589} (2000) 461
[arXiv:hep-th/0006091].

\bibitem{Jurco}
B.~Jurco, S.~Schraml, P.~Schupp and J.~Wess,
Eur.\ Phys.\ J.\ C {\bf 17} (2000) 521
[arXiv:hep-th/0006246].

\bibitem{Bars}
I.~Bars, M.~M.~Sheikh-Jabbari and M.~A.~Vasiliev,
Phys.\ Rev.\ D {\bf 64} (2001) 086004
[arXiv:hep-th/0103209].

\bibitem{Chaichian}
M.~Chaichian, P.~Presnajder, M.~M.~Sheikh-Jabbari and A.~Tureanu,
Phys.\ Lett.\ B {\bf 526} (2002) 132
[arXiv:hep-th/0107037].

\bibitem{Schucker}
T.~Schucker,
\it ``Geometries and forces,'' \rm
arXiv:hep-th/9712095.

\bibitem{StMod}
M.~Chaichian, P.~Presnajder, M.~M.~Sheikh-Jabbari and A.~Tureanu,
\it ``Noncommutative standard model: Model building,'' \rm
arXiv:hep-th/0107055.

\bibitem{iikk}
N.~Ishibashi, S.~Iso, H.~Kawai and Y.~Kitazawa,
Nucl.\ Phys.\ B {\bf 573} (2000) 573
[arXiv:hep-th/9910004].

\bibitem{amns}
J.~Ambjorn, Y.~M.~Makeenko, J.~Nishimura and R.~J.~Szabo,
JHEP {\bf 9911} (1999) 029
[arXiv:hep-th/9911041].

\bibitem{amns2}
J.~Ambjorn, Y.~M.~Makeenko, J.~Nishimura and R.~J.~Szabo,
Phys.\ Lett.\ B {\bf 480} (2000) 399
[arXiv:hep-th/0002158].

\bibitem{amns3}
J.~Ambjorn, Y.~M.~Makeenko, J.~Nishimura and R.~J.~Szabo,
JHEP {\bf 0005} (2000) 023
[arXiv:hep-th/0004147].

\bibitem{Gross}
D.~J.~Gross, A.~Hashimoto and N.~Itzhaki,
Adv.\ Theor.\ Math.\ Phys.\  {\bf 4} (2000) 893
[arXiv:hep-th/0008075].

\bibitem{bv}
A.~Bassetto and F.~Vian,
JHEP {\bf 0210} (2002) 004
[arXiv:hep-th/0207222].

\bibitem{nostro1}
A.~Bassetto, G.~Nardelli and A.~Torrielli,
Nucl.\ Phys.\ B {\bf 617} (2001) 308
[arXiv:hep-th/0107147].

\bibitem{nostro2}
A.~Bassetto, G.~Nardelli and A.~Torrielli,
Phys.\ Rev.\ D {\bf 66} (2002) 085012
[arXiv:hep-th/0205210].

\bibitem{sheikh}
A.~Das and M.M.~Sheikh-Jabbari, 
JHEP {\bf 0106} (2001) 028
[arXiv:hep-th/0103139].

\bibitem{hoo}
G.~'t Hooft, Nucl.\ Phys.\ {\bf B75} (1974) 461.

\bibitem{w}
T.T.~Wu, Phys.\ Lett.\ {\bf 71B} (1977) 142.

\bibitem{m}
S.~Mandelstam, Nucl.\
Phys.\ {\bf B213} (1983) 149. 

\bibitem{le}
G.~Leibbrandt, Phys.\ Rev.\ {\bf D29} (1984) 1699.

\bibitem{bbg}
A.~Bassetto, F.~De Biasio and L.~Griguolo, Phys. Rev. Lett. {\bf 72}
(1994) 3141
[arXiv:hep-th/9402029].

\bibitem{boul}
D.V.~Boulatov, Mod. Phys. Lett. {\bf A9} (1994) 365
[arXiv:hep-th/9310041].

\bibitem{daul}
J-M.~Daul and V. A.~Kazakov, Phys. Lett. {\bf B335} (1994) 371
[arXiv:hep-th/9310165].

\bibitem{test}
A.~Bassetto, G.~Nardelli and R.~Soldati, {\it ``Yang-Mills Theories
in Algebraic Non-covariant Gauges''}, World Scientific, Singapore 1991.

\bibitem{stau}
M.~Staudacher and W.~Krauth, Phys.\ Rev.\ {\bf D57} (1998) 2456
[arXiv:hep-th/9709101].

\bibitem{anlu}
A.~Bassetto and L.~Griguolo, Phys. Lett. {\bf B443} (1998) 325
[arXiv:hep-th/9806037].

\bibitem{QCD}
A.~Bassetto,
\it ``{QCD}: From four to two dimensions,'' \rm
arXiv:hep-th/9809084.

\bibitem{BGV}
A.~Bassetto, L.~Griguolo and F.~Vian, Nucl. Phys. {\bf B559} (1999)
563
[arXiv:hep-th/9906125].

\bibitem{prudni3}
A.P.~Prudnikov, Yu.A.~Brychkov, O.I.~Marichev, 
\it Integrals and Series, \rm Vol. III, Gordon and Breach Science Publishers (1990).

\bibitem{agm}
O.~Aharony, J.~Gomis and T.~Mehen,
JHEP {\bf 0009} (2000) 023
[arXiv:hep-th/0006236].

\bibitem{clz}
C.~S.~Chu, J.~Lukierski and W.~J.~Zakrzewski,
Nucl.\ Phys.\ B {\bf 632} (2002) 219
[arXiv:hep-th/0201144].

\bibitem{abz}
L.~Alvarez-Gaume, J.~L.~Barbon and R.~Zwicky,
JHEP {\bf 0105} (2001) 057
[arXiv:hep-th/0103069].

\bibitem{ssto}
N.~Seiberg, L.~Susskind and N.~Toumbas,
JHEP {\bf 0006} (2000) 021
[arXiv:hep-th/0005040].

\bibitem{gmms}
R.~Gopakumar, J.~M.~Maldacena, S.~Minwalla and A.~Strominger,
JHEP {\bf 0006} (2000) 036
[arXiv:hep-th/0005048].

\bibitem{kp}
S.~Kar and S.~Panda,
JHEP {\bf 0211} (2002) 052
[arXiv:hep-th/0205078].

\bibitem{b}
C.~P.~Burgess,
Nucl.\ Phys.\ B {\bf 294} (1987) 427.

\bibitem{n}
V.~V.~Nesterenko,
Int.\ J.\ Mod.\ Phys.\ A {\bf 4} (1989) 2627.

\bibitem{bp}
C.~Bachas and M.~Porrati,
Phys.\ Lett.\ B {\bf 296} (1992) 77
[arXiv:hep-th/9209032].

\bibitem{ba}
C.~Bachas,
Phys.\ Lett.\ B {\bf 374} (1996) 37
[arXiv:hep-th/9511043].

\bibitem{mio}
A.~Torrielli,
\it ``Cutting rules and perturbative unitarity of noncommutative  electric-type field theories from string theory,'' \rm
arXiv:hep-th/0207148

\bibitem{bcr}
A.~Bilal, C.~S.~Chu and R.~Russo,
Nucl.\ Phys.\ B {\bf 582} (2000) 65
[arXiv:hep-th/0003180].

\bibitem{gkmrs}
J.~Gomis, M.~Kleban, T.~Mehen, M.~Rangamani and S.~H.~Shenker,
JHEP {\bf 0008} (2000) 011
[arXiv:hep-th/0003215].

\bibitem{ad}
O.~Andreev and H.~Dorn,
Nucl.\ Phys.\ B {\bf 583} (2000) 145
[arXiv:hep-th/0003113].

\bibitem{kl}
Y.~Kiem and S.~M.~Lee,
Nucl.\ Phys.\ B {\bf 586} (2000) 303
[arXiv:hep-th/0003145].

\bibitem{l}
H.~Liu and J.~Michelson,
Phys.\ Rev.\ D {\bf 62} (2000) 066003
[arXiv:hep-th/0004013].

\bibitem{crs}
C.~S.~Chu, R.~Russo and S.~Sciuto,
Nucl.\ Phys.\ B {\bf 585} (2000) 193
[arXiv:hep-th/0004183].

\bibitem{p}
J.~Polchinski,
\it ``String Theory''. \rm Vol. 1 Cambridge, UK: Univ. Pr. (1998).

\bibitem{seme}
E.~T.~Akhmedov, P.~DeBoer and G.~W.~Semenoff,
Phys.\ Rev.\ D {\bf 64} (2001) 065005
[arXiv:hep-th/0010003].

\bibitem{har}
J.A.~Harvey, {\it ``Topology of the Gauge Group in Noncommutative
Gauge Theory''}, hep-th/0105242.

\bibitem{sza}
L.D.~Paniak and R.J.~Szabo, {\it ``Instanton Expansion of Noncommutative
Gauge Theory in Two Dimensions''}, hep-th/0203166.

\bibitem{lu}
L.~Griguolo, D.~Seminara and P.~Valtancoli,
JHEP {\bf 0112} (2001) 024
[arXiv:hep-th/0110293].

\bibitem{bieten}
W.~Bietenholz, F.~Hofheinz and J.~Nishimura,
JHEP {\bf 0209} (2002) 009
[arXiv:hep-th/0203151].

\bibitem{prudni}
A.P.~Prudnikov, Yu.A.~Brychkov, O.I.~Marichev, 
\it Integrals and Series, \rm Vol. II, Gordon and Breach Science Publishers (1988).

\bibitem{dgo}
G.~De Risi, G.~Grignani and M.~Orselli,
JHEP {\bf 0212} (2002) 031
[arXiv:hep-th/0211056].

\bibitem{azam1}
M.~Azam,
Phys.\ Rev.\ D {\bf 35} (1987) 2043.

\bibitem{azam2}
M.~Azam,
Phys.\ Rev.\ D {\bf 43} (1991) 4148.

\bibitem{azam3}
M.~Azam,
arXiv:hep-th/0211256.

\bibitem{div1}
P.~Di Vecchia, A.~Lerda, L.~Magnea and R.~Marotta,
Phys.\ Lett.\ B {\bf 351} (1995) 445
[arXiv:hep-th/9502156].

\bibitem{div2}
P.~Di Vecchia, L.~Magnea, A.~Lerda, R.~Russo and R.~Marotta,
Nucl.\ Phys.\ B {\bf 469} (1996) 235
[arXiv:hep-th/9601143].

\bibitem{div3}
P.~Di Vecchia, L.~Magnea, A.~Lerda, R.~Marotta and R.~Russo,
Phys.\ Lett.\ B {\bf 388} (1996) 65
[arXiv:hep-th/9607141].

\bibitem{bern1}
Z.~Bern, D.~A.~Kosower and K.~Roland,
Nucl.\ Phys.\ B {\bf 334} (1990) 309.

\bibitem{bern2}
Z.~Bern and D.~A.~Kosower,
Phys.\ Rev.\ D {\bf 38} (1988) 1888.

\bibitem{bern3}
Z.~Bern and D.~A.~Kosower,
Nucl.\ Phys.\ B {\bf 379} (1992) 451.

\bibitem{Magro}
G.~Magro,
arXiv:quant-ph/0302001.

\bibitem{Witten2d}
E.~Witten,
Commun.\ Math.\ Phys.\  {\bf 141} (1991) 153.








\end{thebibliography}
